# 2024 Roadmap on Magnetic Microscopy Techniques and Their Applications in Materials Science

D. V. Christensen[1,2], U. Staub[3], T. R. Devidas[4], B. Kalisky[4], K. C. Nowack[5,6], J.L. Webb[7], U.L. Andersen[7], A. Huck[7], D. A. Broadway[8,9] K. Wagner[8], P. Maletinsky[8], T. van der Sar[10], C. R. Du[11], A. Yacoby[12], D. Collomb[13] S. Bending[14], A. Oral[15], H. J. Hug[16,8], A.-O. Mandru[16], V. Neu[17], H. W. Schumacher[18], S. Sievers[18], H. Saito[19], A.A. Khajetoorians[20], N. Hauptmann[20], S. Baumann[21], A. Eichler[22], C. L. Degen[22], J. McCord[23,24], M. Vogel[23,24], M. Fiebig[25], P. Fischer[26,27], A. Hierro-Rodriguez[28,29], S. Finizio[3], S. S. Dhesi[30], C. Donnelly[31], Felix Büttner[32,33], O. Kfir[34], W. Hu[35], S. Zayko[36], S. Eisebitt[37,38], B. Pfau[37], R. Frömter[39], M. Kläui[39], F. S. Yasin[40], B. J. McMorran[41], S. Seki[42], X. Yu[40], A. Lubk[17,43], D. Wolf[17], N. Pryds[1], D. Makarov[44], M. Poggio[8].

1 Department of Energy Conversion and Storage, Technical University of Denmark, DK-2800 Kgs. Lyngby, Denmark.

2 Email: Dennis Christensen, dechr@dtu.dk

3 Swiss Light Source, Paul Scherrer Institute, 5232 Villigen PSI, Switzerland

4 Department of Physics and Institute of Nanotechnology and Advanced Materials, Bar-Ilan University, Ramat Gan 5290002, Israel

5 Laboratory of Atomic and Solid-State Physics, Cornell University, Ithaca, New York 14853, USA

6 Kavli Institute at Cornell for Nanoscale Science, Ithaca, New York 14853, USA

7 Centre for Macroscopic Quantum States (bigQ), Department of Physics, Technical University of Denmark, 2800 Kongens Lyngby, Denmark

8 Department of Physics, University of Basel, 4056 Basel, Switzerland

9 School of Science, RMIT University, Melbourne, Victoria, Australia

10 Department of Quantum Nanoscience, Kavli Institute of Nanoscience, Delft University of Technology, Lorentzweg 1, 2628 CJ Delft, The Netherlands.

11 Department of Physics, University of California, San Diego, La Jolla, California 92093, USA

12 Department of Physics, Harvard University, 17 Oxford Street, Cambridge, Massachusetts 02138, USA.

13 Quantum Brilliance GmbH, Tullastraße 72, Freiburg im Breisgau, Germany

14 Department of Physics, University of Bath, Bath BA2 7AY, United Kingdom

15 NanoMagnetics Instruments Ltd, Suite 290, 266 Banbury Road, Oxford OX2 7DL, United Kingdom

16 Empa, Materials Science and Technology, Ueberlandstrasse 129, 8600 Duebendorf, Switzerland

17 Leibniz Institute for Solid State and Materials Research, D-01069 Dresden, Germany

18 Physikalisch-Technische Bundesanstalt, 38116 Braunschweig, Germany

19 Department of Mathematical Science and Electrical-Electronic-Computer Engineering, Graduate School of Engineering Science, Akita University, Akita, Japan

20 Scanning Probe Microscopy Department, Institute for Molecules and Materials, Radboud University, Nijmegen, The Netherlands

21 University of Stuttgart, Institute for Functional Matter and Quantum Technologies, Stuttgart, Germany

22 Laboratory for Solid State Physics, ETH Zürich, 8093 Zürich, Switzerland

23 Department for Materials Science, Kiel University, Kiel, Germany

24Kiel Nano, Surface and Interface Science (KiNSIS), Kiel University, Kiel, Germany

25 Department of Materials, ETH Zurich, Zurich, Switzerland

26 Materials Sciences Division, Lawrence Berkeley National Laboratory, Berkeley CA 94720, USA

27 Department of Physics, University of California Santa Cruz, Santa Cruz CA 95064, USA

28 Departamento de Física, Universidad de Oviedo, 33007 Oviedo, Spain

29 CINN (CSIC-Universidad de Oviedo), 33940, El Entrego, Spain

30 Diamond Light Source, Chilton, Didcot, Oxfordshire, OX11 0DE, UK

31 Max Planck Institute for Chemical Physics of Solids, Dresden, Germany

32 Helmholtz-Zentrum für Materialien und Energie, Berlin, Germany

33 University of Augsburg, Augsburg, Germany

34 School of Electrical Engineering, Tel Aviv University, Tel Aviv 69978, Israel

35 National Synchrotron Light Source II, Brookhaven National Laboratory, Upton, NY 11973

36 Max Planck Institute for Multidisciplinary Sciences, Göttingen, Germany

37 Max Born Institute, Max-Born-Str. 2a, 12489 Berlin, Germany

38 Institute of Optics and Atomic Physics, Technische Universität Berlin, Straße des 17. Juni 135, 10623 Berlin, Germany

39 Institute of Physics, Johannes Gutenberg-University Mainz, 55099 Mainz, Germany

40 RIKEN Center for Emergent Matter Science (CEMS), Wako, Japan

41 Department of Physics, University of Oregon, Eugene, Oregon, USA

42 Department of Applied Physics and Institute of Engineering Innovation, University of Tokyo, Tokyo, Japan

43 Institute of Solid State and Materials Physics, TU Dresden, 01062 Dresden, Germany

44 Helmholtz-Zentrum Dresden-Rossendorf e.V., Institute of Ion Beam Physics and Materials Research, 01328 Dresden, Germany.

**Abstract:**

*Considering the growing interest in magnetic materials for unconventional computing, data storage, and sensor applications, there is active research not only on material synthesis but also characterisation of their properties. In addition to structural and integral magnetic characterisations, imaging of magnetization patterns, current distributions and magnetic fields at nano- and microscale is of major importance to understand the material responses and qualify them for specific applications. In this roadmap, we aim to cover a broad portfolio of techniques to perform nano- and microscale magnetic imaging using SQUIDs, spin center and Hall effect magnetometries, scanning probe microscopies, x-ray- and electron-based methods as well as magnetooptics and nanoMRI. The roadmap is aimed as a single access point of information for experts in the field as well as the young generation of students outlining prospects of the development of magnetic imaging technologies for the upcoming decade with a focus on physics, materials science, and chemistry of planar, 3D and geometrically curved objects of different material classes including 2D materials, complex oxides, semi-metals, multiferroics, skyrmions, antiferromagnets, frustrated magnets, magnetic molecules/nanoparticles, ionic conductors, superconductors, spintronic and spinorbitronic materials.*

# Contents:

# Introduction

Dennis V. Christensen[1], Urs Staub[2], Nini Pryds[1], Denys Makarov[3], Martino Poggio[4]

*[1] Department of Energy Conversion and Storage, Technical University of Denmark, 2800 Kgs. Lyngby, Denmark*
*[2] Swiss Light Source, Paul Scherrer Institute, 5232 Villigen PSI, Switzerland*
*[3] Helmholtz-Zentrum Dresden-Rossendorf e.V., Institute of Ion Beam Physics and Materials Research, 01328 Dresden, Germany*
*[4] Department of Physics, University of Basel, 4056 Basel, Switzerland*

Today's devices rely on the flow of charge and the magnetization of materials for sensing, carrying out computations, and storing data. The continued pressure to make these devices smaller, more efficient, and more powerful is pushing researchers to explore the use of spin currents, quantum transport, nanoscale spin configurations, and various types of quantum bits. Here, magnetic microscopy techniques play a vital role in enabling researchers to visualize, manipulate, and understand magnetic and transport phenomena at the micro- and nanoscale, and eventually optimize materials and devices for specific applications. The broad portfolio of magnetic microscopy techniques allows for spatially resolving a plethora of functional properties, including imaging the magnetic landscape, visualizing complex current distributions, mapping the magnetic susceptibility, and detecting ultra-fast spin dynamics. Key applications in material science range from quantitatively mapping the flow in low-dimensional material systems, spatially resolving the superconducting fluid density, and disclosing the dynamics of spin-waves and spin textures. By utilizing the properties of magnetic fields to traverse through solid materials, several magnetic microscopy techniques offer the possibility to spatially resolve processes buried underneath protective layers or at the interface of heterostructures.

In this roadmap, experts in diverse areas of magnetic microscopy techniques have been invited to provide their views on the current capabilities of the techniques, forecast future technical advances, and assess how such advances may impact material science. Although magnetic microscopy techniques find use in a broad range of fields, this roadmap focuses on their use within material science and physics. The aim is to provide a single access point comprising forward-looking opinions on selected techniques rather than presenting a comprehensive account of all magnetic microscopy techniques. We hope that the roadmap will serve as a valuable resource for experts and students in magnetic imaging techniques and functional materials, enterprises developing or leveraging on the methods, readers from outside the field as well as those entering the field.

The roadmap divides the magnetic microscopy techniques into three major fields, namely techniques based on scanning probe microscopy, light, and electrons:

**Magnetic Microscopy based on Scanning Probes**:
Scanning tunneling microscopy (STM), introduced in 1981, produced the first real-space images of individual atoms on a surface. A boom in scanning probe microscopy (SPM) followed this, leading to various probes now being routinely used to form diverse images of surfaces by rastering the physical probe in close proximity to a sample while making a measurement at each position. In the area of magnetic imaging, some of the techniques combining the highest spatial resolution with the highest sensitivity are based on scanning probes. These include magnetic force microscopy, scanning superconducting quantum interference device (SQUID) microscopy, scanning Hall microscopy, scanning nitrogen-vacancy center microscopy, and spin-polarized scanning tunneling microscopy. These five techniques are covered in the subsequent sections of this roadmap. The techniques work under a wide variety of different conditions, and each is best suited to different types of samples. All are restricted to mapping contrast associated with the magnetic field or magnetization at the surface of a sample. As physically scanning a probe across a sample is typically slow, these techniques are best suited to 100 nm to 100 µm fields of view with resolutions down to the atomic scale in some cases. The spatial resolution is ultimately limited by a combination of the size of the scanning probe tip or sensor and the proximity of the latter to the sample surface. The high sensitivity and excellent spatial resolution enable, e.g., real-space imaging of uncompensated magnetization at the surface of antiferromagnets and visualization of the current distribution in the ballistic, hydrodynamic, or superconducting transport regimes. In addition, scanning probes are a convenient choice for measuring weakly magnetic materials with low magnetization volumes and buried magnetic textures, such as encapsulated 2D magnets.

While there is an astounding number of different types of scanning probe microscopies, the list is bound to grow in the coming years. The development of ever more advanced fabrication techniques, including focused ion beam milling and advanced lithography, allows for patterning ever more complex and compact sensing devices at the apex of scanning probes. These include devices such as the sub-micron-scale graphene Hall bars, robust nanometer-scale SQUIDs, and sharp diamond tips with stable, ultra-shallow nitrogen vacancies, as discussed later in this roadmap. This is complemented by the development of multimodal scanning probes where imaging of magnetic susceptibility and spontaneous magnetic order can be combined with the mapping of temperature profiles, electric fields, and current distributions. The new generation of scanning probes promises to enable new types of contrast, including magnetic contrast, that could help provide microscopic information about length scales, inhomogeneity, and interactions in emerging and poorly understood magnetic materials.

**Light-based Magnetic Microscopy**:

Conventional imaging based on magnetooptics using Kerr or Faraday effects provides easy access to magnetic textures with a spatial resolution of hundreds of nanometers and femtosecond temporal resolution in stroboscopic measurements. Current activities on the field- or current-driven evolution of magnetic textures in not only planar but also 3D magnetic architectures call for major improvements of the magnetooptic microscopies towards enhancement of contrast enabled by plasmon filtering and on-the-fly image analysis using GPU computing assisted by machine learning methods. The improvement in the signal-to-noise ratio should allow for single-shot magnetization dynamics studies of non-repetitive events, which are out of reach at present at short time scales. It is envisioned that combining magnetooptics with near-field microscopies will allow for routine measurements at sub-100 nm spatial resolution. These features will broaden the use of magnetooptic techniques for magnetic topological materials like Weyl semimetals, collinear and non-collinear antiferromagnets, as well as magnetoelectrics.

Frequency doubling of light in matter, termed "second harmonic generation" (SHG), is a powerful method for studying ordered states of matter based on the symmetry change accompanying the emergence of the order. The greatest potential of SHG microscopy is in the investigation of states with compensated magnetization, like antiferromagnets, states with coexisting magnetic and electric order, termed multiferroics, and the spatially resolved imaging of the associated domains. Magnetization processes are resolved with sub-picosecond accuracy, which is particularly useful in developing new magnetoelectronic or spintronic functionalities and devices. Based on the current development of SHG equipment, this technique could become a standard component of sample preparation equipment like deposition chambers. These so-called in situ SHG equipment will provide real-time feedback on growth conditions of magnetically and electrically active materials to facilitate the discovery or optimization of multiferroic and magnetoelectric materials for low-energy storage and computing devices. It is foreseeable that SHG and other nonlinear optical methods will be extended to THz frequencies, opening possibilities to probe elementary excitations in materials resonantly and thus very directly and effectively.

X-rays are powerful probes for magnetism with the strength of being chemically selective to the ions carrying the magnetic moment. This leads to additional sensitivity in both x-ray scattering and absorption processes, allowing the study of small amounts of materials. In addition, it enables very high spatial resolution down to the 10 nm range with sufficient signal strength to probe magnetism. Current imaging techniques cover direct methods such as x-ray photo-emission-electron microscopes (XPEEM) or scanning transmission x-ray microscopes (STXM) as well as indirect methods based on coherent x-ray scattering such as holography, ptychography, and coherent diffraction imaging (CDI). Current developments are concerned with in-situ imaging, imaging of 3-dimensional magnetization distributions in planar and geometrically curved magnets as well as extending x-ray magnetic imaging to antiferromagnets. Strong future improvements are anticipated from the huge enhancement of the x-ray coherence flux fraction obtained from the upgraded diffraction-limited light sources. In addition, the possibility to study dynamic stochastic processes by imaging on ultrafast timescales based on x-ray free-electron lasers and from laser-driven higher harmonic generation sources will give a completely new view of ultrafast magnetization with high spatial resolution.

**Electron-based Magnetic Microscopy**:
Imaging magnetization using electron microscopes allows for magnetic characterization with exceptional spatial resolution down to the atomic scale. The roadmap focuses on three electron-based magnetic imaging techniques, including Lorentz transmission electron microscopy (LTEM), electron holography, and scanning electron microscopy with polarization analysis (SEMPA). LTEM and its reciprocal technique Lorentz scanning transmission electron microscopy (LSTEM) are used to image nanometric magnetic domains or even track the response of a spin texture to external stimuli such as heat, electric currents, or magnetic fields. In addition to further improving the resolution, advances in cryogenic and environmental TEM holders and fast direct detector technology place L(S)TEM at the forefront of magnetic imaging techniques in the decades to come to probe nontrivial spin textures and their dynamics. Recent developments of objective lenses of STEM provide access to imaging of magnetic field distribution at the atomic scale, even in antiferromagnetic materials, which opens new avenues for developing and understanding antiferromagnetic spintronics and multiferroics.

Electron holography allows quantitative mapping of projected in-plane magnetic flux densities of magnetic 2D and 3D textures at spatial resolutions approaching atomic length scales. When combined with electron tomography, it is possible to properly reconstruct all Cartesian components of the flux density in 3D of complex-shaped magnetic architectures at the nanoscale, such as magnetic skyrmions, domain walls, and vortices. Ongoing activities extend this technique towards complex sample manipulation using temperature, currents, electric and magnetic fields, light, and strain to study transient material responses within, e.g., magnetostriction and magnetoionics. In this respect, further development focuses on increasing the time resolution of the electron holography.

Complementary to the transmission-based electron microscopy imaging techniques, SEMPA allows ultimate surface sensitivity to image two-component magnetization vector of magnetic textures at surfaces of ferromagnets and synthetic antiferromagnets. The recent method developments enable measurements of antiferromagnetic materials. The new advances in detectors with appropriate correctors for SEMPA are envisioned to improve the time resolution below 100 ps, provide imaging resolution better than 10 nm and enable studies of magnetic texture evolution in magnetic fields.

By addressing a range of key magnetic microscopy techniques where scanning probes, light, and electrons are used to examine and manipulate magnetization and charge transport in condensed matter, we intend to provide snapshots of the current state and future projections within magnetic imaging techniques. We hope the roadmap will inspire the further use and development of these techniques within material science and physics.

**Acknowledgment**:

D.V.C. acknowledges the support of Novo Nordisk Foundation NERD Programme: New Exploratory Research and Discovery, Superior Grant NNF21OC0068015. D.V.C. and N.P. acknowledge the support of Novo Nordisk Foundation Challenge Programme 2021: Smart nanomaterials for applications in life-science, BIOMAG Grant NNF21OC0066526. N.P. acknowledges funding from the Danish Council for Independent Research Technology and Production Sciences for the DFF- Research Project 3 (grant No 00069B) and the funding from the ERC Advanced "NEXUS" Grant 101054572.

# 1 - Scanning Superconducting Quantum Interference Device (SQUID) Microscopy

*T. R. Devidas[1], B. Kalisky[1], K. C. Nowack[2,3]*
[1]Department of Physics and Institute of Nanotechnology and Advanced Materials, Bar-Ilan University, Ramat Gan 5290002, Israel
[2]Laboratory of Atomic and Solid-State Physics, Cornell University, Ithaca, New York 14853, USA

[3]Kavli Institute at Cornell for Nanoscale Science, Ithaca, New York 14853, USA

**Current State of the Technique**

Scanning SQUID microscopy (SSM) is a versatile technique to study and image magnetic, superconducting and transport properties at micro and nano-scales. This technique enables sensitive, spatially resolved detection of stray magnetic fields above a sample, across a wide range of temperatures.

A DC SQUID sensor consists of a superconducting loop interrupted by two Josephson junctions (JJs). For bias currents exceeding the JJ's critical current, a DC voltage develops across the SQUID. When magnetic flux threads the SQUID loop a screening current in the loop maintains flux quantization within the loop. As a result, the critical current and the voltage across the SQUID are a periodic function of the applied flux with a period equal to the flux quantum, $\phi_0 = h/2e$, with $h$ Planck's constant, $e$ elementary charge [1].

Two key characteristics of a scanning SQUID sensor are flux sensitivity and spatial resolution. However, many other characteristics are of interest as well, including measurement bandwidth, ease-of-use, ease-of-fabrication, and functionality beyond magnetometry. The flux sensitivity is influenced by many factors, such as circuit and junction design, materials and the details of the scanning setup [2][3]. The spatial resolution is largely determined by geometry, i.e., the size of the sensitive area and the distance between the sample and the sensitive area.

There are currently two main approaches to SSM, which differ in geometry and functionality of the SQUID sensor. One approach is to fabricate planar large-scale superconducting devices in which the sensitive area is a small part of the overall SQUID and is spatially separated from the JJs on the chip (SQUID on chip, SOC). These SQUIDs are typically based on Nb/$AlO_x$/Nb trilayer JJ technology [4]. SOCs usually have micrometer-scale pickup loops and include on-chip circuitry such as flux modulation coils allowing for a flux-locked feedback loop to linearize the inherently nonlinear SQUID response, and an on-chip field coil concentric with the sensitive area, enabling local magnetic response measurements. Typical spatial resolution of SOC lies in the micrometer range [2].

An alternative approach to realize a scanning SQUID is to fabricate a small SQUID loop for example at the apex of a hollow quartz tube (SQUID on tip, SOT) [5][6] or, realized more recently, on an AFM cantilever (SQUID on lever, SOL) [7]. The JJs in these SQUIDs are realized in the form of Dayem bridges, which allows the fabrication of loops with sub-100 nm diameter while maintaining µAs of critical current. In SOTs, a quartz tube is pulled into a sharp pipette, and superconducting layers are deposited in three steps to define two superconducting leads connected to a superconducting loop. In SOLs, the loop is created via focused ion beam milling of a continuous superconducting thin film deposited on a blunted AFM cantilever. These geometries are suited for scanning in close proximity to a sample allowing for sub-micron spatial resolution. The SQUIDs have been made from different superconductors including Al, Nb, Pb, In, and MoRe which results in different performances in terms of flux sensitivity and continuous operation in magnetic field ranges up to several Tesla.

The flux sensitivity of SSM sensors ranges from a few hundred $n\phi_0/\sqrt{Hz}$ to a few $\mu\phi_0/\sqrt{Hz}$ depending on the operating frequency and details of the sensor. The sensitivity to different sources of magnetic field then depends on how much flux is coupled into the SQUID, which in turn is determined by the SQUID geometry and the spatial dependence of the magnetic field lines produced by the source [2][8].

The basic operation mode of SSM is magnetometry (Fig. 1d.), which captures the static magnetic flux over a plane close to the sample. Susceptometry map is acquired by recording the local magnetic response of the sample to a varying local magnetic field applied via the on-chip field coil (Fig. 1e.). The spatial distribution of electric current can be mapped by SSM by recording the magnetic fields generated by the current flow in a sample (Fig. 1f.). Operating the last two with alternate excitations allows measuring all modes simultaneously. Mechanical vibrations at a different frequency can be used to improve DC magnetometry sensitivity [9].

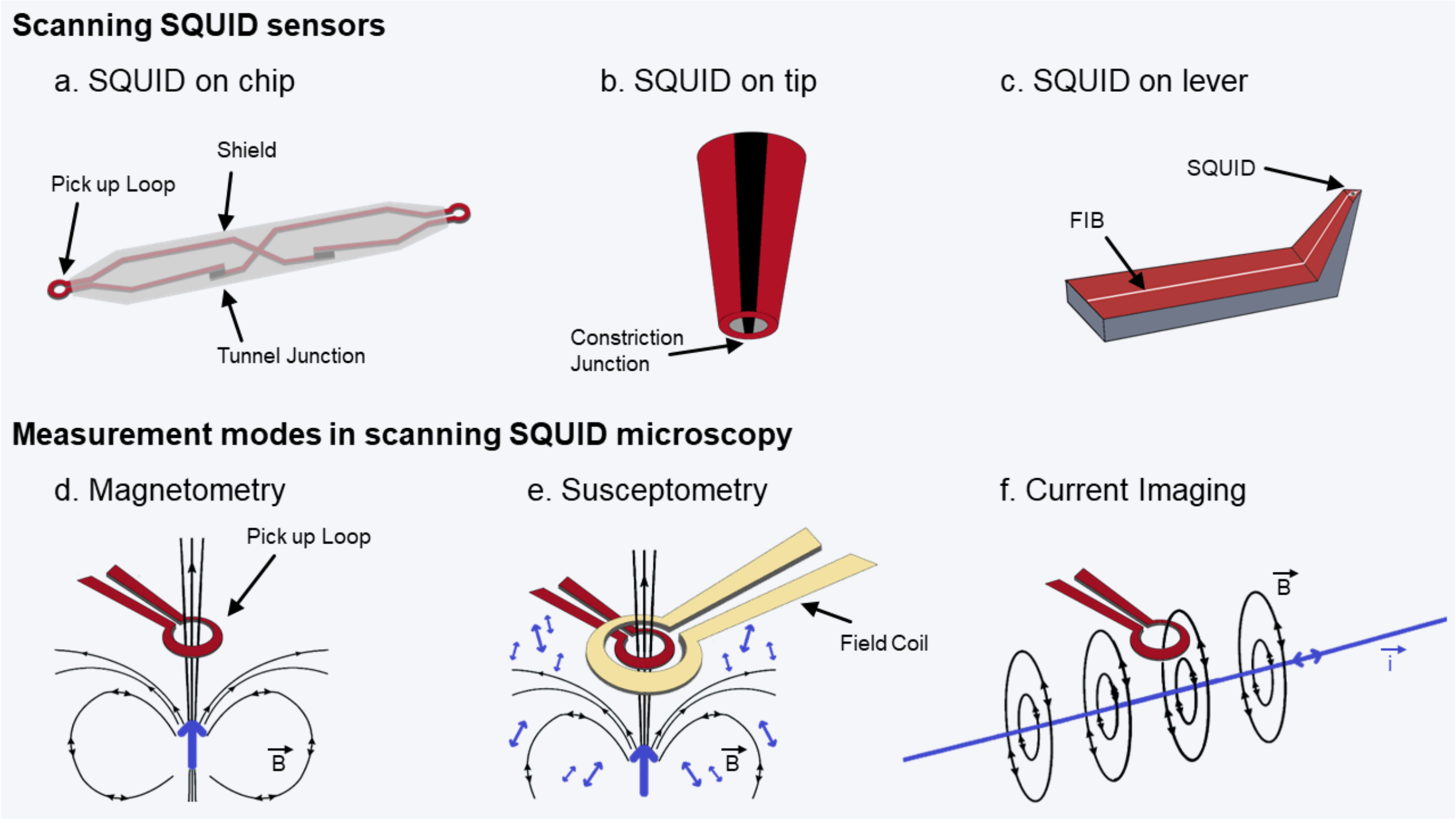

***Figure 1**: Scanning SQUID sensors and operation modes.* ***(a-c)*** *- Illustration of 3 sensor realizations: (a) SQUID on chip (SOC) (b) SQUID on tip (SOT) (c) SQUID on lever (SOL).* ***(d-f)****- Measurement modes in SSM. In (d) magnetometry, the SQUID's pick-up loop captures the magnetic field lines penetrating its area. By rastering the pick-up loop above the sample's surface we generate a map of the local magnetic flux. In (e) susceptometry, current flowing through a single turn field coil concentric with the pick-up loop creates a magnetic field at the sample's surface. The sample's response to this field is sensed by the pick-up loop. By modulating the field coil current we lock into the response, allowing for simultaneous measurement of magnetometry and susceptometry. In (f) current imaging, the magnetic fields generated by the current flow are imaged, allowing for reconstruction of the current density map.*

**Future Advances of the Technique**

SQUID sensors can be fabricated using nanolithography tools and rely on electrical measurement for readout, requiring only a few electrical leads. This makes it straightforward to envision numerous extensions and variations to the technique. Here we give examples of some possible directions.

The current SQUID sensors actively used for SSM have several limitations. SOCs have limited spatial resolution largely due to their geometry and layer structure, limiting the sample-pickup loop distance to around ~300 nm. Additionally, the operation of SOCs is restricted to fields smaller than ~10 mT due to the properties of the JJs. On the other hand, SOTs achieve high spatial resolution of about ~50 nm and can operate in magnetic fields up to a few tesla. However, they do not operate in a flux-locked loop, which can limit the dynamic range and the linear relationship between the measurement signal and the magnetic field at the sample, important for quantitative reconstructions. Furthermore, a powerful capability of SOCs, simultaneous local magnetic response measurements, enabled by an on-chip field coil, is absent in SOTs and SOLs. Mapping the local susceptibility is particularly interesting when studying superconductors.

In terms of fabrication, SOTs are fabricated one by one with limited control over the resulting loop size. The recently developed SOLs achieved high spatial resolution and integration with force microscopy [7]. While the fabrication is more deterministic, each SQUID is fabricated individually with multiple processing steps using a focused ion beam. On the other hand SOCs are fabricated on wafer-scale.

The variety of scanning SQUID sensors covers a wide range of spatial resolution and sensitivity. One desirable future advance is nanoscale scanning SQUIDs that overcome the limitations of existing sensors and combine their advantages allowing for high spatial resolution, high sensitivity, and operation in magnetic fields, while realizing advanced functionality through on-chip circuitry. This will likely require fabrication in a planar geometry using nanolithography and integration with a cantilever enabling scanning close to a sample. Using standard lithography tools would allow for wafer-scale fabrication with nominally identical properties on the same chip - a prerequisite for sharing sensors with the wider community and establishing user facilities. This could make SSM an accessible technique for the wider scientific community.

SQUID sensors can benefit from rapid advancements in superconducting, classical, and quantum information technology, to enhance measurement bandwidth, time resolution, sensitivity, and measurement modes. Possibilities range from utilizing the same cryogenic amplifiers and read-out schemes to implementing a flux-sensitive scanning superconducting quantum bit. Several advances have already been demonstrated, such as the implementation of a dispersive microwave read-out of a scanning RF SQUID [10] and a sampling SQUID sensor integrating a Josephson pulse generator [11]. The dispersive readout relies on shunting the SQUID with an on-chip capacitor to form a LC resonator with a flux-dependent resonance frequency. This enables in principle higher sensitivity and wider measurement bandwidth. The sampling SQUID has demonstrated a ~40 ps time resolution for periodically triggered signals, limited by instrumentation rather than the intrinsic time resolution. Applying these sensors to quantum materials will open up interesting new dimensions. With increased bandwidth SQUIDs could be used to probe magnetic noise complementary to the capabilities of nitrogen-vacancy centers in diamond, particularly in terms of operating temperatures and field range, and achieve unprecedented time resolution for magnetic imaging.

Advancements in SSM can extend beyond traditional magnetometry and magnetic response measurements. One direction is the on-chip integration of SQUID sensors with other on-chip sensors, enabling access to complementary physical properties with a single scanning probe. SQUIDs are well-suited for this due to their compatibility with standard lithography approaches. Another direction is integrating the ability to apply new stimuli or excitations to the sample the SSM sensor and track magnetically detectable responses. For example, instead of a field coil, a local gate can be included, a nanoscale structure to apply local stress, or high-frequency structures to apply microwave excitations. Leveraging the properties of JJs can give access to new signals. The strong dependence of their critical current on temperature can complicate SQUID performance, but can also be leveraged for high-sensitivity local thermometry [12]. Scanning a single JJ can unlock new capabilities. A JJ has flux-sensitive characteristics and emits and absorbs radiation at $\omega = 2eV/\hbar$ where $V$ is the voltage across the junction. Leveraging the latter, JJs have been used as on-chip tunable sources and spectrometers of microwave radiation with operating frequencies exceeding 100 GHz in non-scanning circuits [13]. The incorporation of sensing layers is another approach to expand the capabilities of SQUID and leverage their high sensitivity. For instance, SSM can contactless read out a sensing layer with a strong dependence on the local magnetic properties enabling, for example, highly-sensitive thermometry. Lastly, SQUID capabilities are expected to improve with material and fabrication development, for example, high-$T_c$ SQUIDs [14] will expand the accessible temperature range for SSM.

**Current and Future Use of the Technique for Material Science**

For the past three decades, SSM has been used to explore superconductivity, magnetism, and current distributions in a wide range of samples such as bulk crystals, thin films, interfaces, heterostructures, and nanowires. SSM probes the stray magnetic field above a sample that penetrates any non-magnetic top layer. Therefore buried, top-gated, patterned, lithographically defined and even ionic-gated samples can be studied with SSM. Different measurement modes and capabilities enable, for example, studying magnetic properties such as magnetic textures and susceptibility in magnetic materials and trapped flux, vortices, and diamagnetism in superconductors. Magnetic imaging also allows for the visualization of current distributions in 2D materials. Currents and super-currents can be mapped simultaneously with other electronic properties[2].

Using SSM for probing magnetic properties offers benefits beyond spatial imaging. Its local nature and high sensitivity allow investigation of weak signatures and small sample volumes, outperforming the sensitivity of bulk measurements by several orders of magnitude. Substrate contributions in thin film samples can be determined in-situ via side-by-side measurements in areas where the film is partially removed. This also helps in identification of the substrate's role in thin films and interfaces, where electronic properties of the active layer often stem from the substrate or its structural properties.

In the following, we describe the application of different measurement modes of SSM across different material classes and provide an outlook on future directions throughout. SSM serves as a tool for characterizing multiple material properties, assessing their uniformity, detecting defects, and verifying predicted physical phenomena with magnetic signatures. Moreover, SSM can uncover new and unexpected phenomena. Due to the constraints of this roadmap format, we only highlight a few recent experiments in each area as a starting point for further exploration. We refer the reader to existing reviews for more details and for a complete overview of SSM.

***Probing magnetic materials.*** SSM has been used to probe a variety of magnetic systems. It can map magnetic textures, and gain information even on resolution-limited disordered domain configurations. Isolated moments can be detected due to the high sensitivity and will appear as micro- or nano-scale objects due to the convolution with the sensor's point spread function. Reconstructing the moment configuration from SSM images is typically challenging since the moment distribution is underconstrained, i.e., no unique solution for the magnetization exists given an image of the stray magnetic field. Unique reconstruction is possible in specific cases, such as a two-dimensional magnetization pointing along a fixed axis. Despite these constraints, imaging magnetic structures as a function of temperature, electrostatic gating, strain, and other conditions provides valuable insights into their nature and behavior.

SSM has been applied to study magnetism at complex oxide interfaces, such as revealing the coexistence of magnetism and superconductivity at the $LaAlO_3/SrTiO_3$ interface [15], strain tunable magnetism [16], and studying the emergence of magnetic order in $LaMnO_3/SrTiO_3$ above a critical $LaMnO_3$ thickness [17]. Images as a function of magnetic fields can reveal superparamagnetic and ferromagnetic dynamics at the magnetic reversal, for example, in Cr-doped $Bi_2(Sb,Te)_3$ [18]. SSM is sufficiently sensitive to study magnetism in few-layer van der Waals structures fabricated through mechanical exfoliation, for example, it has enabled imaging of orbital magnetism in twisted BLG associated with a quantum anomalous Hall phase [19].

Local susceptometry can be used to track magnetic phase transition, by mapping the sample's magnetic response as a function of the temperature. This allows estimating the density of responding magnetic moments, and its frequency dependence can probe the dynamics of the moments. For example, an out-of-phase magnetic response in layered antiferromagnet $EuCd_2P_2$ indicated the presence of magnetic fluctuation [20]. As the bandwidth and spatial resolution of SQUID sensors improve, SSM will be able to access more material systems, for example, frustrated magnetic systems. Beyond the traditional use of SSM, local strain tuning of magnetism has been shown by applying a small amount of force to the sample by pressing the SOC's blunt tip into the sample in a controlled manner [16][21] .

***Probing superconducting materials.*** SSM has made significant contributions to the study of superconductors[2][8] . For instance, the direct confirmation of the *d*-wave order parameter in cuprates was facilitated by SSM in a suite of elegant phase-sensitive measurements [22][23] in which spontaneous currents in superconducting rings were probed.

SSM can image the magnetic fields generated by Meissner screening currents, map the positions of superconducting vortices and explore their dynamic behavior. The local magnetic susceptibility in superconducting samples is proportional to the superfluid stiffness, which is related to the superfluid density and the magnetic penetration depth for bulk and 2D superconductors.

SSM can characterize the uniformity of a superconducting thin film and identify the presence of defects. SSM can distinguish different forms of spatial variation of the superfluid stiffness and critical temperature that may cause a broadened resistive transition. Measurements of the superfluid stiffness as a function of temperature have been performed on the micrometer scale and as a function of tuning parameters such as electrostatic gating. Through modeling, the superfluid stiffness can be extracted quantitatively. The low-temperature value, and the onset of the transition, and its temperature dependence can give insights on whether the superconductor realizes the clean or dirty limit, if nodes and multiple gaps exist, and if a clear Berezinskii-Kosterlitz-Thouless transition is present. The imaging of vortices can provide insights into unconventional superconducting states, and advance understanding of flux trapping properties of conventional superconducting films and structures, which is important for applications including classical and quantum superconducting computing.

Again, we highlight a few recent examples. SSM can probe dynamics of superconducting samples, e.g. fast dynamics and flow of vortices was observed in a Pb film [24], whereas fluctuations in superfluid density near the critical temperature were probed in the superconductor NbTiN [25]. In microstructures of heavy-fermion superconductor $CeIrIn_5$, SSM revealed that differential thermal contraction between the structure and the substrate combined with strain sensitivity of $CeIrIn_5$ causes a large-scale non-uniform superconducting transition [26]. SSM is sufficiently sensitive to small sample volumes and low superfluid density, such that it allowed measurements of the superfluid stiffness of $LaAlO_3/SrTiO_3$ [15] and of micrometer sized samples of atomically thin ionic gated $MoS_2$ [27].

SSM is a valuable tool for designing phase-sensitive measurements of unconventional superconducting order parameters [22][23] and for determining current-phase relationships (CPR) of exotic JJs. In these measurements, a JJ is embedded in a superconducting ring formed by a s-wave superconductor. Magnetic flux through the ring imposes a phase-bias across the JJ causing a supercurrent in the ring, which can be magnetically detected with the SQUID. To couple flux into the ring an on-chip field coil is used, enabling high bandwidth and rejection of background signals. As examples, the CPR of a JJ based on a 2D topological insulator [28] and of a few mode InAs nanowire JJ [29] have been measured.

Finally, SSM is also suitable for the discovery of unexpected phenomena with the discovery of a hidden magnetic phase above the superconducting transition temperature of the van der Waals superconductor 4Hb-$TaS_2$ [30] and vortex excitation that carry a temperature dependent fraction of the flux quantum [31] being recent examples.

With improved spatial resolution both for magnetometry and susceptometry, imaging supercurrent distributions in unconventional JJs and the proximitized superfluid stiffness will be feasible. Another direction is that atomic scale measurements of the superconducting gap and vortices by STM could be correlated directly with local SSM measurements.

***Visualizing current distributions.*** Electric currents produce magnetic fields. A 2D current distribution can be uniquely reconstructed from an image of one component of this stray magnetic field. This principle has been applied to visualize current distributions in various materials. In fact, one of the initial applications of SSM was troubleshooting failing integrated circuits through current imaging. In the context of materials research, visualizing current distributions complements electronic transport measurements, which are often the first way to characterize new conducting materials. Transport measurements average over the sample volume sometimes in unexpected ways, in particular, when non-uniformities in a sample lead to real-space variations in the local conductivity. Visualizing the current distribution has revealed surprises and explanations for unusual transport behavior that was previously difficult to understand. A number of electronic phenomena lead to direct signatures in the current distribution with viscous electron flow, bulk versus edge currents in topological phases of matter, and electron guiding and focusing being examples. This measurement mode is not unique for SSM - for example, more recently NV centers in diamond are employed to visualize current distributions based on the same principle.

Here we highlight a few recent examples of using SSM to visualize current distributions. SSM showed that a transport current flows along the edge of a quantum spin Hall insulator [32], while finding that the current flows through the bulk in quantum anomalous Hall insulator Cr-doped $Bi_2(Sb,Te)_3$ [33]. At the $LaAlO_3/SrTiO_3$ interface, the role of structural domains was revealed by visualizing their impact on the local current distribution [34] and on the universality of the metal-insulator transition in $SrTiO_3$-based interfaces [35].The presence of vortices in the viscous electron flow regime was verified by direct visualization [36]. In addition to transport currents, equilibrium currents can exist in samples.

Coexisting topological and non-topological currents have been visualized in the quantum Hall regime in graphene [37], as well as equilibrium currents arising from twist-angle disorder in magic-angle twisted bilayer graphene [38]. Imaging of how current in superconducting wire networks [39] reveals non-uniformity of the current distribution and the role of disorder close to the superconducting transition temperature.

### *Some concluding remarks*

Scanning SQUID microscopy (SSM) is a powerful and versatile technique that has been used for more than three decades to provide local magnetic measurements of quantum materials. The technique enables sensitive, spatially resolved detection of stray magnetic fields above a sample, at low, but variable temperatures. We suggest that advancements in SSM can extend beyond traditional magnetometry and magnetic response measurements, and can be used to access complementary physical properties with a single scanning probe. We hope that SSM becomes more widely accessible in the future to accelerate the development of new quantum materials.

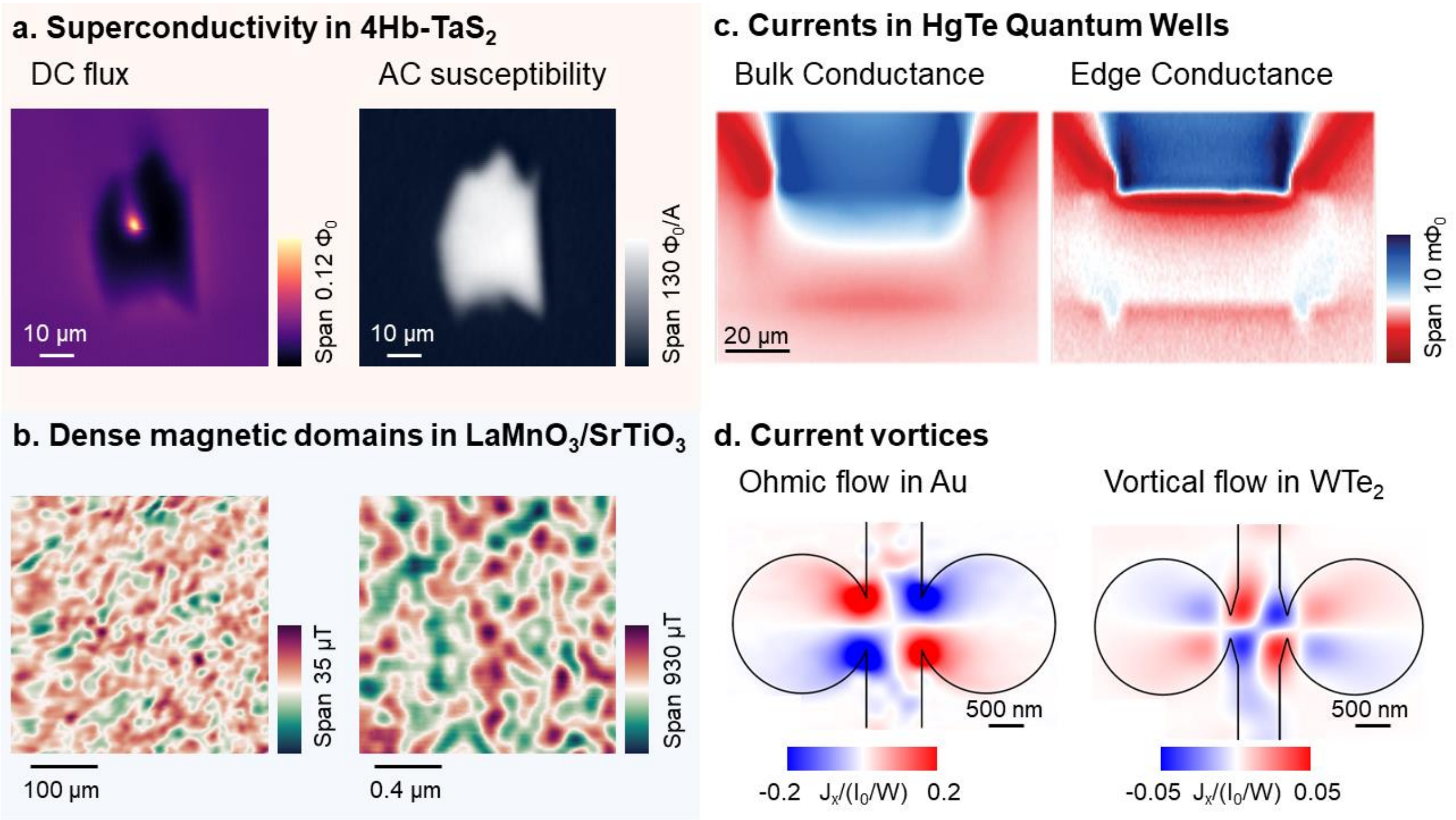

***Figure 2**: (a) SSM on a superconducting flake of 4Hb-TaS$_{2[30]}$. The DC flux image shows the Meissner effect and a vortex in a superconducting flake. The AC susceptibility images the local diamagnetic magnetic response. (b) Magnetic domains in LaMnO$_3$/SrTiO$_3$ heterostructure imaged by a SOC (left) [17] and SOT (right) [40]. Note the different scalebars. (c) SSM imaging of magnetic fields produced by a transport current flowing in a HgTe quantum well Hall bar [32]. The images show conduction through the bulk (left) when the system is in a metallic regime, and conduction along the edge (right) when the bulk is tuned into an insulating regime via electrostatic gating. (d) Current density $J_x(x,y)$ in Au (left) and WTe$_2$(right), reconstructed from $B_z(x,y)$ measured using SOT [36]. The current shows laminar flow in Au, and a closed loop vortical flow, indicative of a hydrodynamic nature, in WTe$_2$.*

**Acknowledgment:**
T.R.D and B.K. were supported by European Research Council Grant No. ERC-2019-COG-866236 and Israeli Science Foundation Grant No. ISF-228/22.

# 2. Nitrogen Vacancy Magnetometry

## 2a. Imaging electronic transport in materials using colour centers

*J.L. Webb[1], U.L. Andersen[1], A. Huck[1]*

[1] Centre for Macroscopic Quantum States (bigQ), Department of Physics, Technical University of Denmark, 2800 Kongens Lyngby, Denmark

**Current State of the Technique**

Optically active point defects (colour centres) in the solid state offer an excellent platform for imaging electrical transport in a wide range of materials, ranging from biomaterials to inorganic 2D van der Waals heterostructures. Acting as individual atomic scale sensors, colour centres provide high sensitivity to the local environment, high spatial resolution (nano-meter/micro-meter scale for single/ensemble sensors) as well as fast readout and initialisation (both sub-µs) for time-resolved measurements. The principle of using colour centres to study electronic transport is well established in three material systems: boron vacancies in hexagonal boron nitride (hBN), divancancies in silicon carbide (SiC) and in particular nitrogen-vacancy (NV) centres in diamond. Since the original idea of a spin microscope with optical readout [1], NV-based sensing [2, 3], and the first demonstration of microscopic imaging [4], the field has rapidly developed, and offers strong potential for future materials science studies.

Sensing and imaging with colour centres relies on the well-established technique of optically detected magnetic resonance (ODMR) spectroscopy. In this process, local electric or magnetic fields modulate the defect fluorescence emission when illuminated by pump laser light and a resonant field in the radiofrequency to microwave range. A more comprehensive discussion on the sensing principle and limitations is extensively covered in [5] and [6].

Interest in colour centre imaging has primarily been driven by its ability to image vector magnetic flux density (B) or electric field (E). For magnetic field imaging, this attribute is particularly appealing for systems without the presence of any material magnetisation, where alternative materials techniques such as Kerr microscopy or magnetic force microscopy (MFM) cannot be used. Another intriguing advantage is that field imaging with colour centre can be entirely passive, requiring no mechanical contact of the target material. When illumination on target sample is negligible, this allows electronic transport (charge or spin) within a material to be studied by induced electric/magnetic field without risk of disrupting sensitive effects via mechanisms such as heating or inducing photoelectric charge.

State-of-the-art sensing of electrical transport can be divided into ensemble and single defect methods. Ensemble methods utilise a dense ensemble of colour centres within a bulk host, located in close proximity to the material to be imaged, with the defect fluorescence emission captured by a microscope objective and a CCD/CMOS camera. This technique enables capturing transport properties over a wide field of view with diffraction-limited spatial resolution (approximately at the micron level), with the temporal resolution by camera frame rate. In contrast, single NV methods use just a single colour centre, hosted either in a fixed host material or in a scannable tip. Imaging can be achieved by translating either tip or sample in a raster pattern while recording data from the colour centre using a sensitive photodetector. For fast transport processes, the single defect method effectively offers only a point probe measurement due to the slow rate of translating either probe or sample. However, for slow processes, repeating signals that can be averaged or variance detected, the method affords superior spatial resolution, higher proximity to the target material, greater resistance to environmental decoherence and faster readout.

**Future Advances of the Technique**

High-quality imaging requires a strong defect centre optical response to the target field, strong resistance to decoherence while maintaining sensitivity to environmental changes, and a minimum level of measurement noise. Achieving these requires in turn a stable, contamination-free host material, efficient defect generation, effective defect control, efficient optical readout, and some degree of noise rejection through shielding or filtering.

Future advances in these areas can be divided into materials and process development. This includes both host and defect generation, improvements in coherent control and readout of the defects and methodologies to minimise measurement noise.

Current efforts in host material engineering seek to mitigate factors such as host material strain and contamination and charging, that adversely affect sensing performance, particularly for colour centre ensembles. An important aspect here is isotopic purification, which removes isotopes (e.g. 13C) with intrinsic spin that would otherwise cause decoherence and reduce sensitivity. For diamond, $^{12}C$ enriched substrates have begun to be commercialised. A significant recent advance is the achievement of isotopic purification in SiC [7], yielding up to a 50-fold improvement in coherence times. Another promising future route is material growth with preferentially aligned colour centres, ensuring a uniform crystallographic orientation for all colour centres in an ensemble, which optimises the response (in terms of ODMR contrast) [8].

For control and readout, methods derived from NMR/ESR using pulsed control fields are being implemented for imaging experiments to improve the control of the colour centre state. This includes work on optimal control such as microwave pulse shaping [9] and advanced pulse sequencing. The aim is to achieve noise rejection and overcome host crystal inhomogeneity, offered by the double quantum 4-Ramsey technique [10]. Future work on using dynamical decoupling sequences for AC sensing holds potential for substantial improvement. Similarly, progress in the transmission of microwave nearfield antenna signals to the material could yield increased Rabi frequencies and more homogenous control.

Minimisation of measurement and background noise is key to optimal imaging performance. This noise can originate from the system itself (e.g. thermal), background fields or from control/readout. Critical to noise minimisation are the development of common mode rejection methods, use of mu-metal enclosures and/or frequency and time domain filtering methods. Widefield imaging at low frame rates and relying on temporal averaging is especially susceptible to noise sources, which for conventional CCD/CMOS cameras can cause the sensitivity to plateau when time averaging across multiple acquisition frames. Future use of cameras with in-built lock-in amplification [11], balanced detection techniques [12] or using the optically excited states [13] could prove successful in yielding high-SNR imaging.

Ultimately, there remains two fundamental challenges for colour centre imaging. The first is the low efficiency of the colour centre systems, regarding the ratio of the input optical and microwave power to usable signal output. This is due to a) the necessity to keep colour centre density low (to avoid defect-defect interaction), resulting in an overall low light absorption cross section and b) due to difficulty in efficiently collecting fluorescence emission. The latter is particularly problematic for host crystals with high refractive index such as SiC and diamond. The second substantial problem is the high level of non-signal background fluorescence emission, adding a high level of optical shot noise. It would instead be desirable to have a system emitting light only when activated by a microwave field resonant with the colour centre energy levels.

Although these challenges could be tackled by material engineering (e.g. larger substrates) or through optical engineering (e.g. solid immersion lenses), it would be advantageous to explore alternative systems. This may include defects hosted in 2D materials, avoiding the refractive index issue and potentially offering strong, visible-range emission suited to Si-based imaging detectors. Here, ongoing ab initio studies utilising automated machine-learning based searches [14] may identify promising optically active defects. This could also include hybrid systems, such as the introduction of additional dopants or overgrowth on diamond in order to increase optical response or suppress decoherence effects [15].

**Current and Future Use of the Technique for Material Science**

To date, the majority of experiments for materials science investigations have relied on widefield imaging using a defect ensemble, as this provides the simplest method to realise high spatial resolution (on the µm- scale) and wide field of view (µm-mm scale). A particular focus has been the imaging of electrical current transport in microelectronic devices by induced magnetic field, for both fundamental study and for applications such as circuit inspection.

Imaging can be achieved with sensing bandwidths up to several kHz sub-ms time resolution. Although this may not be fast enough to capture in real-time electronic signal propagation in microdevices, temporal averaging can allow facilitate the imaging of average current propagation patterns. Studies have included inspection of capped and uncapped integrated circuits [16], and in particular study of current flow in novel devices made from 2D materials such as graphene [17,18] or transition metal dichalcogenides (TMDs) including imaging of photoinduced current [19]. 2D materials are ideally suited for colour center imaging as transport is confined to a thin layer which can be in high proximity to the colour centre host. Their flatness also makes them highly suitable for study with NV-based scanning probe microscopy (SPM) systems. Due to the rapid technical development of such SPM systems and their potential for enhanced spatial resolution, it is likely these will play an increasing role in future materials science studies.

Another intriguing application of colour centre imaging is its use in exploring more exotic electronic transport phenomena, such as spin transport or superconductivity. Imaging is particularly suited to superconducting materials with low critical currents and accessible magnetic phase transitions, such as the Fe-based superconductors [20]. NV-based imaging has been successfully used to image spin transport phenomena such as spin waves induced in yttrium iron garnet (YIG) [21]. Looking forward, the development of low temperature NV-SPM systems, capable of probing both electric and magnetic field at mK temperatures, is of special interest [22]. These systems may prove useful in both the study of both fundamental phenomena (e.g. topological phases) as a complement to traditional transport methods and for applications such as device characterisation, including superconducting qubits for quantum computing.

Finally, colour centre imaging has demonstrated significant potential in recording transport in biomaterials, including electric impulses (e.g. action potentials) in living tissue, enabled by the high degree of biocompatibility of the defect host materials, particularly diamond. Proposals have been put forward to image electrical current transport in living neural networks, for fundamental studies in biology [23]. Compared to existing techniques (e.g. GEVIs), this offers the advantage of requiring no biological modification or use of potentially toxic fluorescent dyes. Although this application has yet to be realised, it could be achieved with modest future improvement in sensitivity. Another relevant development for biomaterials is the use of colour centres in biocompatible nanomaterials (e.g. nanodiamonds) as highly stable fluorescence markers [24] and nanoscale biosensors [25]. Combined with an imaging method such as scanning confocal microscopy using a spinning (Nipkow) disc system, imaging a distribution of colour centre markers could deliver extremely high-quality, sub-wavelength information on both electronic and thermal transport in biomaterials.

**Acknowledgements**

We acknowledge financial support from the Novo Nordisk Foundation for the Biomag project (NF21OC0066526) within the challenge programme and the Danish National Research Foundation (DNRF) through the center for Macroscopic Quantum States (bigQ, Grant No. DNRF0142).

# 2b. Imaging magnetism using nitrogen vacancy magnetometry

*David A. Broadway[1,4], Kai Wagner[2], Patrick Maletinsky[3]*

[1] Department of Physics, University of Basel, Basel, Switzerland https://orcid.org/0000-0002-7375-8766
[2] Department of Physics, University of Basel, Basel, Switzerland https://orcid.org/0000-0003-4252-0584
[3] Department of Physics, University of Basel, Basel, Switzerland https://orcid.org/0000-0003-1699-388X
[4] School of Science, RMIT University, Melbourne, Victoria, Australia

**Current State of the Technique**

The Nitrogen Vacancy (NV) centre in diamond is an atomic scale lattice defect that host an electronic spin that allows for optical spin initialization and readout. Together with coherent spin manipulation using microwave-frequency (MW) driving fields and operation even under ambient conditions, this renders the NV a highly interesting spin system that finds applications in various quantum technologies and first and foremost quantum sensing and nanoscale magnetometry.

Such nanoscale magnetic imaging is based on placing the NV centre spin in close proximity to a sample of interest, in order to measure the stray magnetic fields generated by the sample. There are various experimental approaches to achieve this goal, each with their own advantages and disadvantages, the most widely used being atomic force microscopy using diamond tips hosting NV centre spins (NV-AFM, Fig. 1a) [1] and wide-field optical imaging using the "quantum diamond microscope" (QDM, Fig. 1b) [2]. A highly noteworthy variant of the QDM employs NV centres embedded diamond anvil cells (NV-DAC, Fig. 1c) [3] for magnetic sensing and imaging under high pressures. A key asset of all these techniques is their applicability over a vast range of conditions, operating from the lowest achievable temperatures up to 1000 K [4] and at pressures from ambient to > 10 GPa [3].

In the following, we list and briefly discuss some key technical performance characteristics of NV based magnetic imaging modalities:

**Measurement protocol**
All mentioned techniques detect magnetic fields through the Zeeman effect, building on optical spin initialisation and readout, together with MW driving. The most commonly employed approach is optically detected magnetic resonance (ODMR, Fig. 1d), but more refined method using pulsed, coherent NV driving exist and are described in excellent prior reviews [5].

**Reconstruction of the source**
The main goal of NV-based magnetic imaging is to infer information about the magnetic field source, e.g. currents or magnetisation distributions in the sample under study. The required reconstruction from the imaged magnetic field (Fig. 1e) is an ill-posed problem and as such presents major difficulties in the reconstruction process. Various methods have been discussed in recent years for how to best address this difficult inverse problem [6, 28].

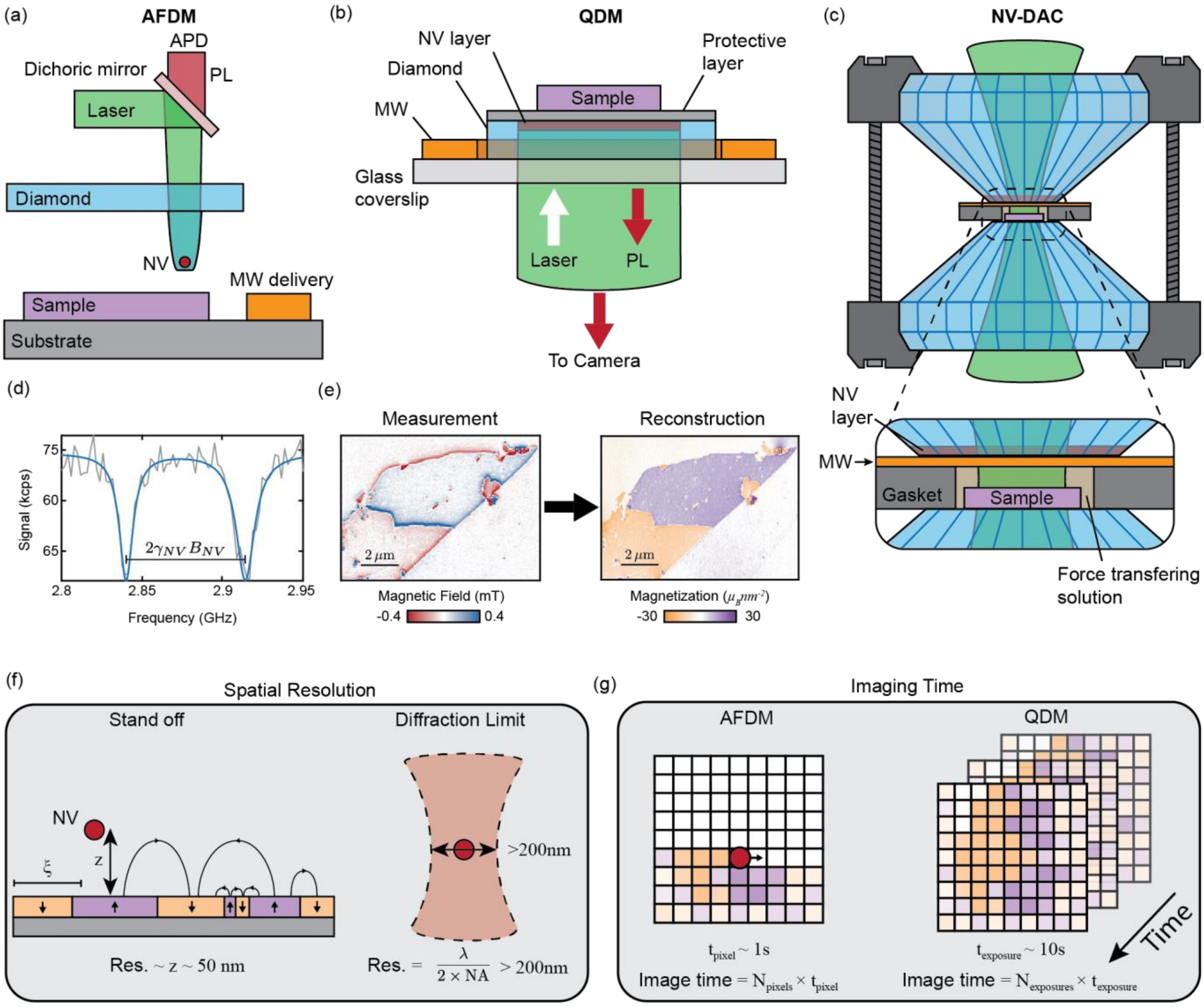

***Figure 1**: Principle of the measurement techniques. (a) Schematic of an atomic force diamond microscope where an all-diamond tip is scanned across a sample of interest. At each point the magnetic field is imaged via the combination of detecting the NV spins PL and microwave (MW) driving. (b) Schematic of a quantum diamond microscope (QDM) where a slab of diamond is used to image a sample that is placed on or near the diamond. The magnetic image is then captured via a camera in a widefield imaging technique. (c) Schematic of a NV diamond anvil cell (NV-DAC) where an NV layer is embedded into a diamond anvil cell for high pressure measurements of magnetic fields. (d) Example of optically detected magnetic resonance (ODMR) where the NV spins sublevels are detected, which infers the magnetic field at the NV via the Zeeman effect. (e) Example of a reconstruction from a magnetic field image that is measured into a magnetisation image. (f) Spatial limitation of the technique where stand-off limits NV-AFM and diffraction limits QDM and NV-DAC. (g)*

**Spatial Resolution**

Spatial resolution in NV magnetometry is ultimately limited by standoff distance between the NV spin and the sample (Fig. 1f) [6]. For NV-AFM with single NV centres, this limit in spatial resolution is effectively realised and NV-sample distances $z_{NV}$ in the range of 20-70 nm have been reported [7]. Conversely, in the QDM, spatial resolution is determined by the diffraction limit, since ensembles of NV centres are addressed using a conventional optical microscopy. While in principle it should be possible to achieve a resolution of ≈200 nm, in practice this is not achieved due to optical aberrations, leading to a resolution in the range of 400 nm [8].

**Imaging Time**

Imaging time is determined by the desired signal-to-noise ratio, whose scaling with time is determined by the sensitivity. The latter is typically on the order of 1 $\mu T/Hz^{½}$ for single NVs and improves with $N^{½}$, if a number of N NVs are employed simultaneously. In NV-AFM, data needs to be acquired pixel-by-pixel, and the imaging time therefore scales linearly with the number of pixels. Conversely, in QDM, all pixels of an image are acquired simultaneously, which leads to a significant speedup in acquisition time (Fig. 1g). In addition, NV-AFM typically employs single NVs for sensing, while in the QDM, several NVs are interrogated in each pixel. As a result, NV-AFM is a comparatively slow technique (acquisition times of several minutes to hours) with high spatial resolution, while QDM has a limited resolution, but allows for rapid data acquisition times on the order of seconds to minutes [8].

**Future Advances of the Technique**

There are several key improvements that can be made to further enhance the sensitivity, operating range and utility of Nitrogen vacancy (NV) scanning magnetometry in the future.

**Spatial Resolution:**

The main fundamental limit to NV spatial resolution is the NV-sample distance $z_{NV}$, mentioned above. While current implementations of the NV-AFM realize values of $z_{NV}$ in the range of 10's of nm, efforts are underway to achieve single-digit nanometre values of $z_{NV}$ and therefore in spatial resolution. For this, two challenges need to be addressed. First, NVs need to be stabilized close to the surface, without significantly degrading their optical, spin, and charge-properties – a formidable and still open challenge in diamond materials engineering. Second, the geometry of all-diamond scanning probes should be adapted to allow for such small values of $z_{NV}$. Current realizations of diamond tips consist of cylindrical tips of ~200 nm diameter, which pose a limitation on its own with regard to optimizing $z_{NV}$. However, interesting new approaches are on the horizon [9] that may allow for the realization of ultrasharp diamond tips for the realization of single-digit-nm resolution in NV-AFM.

**Sensitivity**:

Sensitivity in NV magnetometry is primarily determined by the NV spin coherence time, $T_{coh}$, and the efficiency of optical addressing of the NV spin, as characterized, e.g. by the photon count-rate $I_{PL}$ of NV photoluminescence. NV magnetic field sensitivity scales with $1/(T_{coh}*I_{PL})^{½}$, which motivates improvements in NV optical addressing (e.g. though photonic engineering) and NV coherence times (e.g. by materials engineering or advanced, pulsed quantum sensing protocols). Both aspects have been discussed in detail in past reviews [5].

An additional, sometimes overlooked aspect is the control over the NV charge state. Indeed, only the negatively charged NV centre, $NV^-$, is directly applicable to magnetometry, while other charge states (i.e. $NV^+$ and $NV^0$) exist but should be avoided for magnetometry. Charge state stability directly translates to enhanced magnetic field sensing capabilities, as only the single negatively charged state is utilized for sensing. Approaches for stabilizing the NV charge state have been demonstrated (e.g. through diamond surface modifications [10] or multi-color laser illumination) and are subject to ongoing research in particular with respect to the required nano-fabrication for scanning probes [12]. The different possible NV charge states, however, also offer interesting opportunities for NV spin-to-charge conversion and thereby efficient (even single-shot) NV spin readout [13, 14], or photoelectric readout of the NV spin state [29, 30].

**Real-time spin and charge control**:

Advanced electronic components for NV spin control, such as Field Programmable Gate Arrays (FPGAs) offer another ongoing avenue for future improvements in NV sensing capabilities. For example, using such hardware, it was recently shown that the NV charge state can be deterministically initialized in real-time [15]. Furthermore, NV spin-based measurement sequence can be adapted on the fly length providing efficient, real-time tracking of the instantaneous NV spin transition frequency [16]. Real-time computation can also aid in targeting measurement points of expected higher information gravity via Bayesian updating/sampling and speed-up towards live quantum optimal control [17].

All the above-mentioned aspects closely relate to an improving the spatial resolution, as stabilizing NV centers close to the surface has the potential of reaching the single-digit nanometer scale. This will not only increase all detected signals from a target sample. therefore, the aforementioned surface modifications and control schemes will take a key factor with particular importance for fundamental physical questions on the nanometre-scale.

**Multimodal sensing operation**:
Next to the excellent sensitivities in measuring static and AC magnetic fields, the NV center grants access to several other quantities, such as electric fields or temperature. This is complemented by spectroscopy of magnetic [18] or electrical noise [19].

The possibility to sense several different physical quantities using the same sensing platform and conditions is another key asset of NV based sensing that can help develop a better microscopic understanding of physical processes in samples under study – a prime example is the simultaneous addressing of ferroelectric and magnetic ordering in multiferroic materials (see below). While such multimodal imaging typically increases measurement time and protocol complexity, research is underway to further advance and simplify such multimodal NV imaging.

**Current and future use of NV magnetometry for material science**

Scanning NV magnetometry has the potential to significantly impact the field of material science in the coming years. This was already shown with several use-cases and key publications over the last ten years, and the resulting impact is poised to grow further in the future. The power of NV magnetometry in this regard lies not only in its high sensitivity and nanoscale spatial resolution, but for many applications also in the quantitative and minimally invasive nature of the technique. These combined properties open up new possibilities for the study and development of advanced magnetic materials and devices. In this section, we will discuss how NV magnetometry impacts several areas of application, including 2D magnetic van der Waals materials and their heterostructures, antiferromagnets and multiferroics, Skyrmion hosting magnetic materials, materials with engineered interfacial Dzyaloshinskii-Moriya interaction (DMI), superconductors, and industry applications such as failure analysis and novel memory devices.

2D magnetic van der Waals materials have emerged as a promising platform for spintronics and magnonics, with potential applications in data storage, processing, and communication. Scanning NV magnetometry offers a unique opportunity to probe these materials' magnetic properties. In fact, scanning NV magnetometry is so far the only technique that allowed for nanoscale imaging of spin-textures in atomic monolayers of a vdW magnet [20] (Fig. 2a). Furthermore, the technique allows for addressing magnetism in vdW heterostructures (such as moiré stacks or gated magnets) and thereby extract, e.g., information about interlayer exchange couplings. This deeper understanding could lead to the development of new vdW magnetic materials and devices with tailored properties, such as tunable magnetism, spin transport, and magnon-mediated energy transfer.

Antiferromagnets and multiferroics are another class of materials, where NV magnetometry already showed impact and will continue to do so. Antiferromagnets exhibit magnetic ordering but (near) zero net magnetization and as such are hard to address by conventional experimental techniques. However, weak uncompensated magnetic moments are known to occur when bulk symmetries of an antiferromagnet are broken, i.e. on surfaces or domain walls. The resulting weak magnetic stray fields enable access to nanoscale properties of antiferromagnets by exploiting the sensitivity offered by NV magnetometry. Multiferroics are materials that simultaneously exhibit ferroelectric and magnetic ordering, offering the possibility of controlling magnetism with electric fields and vice versa. Oftentimes, such as for the prototypical multiferroic $BiFeO_3$, the magnetic order in multiferroics is of antiferromagnetic nature. Scanning NV magnetometry then enables the observation and control of complex magnetic textures in these materials [23, 21, 22] (Fig. 2b), ultimately leading to the development of novel antiferromagnetic and multiferroic devices with improved energy efficiency and new functionalities.

Skyrmions, topologically protected magnetic quasiparticles, have garnered significant interest due to their small size, stability, and potential for low-power spintronic applications. Both bulk and thin-film Skyrmion hosting magnetic materials can be studied and manipulated using scanning NV magnetometry [24] (Fig. 2c), which can provide crucial insights into the formation, stability, and manipulation of Skyrmions. This knowledge could pave the way for the development of Skyrmion-based devices, such as racetrack memories and logic gates, that can operate at ultralow power consumption and high speed.

Magnetic multilayer stacks that induce interfacial DMI, a chiral exchange interaction that stabilizes non-collinear magnetic textures, form another area where NV centre magnetometry can make a significant impact. By studying materials with engineered interfacial DMI, researchers can not only quantify the strength of iDMI [25] (Fig. 2d), but also gain insights into the complex interplay between magnetic moments and crystal lattice structures. This understanding could lead to the development of advanced materials with tailored DMI properties for use in spintronic and magnonic applications.

Superconductors and superconducting devices also stand to benefit from advances in scanning NV centre magnetometry [26] (Fig. 2d). By probing the local magnetic fields in superconducting materials, researchers can gain a better understanding of the mechanisms behind superconductivity, vortices and vortex dynamics, ultimately leading to the development of improved superconducting devices, such as superconducting quantum interference devices.

Finally, NV centre magnetometry's nanoscale sensitivity and non-destructive nature make it an ideal tool for various industry applications, including failure analysis and the development novel memory devices, such as magnetic random-access memories (MRAM) (Fig. 2e). In failure analysis, the ability to identify and localize defects in magnetic materials and devices can lead to improved quality control, reliability, and lifetime. In the realm of novel memory devices, NV magnetometry could enable the development of high-density, energy-efficient storage solutions based on new

**Acknowledgements**
We acknowledge financial support from the Swiss Nanoscience Institute (SNI), the National Centre of Competence in Research (NCCR) Quantum Science and Technology (QSIT), a competence center funded by the Swiss National Science Foundation (SNF), by SNF project No. 188521, and by the ERC consolidator grant project QS2DM. D.A.B. acknowledges support through an Australian Research Council DECRA Fellowship (grant no. DE230100192)

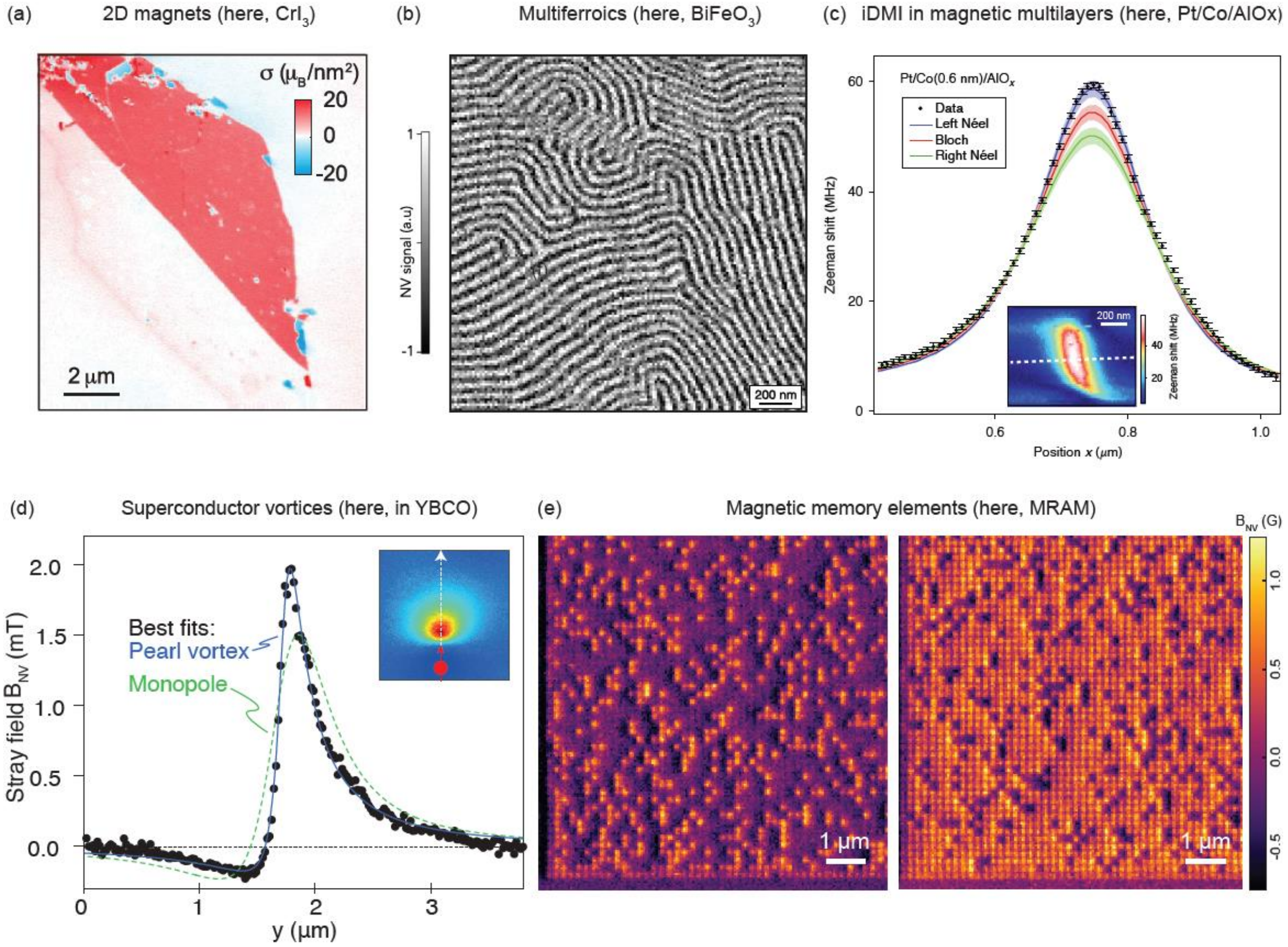

***Figure 2****: Application examples of scanning NV magnetometry imaging. (a) e of an atomically thin "van der Waals" magnet – here, $CrI_3$. The image shows a quantitative map of sample magnetization for a trilayer sample, with an adjacent bilayer of zero net magnetization [20] (b) NV magnetometry image of domain patterns of the spin-cycloid on $BiFeO_3$ [27] (c) Stray field profiles across a domain wall (inset) of a multilayer magnetic sample with iDMI. The quantitative fit to the data allows for the determination of the nature and chirality of the domain wall [25] (d) Stray field image across a vortex in the high-temperature superconductor YBCO (inset). The quantitative fit to the data allows for the discrimination of different physical models describing the stray field (and thereby supercurrent distribution) of the vortex [26]. (e) NV-AFM image of bit patterns measured on a state-of-the art MRAM chip. The two images whose the bit pattern after different switching procedures and their comparison allows for an assessment of bit operation on the single bit level (credits, Qnami AG, application lab).*

# 2c. Imaging magnetic dynamics using spins in diamond

*Toeno van der Sar[1], Chunhui Rita Du[2], Amir Yacoby[3]*

[1]Department of Quantum Nanoscience, Kavli Institute of Nanoscience, Delft University of Technology, Lorentzweg 1, 2628 CJ Delft, The Netherlands.
[2]Department of Physics, University of California, San Diego, La Jolla, California 92093, USA
[3]Department of Physics, Harvard University, 17 Oxford Street, Cambridge, Massachusetts 02138, USA.

**Current State of the Technique**

Condensed matter physics thrives on the interplay between discoveries and the development of new measurement capabilities that are enabled by such discoveries. While in some cases this development involves measuring new physical quantities, often it simply involves using novel physics to report new quantities of interest. For example, global positioning systems (GPS) report position accurately through the measurement of time and uses general relativity to connect the measured quantity to the reported one. Therefore, by using established physics one can report quantities that are very hard to measure.

Magnetometers based on nitrogen-vacancy (NV) spins in diamond (Fig. 1a-b) measure magnetic fields in a rather broad range of frequencies (DC to several GHz) and with high spatial resolution (down to ~50 nm) [1]. Recent developments in NV center magnetometers have made considerable impact in our ability to explore quantum matter through the use of novel physics that enables reporting quantities that are otherwise hard to measure. One such example is the use of the fluctuation dissipation theorem to connect the NV spin relaxation to the spin chemical potential in a magnetic system [2], [3] – an important quantity governing spin transport. The spin excitations in a magnet result in magnetic field fluctuations that leads to relaxation of a nearby NV sensor and thereby enables quantitative determination of the local chemical potential [2].

Another capability enabled by NV magnetometers is the use of spin waves in scattering experiments. Such experiments consist of launching coherent waves or particles with well-defined energy and momentum at a target. Through the measurement of the amplitude and phase of the scattered signal one can reconstruct the target's structure factor [4] and probe its magnetic susceptibility [5]. NV spins can measure the amplitude [6] and phase [4], [7] of propagating spin waves through the magnetic field they generate in time and in space (Fig. 1d-g). A key advantage of the technique is that it allows probing the coherent interactions of the spin waves with target materials and the spin textures and electric currents they host.

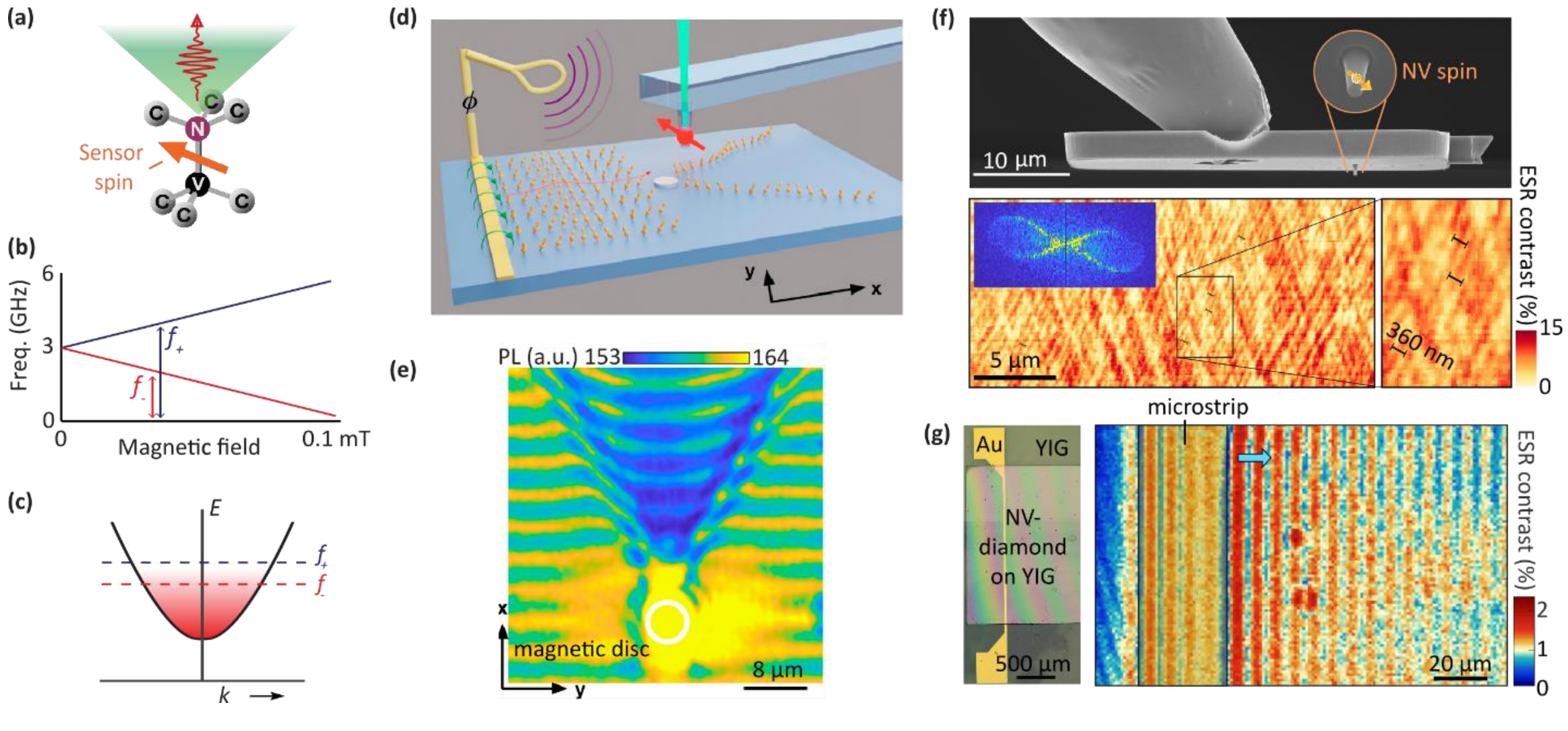

***Figure 1: Imaging coherent spin-wave dynamics using NV spins in diamond.*** *a) The nitrogen-vacancy (NV) center is an atomic defect in the diamond carbon lattice with an S=1 electron spin that can be read out through its spin-dependent photoluminescence. b) NV electron spin resonance (ESR) frequencies as a function of magnetic field. Microwave fields resonant with an ESR frequency can drive coherent spin rotations or induce enhanced relaxation between the spin states. c)*

*Tuning the NV frequencies to lie within the spin-wave band enables phase-sensitive, resonant spin-wave imaging. d-e) Scanning-NV imaging of spin-wave scattering on a magnetic disc. Figures from* [4] *licensed under CC BY-NC-ND 4.0. f) Nanoscale microwave magnetic imaging of scattered spin waves using a scanning diamond tip. Figure from* [8] *under license CC BY 4.0. g) Left: NV diamond placed on a magnetic (YIG) thin film, from* [7] *licensed under CC BY-NC 4.0. Right: NV-ensemble imaging of spin waves underneath a metal microstrip, enabling in-situ measurements of metal-induced spin-wave damping. Figure from* [5], *licensed under CC BY-NC-ND 4.0.*

**Future Advances of the Technique**

We expect the future advances of the technique to focus on studying new magnetic materials, on the interaction of spin waves with spin textures and electric currents, and on expanding the detectable frequency and wavelength range. In current NV-based spin-wave detection schemes, two situations can be distinguished. In the first, the NV spin resonance frequencies lie within the spin-wave band (Fig. 1c), such that spin waves resonant with the NV can directly drive NV spin transitions. In the second, the NV spin resonance frequencies lie below the band (next section), such that the NV is sensitive to longitudinal magnetization dynamics generated by spin-wave mixing [3]. Both situations provide excellent new opportunities for probing magnetic materials and spin-wave – material interactions. Here we describe the current focus of efforts to advance these techniques.

With phase-sensitive imaging of NV-resonant spin waves established (e.g. Fig. 1(d-g)), a key goal for realizing broad applicability of the technique is to develop protocols that will enable imaging of spin waves far detuned from the NV resonance frequency. Off-resonant spin-wave detection using NV magnetometers would enable high-resolution probing of spin-wave physics in the wide range of magnetic materials with strong saturation magnetizations or magnetic anisotropy, such as ultrathin van der Waals magnets [9]. We discuss two approaches to meet this challenge.

The first approach focuses on developing NV spin-control techniques that render the NV spin sensitive to microwave signals that are far detuned from the ESR frequency. The development of such protocols is an active field of research, with state-of-the-art methods having already demonstrated detection capabilities at ~100 MHz detuning from the NV ESR frequency [10]. However, detecting signals that are detuned by multiple gigahertz, as required for many magnetic materials of interest, remains a challenge as the sensing sensitivity typically decreases with increased detuning. As such, there remains a need for new NV-protocols enabling broadband microwave detection.

The second approach focuses on converting the intra-band spin-wave excitations to magnetic dynamics at frequencies that are detectable with standard NV sensing protocols. On the one hand, mixing between *thermal spin waves* leads to NV-resonant longitudinal spin *noise* that provides access to the magnetization of a broad range of magnetic materials (next section). On the other hand, mixing between *microwave-driven* spin waves results in *coherent* dynamics of the longitudinal magnetization [11]. Broadband detection of the difference-frequency signals is possible by flipping the NV spin at a rate that matches the difference frequency using AC magnetometry / dynamical decoupling protocols [1]. We anticipate that the driven spin-wave mixing processes will enable mapping spin-wave dispersions even at large detuning with the NV frequency, as the frequency- and wavelength-difference of the driven primary spin waves directly determines the longitudinal magnetization dynamics.

A key drive in the field spintronics is the prospect of developing nanoscale spin-wave logic devices. This would benefit from spin-wave imaging techniques with nanoscale resolution, motivating a push of the spatial resolution of NV-imaging of magnetization dynamics down to the few-nanometer scale. The challenge is that the spin-wave frequency quickly increases to the >10-GHz regime for wavelengths below ~100 nm because of the exchange interaction. As such, the required matching of NV- and spin-wave frequencies (Fig. 1(c)) limits the smallest detectable wavelengths. Moreover, the exponential decay of the spin-wave stray fields puts a strenuous requirement on the NV-sample distance: For instance, the stray field of a 50 nm spin wave decays exponentially on a 8 nm scale. While scanning-NV magnetometers readily achieve NV-sample distances of ~50 nm (see, e.g. [9]), few-nm NV-sample distances have until now only reliably been obtained for samples directly deposited onto diamonds with shallow NV centers. Such shallow NV centers enable the detection of few-nm spin waves with a diffraction limited spatial resolution, and as such we anticipate their future use for probing nanoscale spin-wave physics.

**Current and Future Use of the Technique for Material Science**

After the pioneering study on YIG [12], NV magnetometry has been successfully applied to detect spin-fluctuations in other magnetic materials such as permalloy [13], the ferromagnetic insulator nickel-zinc-aluminum-ferrite [14], the antiferromagnetic insulator $\alpha$-$Fe_2O_3$ [15], a synthetic cobalt-based antiferromagnet [16], the topological magnet $MnBi_2Te_4(Bi_2Te_3)_n$ [17], and the low-dimensional magnet $Fe_3GeTe_2$ [18]. For these and many other materials, the magnon band lies above the NV ESR frequencies (Fig. 2(a)) such that the resonant detection methods of Fig. 1 are not applicable. Instead, the NV sensing of magnon dynamics relies on detecting the fluctuations of the longitudinal spin density [3]. The longitudinal spin noise is characterized by the two-magnon scattering processes, where a magnon with a frequency $f + f_{2m}$ can undergo a transition to another magnon with a frequency $f$, emitting magnetic noise at frequency $f_{2m}$ as illustrated in Fig. 2(a). In contrast to the transverse spin noise with a minimum frequency determined by the magnon band gap, the frequency bandwidth of its longitudinal counterpart starts from zero, which can be comfortably addressed by NV centers in the low frequency (field) regime.

The longitudinal spin noise is fundamentally related to diffusive spin transport behaviors via Brownian motion [15], which is further connected to the underlying magnetic parameters of the material systems studied. Building on this new measurement platform, one of us used NV relaxometry to show that the intrinsic spin diffusion constant and longitudinal dynamic spin susceptibility of antiferromagnetic insulator α-$Fe_2O_3$ and van der Waals magnet $MnBi_4Te_7$ could be directly measured in the absence of external spin biases [Figs. 2(b)-2(d)]. The measurement of longitudinal spin noise provides an appealing way to explore local magnetic phase transitions and domain wall motions in magnetic materials. In addition to NV centers, it is also worth mentioning that spin defects hosted by two-dimensional materials, such as hexagonal boron nitride (hBN), can also be utilized to perform spin relaxometry measurements to detect longitudinal spin noise as demonstrated in $Fe_3GeTe_2$/hBN based van der Waals heterostructures very recently [Figs. 2(e)-2(f)].

Moving forward, promising directions include extending the presented spin-relaxometry sensing platform to magnetic quantum states of matter, such as spin liquids with frustrated magnetic ordering [19], Moiré magnetism with stacking engineered atomic registry and lattice interactions [20], and quantum anomalous Hall insulators with unconventional current flow patterns. We share the optimism that quantum magnetometry techniques could provide previously inaccessible information to explore unconventional spin dynamics at the forefront of condensed matter physics.

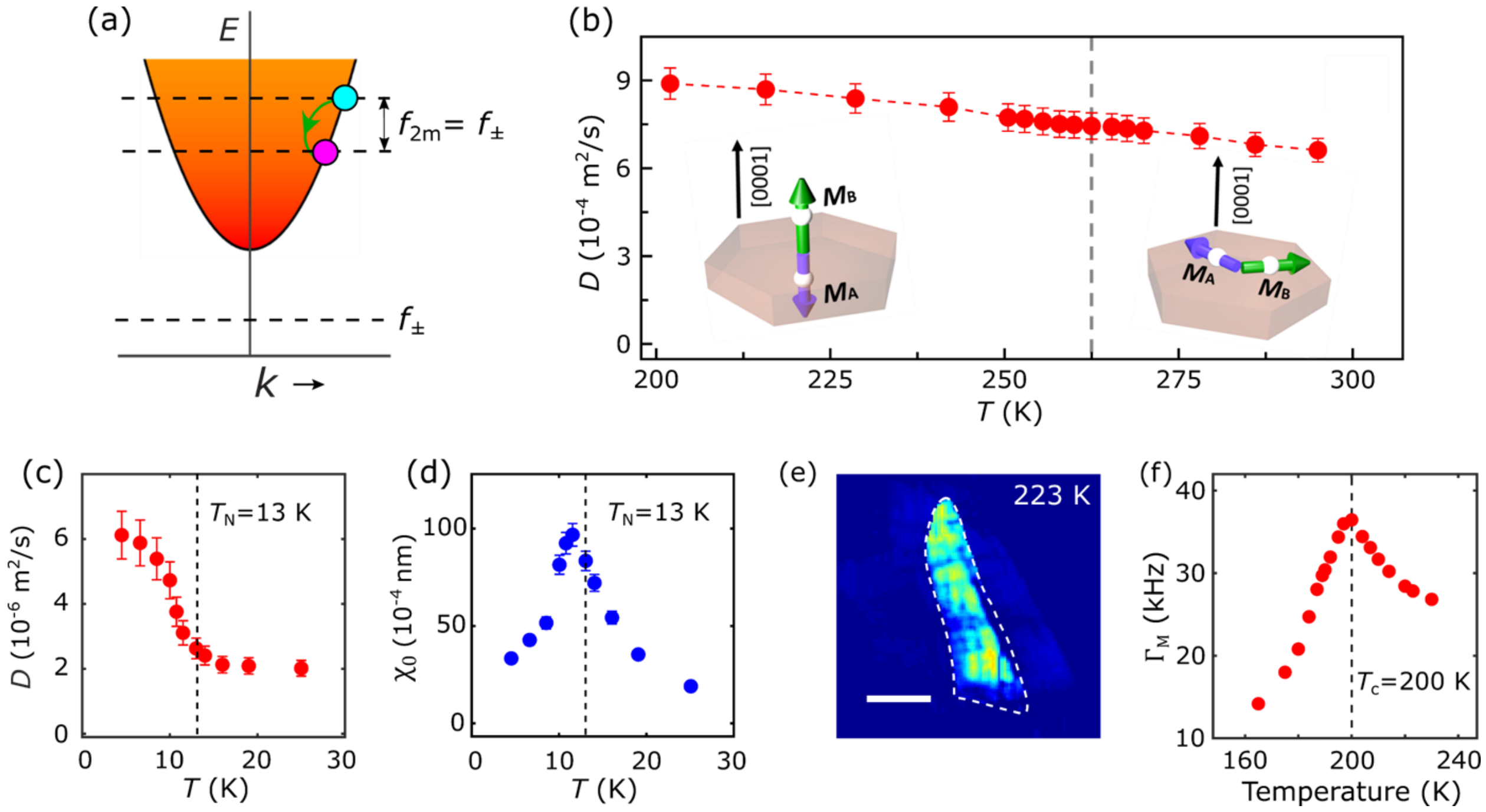

***Figure 2**: NV sensing and imaging of longitudinal magnetization fluctuations.* *(a) Schematic of the NV-resonant longitudinal spin fluctuations generated by mixing between thermal spin waves. (b) The intrinsic spin diffusion constant D of α-$Fe_2O_3$ measured as a function of temperature T. Insets: schematic views of the magnetic order of α-$Fe_2O_3$ below and above the Morin transition temperature. Figures (a)-(b) from* [10] *licensed under CC BY-NC-4.0. (c)-(d) Temperature dependence of intrinsic spin diffusion constant D (c) and longitudinal static magnetic susceptibility $X_0$ (d) of $MnBi_4Te_7$ measured by NV relaxometry techniques. Figures (c)-(d) are adapted with permission from* [17]. *Copyright © 2022 American Chemical Society. (e) Imaging longitudinal spin fluctuations in $Fe_3GeTe_2$ using spin defects in a boron-nitride capping layer. The white dashed line outlines the $Fe_3GeTe_2$ flake. The scale bar is 5 µm. (f) Temperature dependence of spin relaxation ($\Gamma_M$) of boron-nitride spin defects driven by the longitudinal spin fluctuations in a proximal $Fe_3GeTe_2$ flake. The black dashed line marks the Curie temperature of $Fe_3GeTe_2$. Figures (e)-(f) taken from* [18], *licensed under CC BY 4.0.*

**Acknowledgements**

C. R. D acknowledges the support from the Air Force Office of Scientific Research under award No. FA9550-21-1-0125. T. S. acknowledges support from the Dutch Research Council (NWO) under awards VI.Vidi.193.077, NWA.1160.18.208, and OCENW.XL21.XL21.058. A. Y. Acknowledges support from the ARO grant W911NF-22-1-0248, by the Quantum Science Center (QSC), a National Quantum Information Science Research Center of the U.S. Department of Energy (DOE), Gordon and Betty Moore Foundations EPiQS Initiative through Grant No GBMF9468, the STC Center for Integrated Quantum Materials NSF Grant No. DMR-1231319, and the Aspen Center of Physics supported by NSF grant PHY-1607611

# 3 Scanning Hall Microscopy

*David Collomb[1], Simon Bending[2], Ahmet Oral[3]*

[1] Quantum Brilliance GmbH, Tullastraße 72, Freiburg im Breisgau, Germany [0000-0001-5591-8802]
[2] Department of Physics, University of Bath, Bath BA2 7AY, United Kingdom [0000-0002-4474-2554]
[3] NanoMagnetics Instruments Ltd, Suite 290, 266 Banbury Road, Oxford OX2 7DL, United Kingdom [0000-0001-8972-4862]

**Current State of the Technique**

There are many complementary approaches to magnetic imaging and scanning Hall-effect sensors are often chosen because of their high magnetic field sensitivities and wide field range, quantitative linear response, non-invasive performance and fabrication versatility. This allows them to be employed in applications where other semi-quantitative and potentially invasive sensor types with higher spatial resolution, such as magnetic force microscopy (MFM) cantilevers, may not meet requirements. They are also much more compact and easier to use than the recently developed diamond Nitrogen vacancy (NV) centre magnetic microscope, which requires the precise fabrication of a single crystal diamond with an NV centre at the apex of an AFM tip as well as additional laser and microwave excitation. Although the spatial resolution of NV microscopy can in principle approach the atomic scale, imaging is relatively slow due to the need to average the signal at each pixel for typically for a few seconds, and its application is generally limited to low magnetic fields up to a few milliTeslas [1], while SHM is capable of operation up to several Teslas.

Magnetic imaging with Hall sensors dates back over 60 years, driven by their quantitative outputs and non-invasive operation. State-of-the-art Scanning Hall Microscopy (SHM) generally places a Hall sensor within a few micrometres of a Scanning Tunnelling Microscope (STM) tip patterned over a deep etched mesa, forming a planar device as shown in Figure 1. This is fixed to a fine positioning piezoelectric tube at a shallow angle with the tip closest to the sample surface. The SHM sensor can be brought into sub-nanometre tunnel contact with a sample and then scanned over the surface, collecting the Hall voltage at each pixel. By knowing the Hall coefficient of the sensor, calibrated magnetic field maps at the surface can be constructed. Early examples used Hall sensors a few micrometres in size patterned in thin films of Bi, InAs or epitaxial GaAs. The invention of modulation-doped semiconductor heterostructures [2] led to a renaissance of this field [3, 4] due to their low 2D carrier densities and very high carrier mobilities at low temperatures giving rise to minimum detectable fields, $B_{min}$, of <0.1μT $Hz^{-0.5}$ for 1μm sensors at T~4K. GaAs heterostructures remain the choice for cryogenic imaging but their mobilities deteriorate dramatically at room temperature and edge wall depletion limits their minimum dimensions to ≥100nm [5]. To reduce the spatial resolution 50nm Bi Hall probes with $B_{min}$ ~ 0.08 $mTHz^{-0.5}$ have been patterned in a polycrystalline film using a $Ga^{+}$ focused ion beam [6], however these suffer from poor mechanical and chemical stability. Narrow gap semiconductor sensors have also been widely explored due to their much higher 300K mobilities. For example, a 0.5μm InSb probe has been demonstrated with $B_{min}$ ~0.72$\mu THz^{-0.5}$ [7]. Although challenging, significant progress has been made in the growth of epitaxial InSb (and other III-V alloy) heterostructures in recent years allowing higher spatial resolutions to be achieved. The isolation of graphene in 2004 initiated a second renaissance in this field. The very low, tuneable carrier density of this atomically thin material and the extremely high 300K mobilities arising from its unique band structure, make it ideal for nanoscale Hall sensor fabrication [8]. Imaging with 1.5μm CVD graphene sensors has been demonstrated with $B_{min}$~ 20$\mu T/Hz^{0.5}$ [9] while 85nm Hall cross devices with $B_{min}$~ 59$\mu T/Hz^{0.5}$ were subsequently reported [10], both operated at 300K.

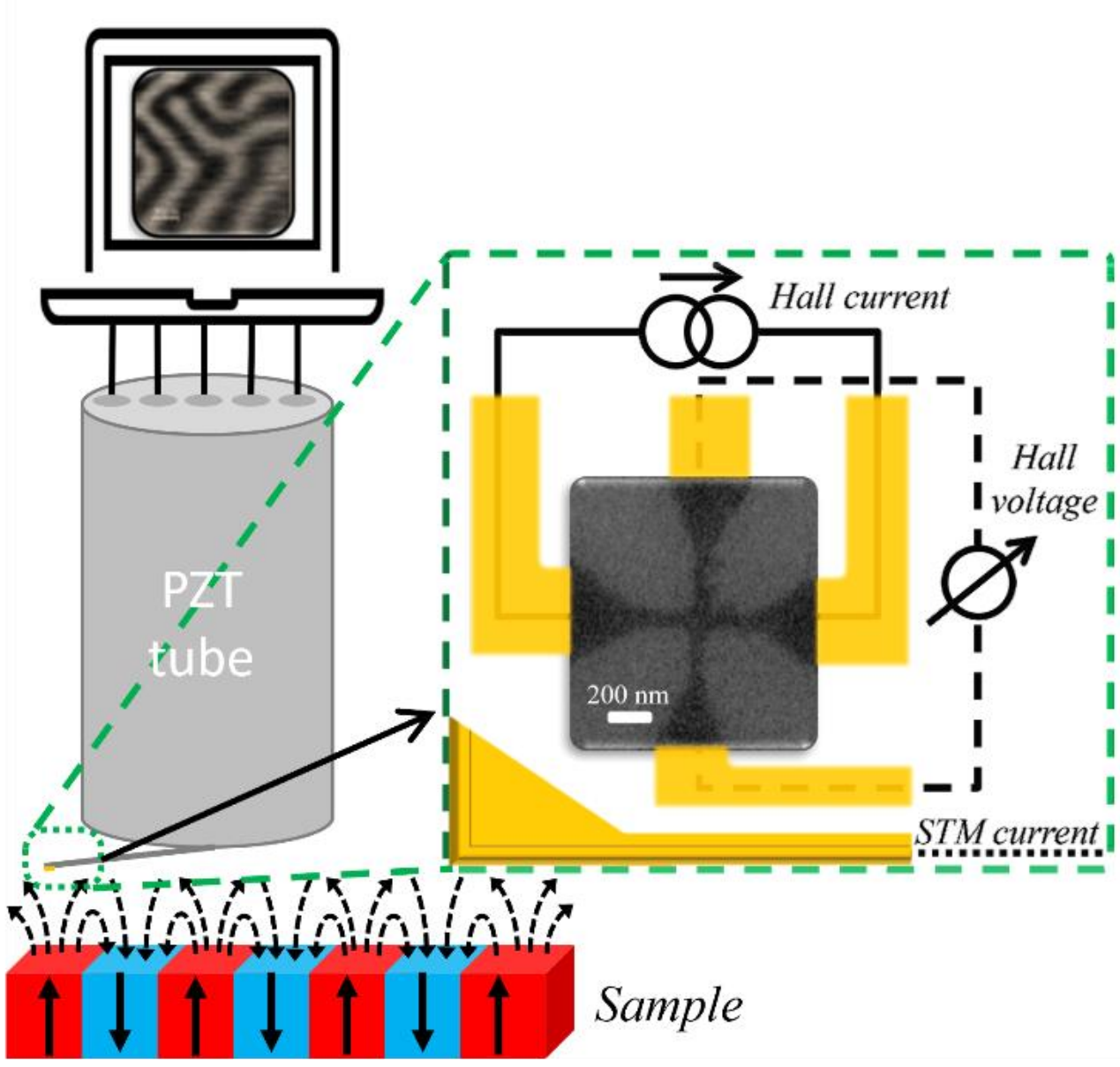

***Figure 1****: Schematic diagram of an SHM showing the planar Hall sensor with a gold STM tip positioned below a 4-quadrant piezoelectric tube. The example image in the top left shows magnetic domains in a ferrimagnetic Yttrium-iron-garnet (YIG) thin film [33]. Framed by green dashed lines is shown a diagram of a typical planar sensor with the Hall probe in close proximity to the apex of an STM tip formed via the deposition of gold over the edge of an etched mesa. The scanning electron micrograph shows an 85 nm wire-width CVD graphene Hall sensor [10].*

**Future Advances of the Technique**

The two foremost challenges in the field of Hall microscopy relate to the search for new materials to enable high performance operation at room temperature and the fabrication of sensors with deep sub-micron spatial resolution without severe degradation of minimum detectable fields. High magnetic field resolution requires a large Hall coefficient (low carrier density) and low Johnson and 1/f noise. These criteria are not satisfied in GaAs/AlGaAs heterostructures at 300K where the carrier mobility is low and 1/f noise dominates the output at low drive currents. Moreover, surface state pinning of the chemical potential leads to sidewall depletion, limiting the minimum practical size of sensors to ~100nm [11]. Although narrow gap III-V semiconductors exhibit much higher 300K carrier mobilities and have surface states that generally lie in the conduction band and do not lead to carrier depletion, they are typically patterned in films with thicknesses in the range in 200-500nm, leaving little scope for improved spatial resolution. The growth of epitaxial InSb heterostructures is challenging but in recent years modulation-doped InSb/$Al_xIn_{1-x}Sb$ quantum well structures have been realised with 300K mobilities up to 51,000 $cm^2/Vs$ for carrier concentrations as low as $5.8 \times 10^{11}$ $cm^{-2}$ [12]. These structures are typically based on ~30nm wide InSb quantum wells that lie ~50nm below the surface of the wafer and promise excellent performance for Hall microscopy applications with resolutions down to ~100nm. There exist many III-V material options and epitaxial heterostructure sensors with other alloy combinations have also been explored including InAs/$Al_xGa_{1-x}Sb$ [13,14], $Al_xIn_{1-x}Sb$/ $Al_yIn_{1-y}Sb$ [15], as well as AlGaAs⁄InGaAs⁄GaAs [16, 17].

Graphene's unique band structure yields very high 300K carrier mobilities while its single atomic layer form places few limits on nanoscale patterning. The different sources of graphene include chemical vapour deposition (CVD) graphene [18-21], epitaxially grown graphene on SiC [22] and mechanically exfoliated graphene [23]. CVD graphene is cheap and readily available in very large areas and despite its relatively defective granular structure still exhibits quite high mobilities. Nanoscale CVD graphene Hall sensors have been demonstrated with sizes as small as 85nm with excellent $B_{min}$ ~59 μT/ $Hz^{0.5}$ at 300K, patterned on Si/$SiO_2$ substrates by electron beam lithography and dry etching [10]. These devices could be tuned with a back gate allowing $B_{min}$ to be optimised close to the charge neutrality point with scope for further reduction in sensor size down to ~50nm when a rapid decrease in mobility due to edge scattering is expected to set in. CVD bilayer graphene Hall sensors have also been shown to exhibit higher sensitivity than comparable monolayer devices in the absence of substrate bias [24]. Epitaxial growth of graphene on SiC can achieve higher mobilities than CVD material, albeit at a higher cost and lacking the ability to tune the carrier concentration with a back gate. The highest carrier mobilities have been demonstrated in mechanically exfoliated graphene devices. The mobility of structures deposited on Si/$SiO_2$ substrates is limited by scattering at charge centres in the adjacent oxide layer. This can be suppressed by the encapsulation of graphene in two layers of hexagonal boron nitride (hBN) [25] when 300K mobilities as high as $10m^2/Vs$ have been demonstrated. This represents a very exciting area for future developments and $B_{min}$ as low as $0.7\mu T/Hz^{0.5}$ has been estimated in $1\mu m$ encapsulated sensors at 300K [26]. It remains to be seen if high mobilities can be retained in nanoscale devices, but from a fabrication perspective sizes comparable to the 10-20nm thickness of hBN capping layers should be achievable. Moreover, breakdown current densities for graphene nanoribbons in excess of $10^{12}A/m^2$ have also been demonstrated [27] and just working at 10% of this value would represent $30\mu A$ through a $1\mu m$ wide graphene strip with potentially achievable noise levels in the range $B_{min}$~0.01-0.1 $\mu T/Hz^{0.5}$ at 300K. The very high carrier group velocity associated with the linear electronic dispersion of graphene also enables very high frequency operation and graphene FETs working at up to 46GHz have been demonstrated [28]. Figure 2 summarises key milestones made in Hall sensor technology, with the recent developments in graphene and its encapsulation showing promise to lead the technique to higher spatial and field resolutions.

Scanning Hall microscopy requires the integration of a second sensor to control the scan height during imaging and/or map the surface topography. The combination of the two complementary data sets yields new information about the sample. The most common secondary sensor is a scanning tunnelling microscopy (STM) tip microfabricated as close to the active Hall element as possible. However, these cannot be used with insulating samples, driving the development of sensors with atomic force tracking. Hall sensors have been integrated onto micromachined cantilevers with piezoresistive deflection detection [29]. Simpler approaches to force tracking have been demonstrated whereby a nanoscale GaAs/AlGaAs Hall sensor was either glued to a GaAs cantilever sandwiched between two piezo plates [30] or to one of the tines of a quartz tuning fork [31]. There have also been several attempts to pattern Hall sensors directly on the end of AFM tips, although this has proved to be very challenging [32]. However, one can anticipate new developments in these areas in order to optimise the imaging spatial resolution.

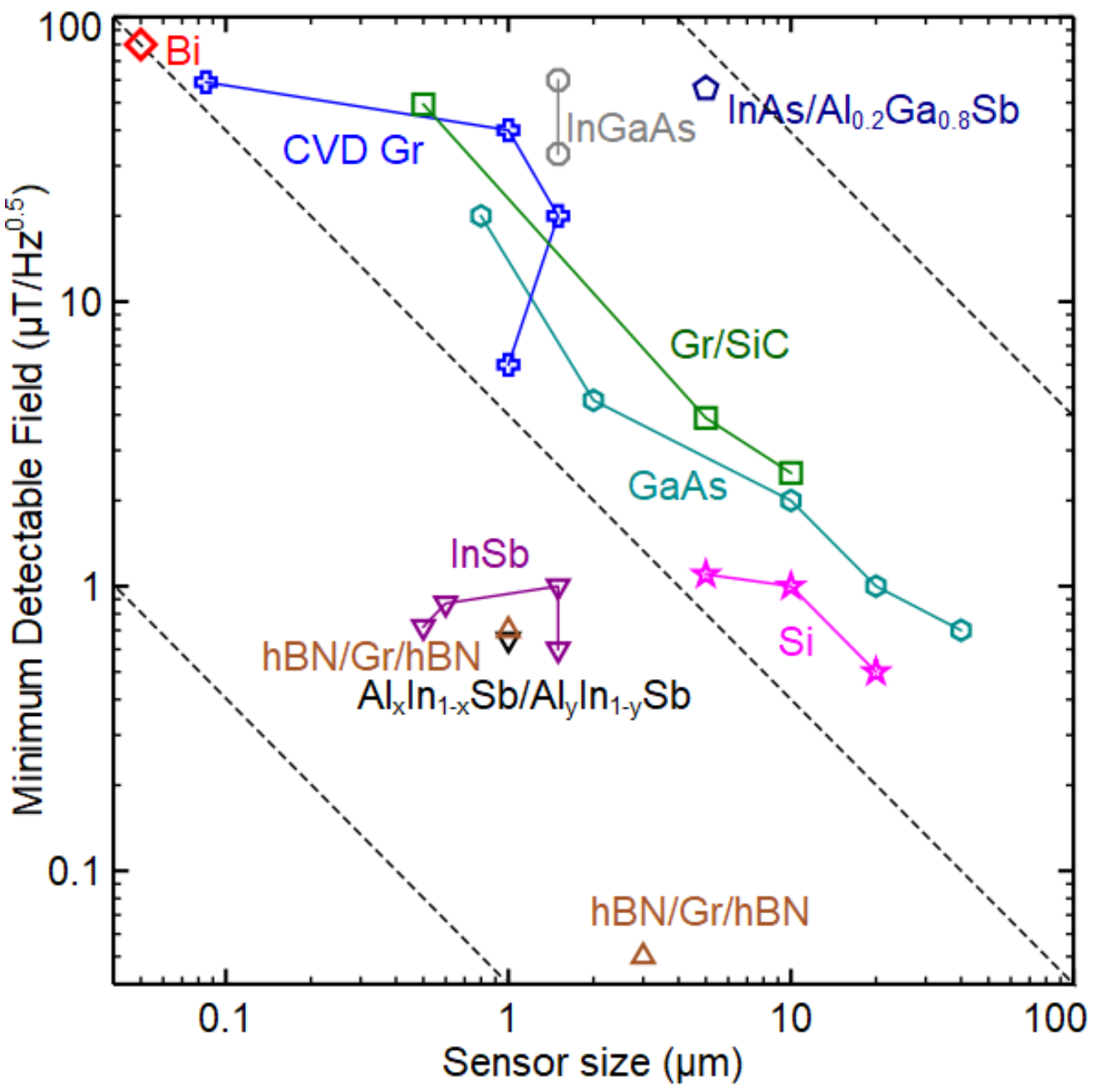

***Figure 2**: Minimum detectable field as a function of size at 300K reported for Hall sensors fabricated from various materials. Data were taken from the following sources. Bi [6], CVD Gr [9, 10], InGaAs [16, 17], InAs/$Al_{0.2}Ga_{0.8}Sb$ [13], Gr/SiC [22], GaAs [5], InSb [7, 14, 42], Si [5], hBN/Gr/hBN [26, 43], $Al_xIn_{1-x}Sb/Al_yIn_{1-y}Sb$ [15].*

## Current and Future Use of the Technique for Material Science

Unconventional type II cuprate high temperature superconductors (HTS) are notable for having a superconducting transition temperature, $T_c$, above 77K without the need to exceed atmospheric pressure. These have generated significant interest not just for their fascinating physics, but also for utilisation in energy-based applications such as lossless transfer of electricity from remote sources to their point of consumption. A major challenge in this development involves maximising the critical current density of manufactured HTS tapes through optimising the flux pinning without degrading $T_c$. SHM is well positioned to study these materials due to its quantitative nature, as well as its variable temperature operation from milli-Kelvin temperatures to well above the highest cuprate $T_c$. By mapping the pinning of vortices to defect sites as well as quantitatively measuring the temperature dependent vortex profiles, SHM provides critical feedback for the optimisation of flux pinning [33, 34, 35, 36]. The miniaturisation of Hall probes down to the nanoscale using graphene will enable an expansion of this research by improving the correlation between vortices and nanoscale pinning sites. This is in addition to being able to access higher magnetic field and applied current regimes where higher resolution is required to resolve the densely packed vortices.

Of particular interest for materials science today are magnetic Skyrmions. These topologically protected magnetic features, which can be as small as 1 nm in diameter, are predicted to represent discrete energetically favourable magnetic states. This has led Skyrmion-based logic and memory devices such as nanoscale racetrack-type technology or ferroelectrically tuneable Skyrmions in oxide heterostructures [37, 38] to be envisioned. Which materials Skyrmions manifest in and how they behave in their host material are questions not yet fully understood. To tackle these Skyrmions need to be visualised via real-space imaging, where the ability to resolve Skyrmions with sizes up to 100 nm is required. The recent improvements in this resolution range using graphene sensors demonstrates SHM's ability to play an important role in this field in the future, especially in materials that exhibit Skyrmions at room temperature.

By inverting the conventional set-up, whereby the Hall probe remains stationary and the magnetic object now rasters above the probe, one can use SHM to characterise an MFM tip's stray field, which allows for quantitative information to be extracted out of the otherwise qualitative MFM images [39]. Here, the tip is scanned over the active Hall probe and at each scan pixel a corresponding Hall voltage is measured. While the measured Hall voltage is dependent on the tip-sensor displacement, the magnetic moment of the tip, and the tip oscillation amplitude, an additional unwanted contribution arises from the electrostatic interaction between tip and probe. This unwanted interaction can be removed by also operating in a force modulated Kelvin probe force mode. With the recent advances in graphene-based Hall probes, this technique could be readily rolled out for calibration of commercially available AFM systems, allowing a user to unlock with ease quantitative MFM with high spatial resolution.

The growing complexity and scale of industrial manufacturing processes and infrastructure from metal bridges to pipework has placed a critical demand for a new arsenal of non-destructive testing (NDT) tools. By exploiting the continued miniaturisation of sensors, NDT tools can become capable of mapping defects and structural changes down to the nanoscale. Scanning Hall probes have also been used as an NDT tool through Scanning Hall Susceptometry. For example, a 200 µm spatial resolution SHM system was combined with a small electromagnet for the monitoring of heat-resistant steels as they age, where the microstructural changes correlated with a change in magnetic response [40]. However, such designs have relied upon micrometre resolution Hall probes, as well as using an excitation coil surrounding the sample. To advance the technique and its applications, nanometre spatial resolutions and the move to 'local' field excitations are desirable. By combining the recent developments in graphene-based probes with previous work integrating a field excitation coil and a Hall probe on the apex of an AFM tip [32], SHM could become a routine inspection tool in various manufacturing processes such as for the control of magnetic storage media or structural monitoring of metallic artefacts.

**Acknowledgements**

D.C. and S.J.B. acknowledge financial support from EPSRC in the UK under grant number EP/R007160/1 and the Superqumap COST Action CA-21144. DC was also supported by a PhD studentship from Lloyds Register Foundation ICON (award nos. G0086).

# 4 – Magnetic Force Microscopy

*Hans J. Hug[1,2], Andrada-Oana Mandru[1], Volker Neu[3], Hans Werner Schumacher[4], Sibylle Sievers[4], and Hitoshi Saito[5]*

[1] Empa, Materials Science and Technology, Ueberlandstrasse 129, 8600 Duebendorf, Switzerland [ORCID: 0000-0002-1281-4138 (HJH), 0000-0001-8778-8571 (AOM)]
[2] Department of Physics, University of Basel, Klingelbergstrasse 82, 4056 Basel, Switzerland
[3] Leibniz Institute for Solid State and Materials Research, Helmholtzstr. 20, D-01069 Dresden, Germany [0000-0001-9170-1280]
[4] Physikalisch-Technische Bundesanstalt, 38116 Braunschweig, Germany [ORCID: 0000-0002-6848-8984 (SS), 0000-0001-7100-682X (HWS) ]
[5] Department of Mathematical Science and Electrical-Electronic-Computer Engineering, Graduate School of Engineering Science, Akita University, Akita, Japan [ORCID: 0009-0000-9554-3590]

**Current State of the Technique**

Magnetic force microscopy (MFM) is a laboratory-based scanning probe technique that uses a magnetic tip on an oscillating cantilever to image magnetic stray fields emanating from a sample with a spatial resolution down to 10 nm [1]. MFM is a versatile tool for characterizing materials as it can simultaneously map the topography and local Kelvin potential [2] apart from the stray field, requires minimal sample preparation and can image magnetic layers covered by oxidation protection layers or thinner top contacts. Additionally, MFM can be performed under various conditions, including high magnetic fields and variable and cryogenic temperatures [3], and in liquid environments [4].

Typically, MFM is performed under ambient conditions using a line-by-line two-pass measurement method (Fig. 1). During the first pass, the tip intermittently contacts the sample while utilizing a feedback to trace the topography. In the second pass, the tip is lifted off the surface to predominantly measure the longer-ranged tip-sample interaction forces. These include the desired magnetic force on the tip that can be expressed as the convolution between the tip magnetization $\mathbf{M}_t$ and the derivative of the sample stray field $\mathbf{H}_s$ [1]:

$$\mathbf{F}_t(\mathbf{r},z) = \mu_0 \int_{V_t} [\mathbf{M}_t(\mathbf{r}',z') \cdot \boldsymbol{\nabla}]\mathbf{H}_s(\boldsymbol{r}+\boldsymbol{r}',z+z')\mathrm{d}\boldsymbol{r}'\mathrm{d}z' \qquad [1]$$

Magnetically-hard tips, where the magnetization $\mathbf{M}_t$ can be considered as fixed, are generally used. The force acting on the tip is detected via the phase shift signal of the oscillating cantil$\Delta\phi \propto dF_{\mathbf{n}}/d\mathbf{n}|_{\mathrm{eff}}$ , where **n** represents the canted cantilever normal and also the canted direction of the oscillating cantilever. The index "$|_{\mathrm{eff}}$" indicates that the derivative $\mathrm{d}F_{\mathbf{n}}/\mathrm{d}\mathbf{n}|_{\mathrm{eff}}$ is a weighted average of the local derivative over the oscillation amplitude of the tip [1].

Note that since $\mathbf{F}_t$ arises from a sample field-tip magnetization convolution, it depends on the spatial wavelength of the stray field pattern. Without calibration, the interpretation of results remains qualitative. It is also worth noting that an increased magnetic tip volume (e.g. thicker coating) increases the signal amplitude as per eq. 1. However, this comes at the cost of spatial resolution, and an increased risk of an alteration of the sample's micromagnetic state by the stray field of the tip.

Therefore, further advancements in MFM require improvements in signal-to-noise ratio, measurement functionality, data analysis, tip calibration procedures, and the development of more advanced magnetic probes.

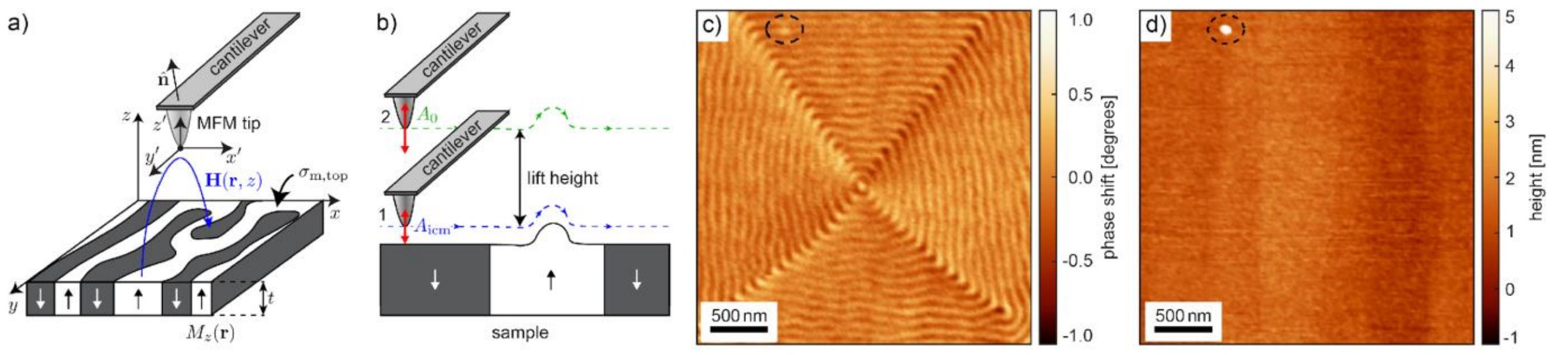

**Figure 1**:
a) coordinate systems and cantilever with normal vector $\hat{n}$ canted towards the surface of the magnetic thin film sample of thickness $t$ and an up/down domain pattern generating magnetic surface charge densities $\sigma_{m,top}(r)$, $\sigma_{m,bottom}(r)$, at the top and bottom surfaces of the film, respectively, and possibly magnetic volume charges, $\rho_m = -\mathrm{div}M$, all being a source of the magnetic field **H**(**r**,$z$).
b) lift-mode operation: each scanline is scanned twice. First the topography is recorded in the intermittent contact mode with an oscillation amplitude setpoint $A_{icm}$ that is smaller than the free oscillation amplitude $A_0$. The magnetic tip-sample interaction is then measured in the second scan which is performed at a pre-set lift height of the tip (dashed green line) above the previously recorded topography scan line (dashed blue line).
c) MFM image of weak stripe domains around a vortex core in a 40 nm thin epitaxial Fe-Co-C film with strain-induced perpendicular magnetic anisotropy [5]. d) topography data. Note that the small particle highlighted by the dashed ellipse also generates contrast in the "magnetic image" [dashed ellipse in panel c)] demonstrating that a complete separation between magnetic and topographical contrast contributions requires more advanced techniques as discussed in the following sections.

## Future Advances of the Technique

The following paragraphs detail proven methods to unlock the complete potential of Magnetic Force Microscopy (MFM). Presently, these methods are adopted by only a few research groups focused on advancing MFM. To make these advancements accessible to a broader research community, there is a need for instruments that are commercially available, along with standardized metrology procedures, software, and MFM tips that incorporate these advanced techniques.

*Improved Sensitivity and Advanced Operation Modes:* Working at minimum tip-sample distance during scanning is key to a large signal-to-noise ratio. Furthermore, sensitivities improve by up to x100 by operation in vacuum, using cantilevers coated only at the tip-end and by using softer silicon cantilevers. Fields down to a few tens of $\frac{\mu T}{\sqrt{Hz}}$ [6] can then be measured even with low magnetic moment tips [Fig 2a)]. However, more complex instrumentation, such as a phase-locked loop to drive the cantilever on resonance, along with non-contact operation and tip-sample distance control modes, are needed. An operational mode that provides extremely robust non-contact tip-sample distance control is based on the modulation of the tip-sample electric potential [7]. This leads to sidebands at $f_0 \pm f_{mod}$ and $f_0 \pm 2f_{mod}$, which can be used to map the local Kelvin potential (KPFM) and for a (slow) control of the tip-sample distance. This enables quantitative comparison of data acquired on different samples, as needed for quantitative MFM as an example. Differential imaging [Fig 2b)] [3] and KPFM techniques can be used to disentangle various contributions to the measured contrast [Fig 2c)]. In addition, a further miniaturization of the cantilever or the use of nanowire cantilevers is a promising approach to optimize the stiffness-to-resonance frequency ratio and thus the force sensitivity.

*Tip-response Calibration and Quantitative Measurements*: A calibration of the tip's response to a range of spatial stray field wavelengths [8] [9] [Fig. 2d)] allows obtaining the vector magnetic field [Fig. 2e)] from the measured MFM signal [8] or testing the fidelity of candidate micromagnetic structures [Fig. 2f)] [3]. Such a tip calibration can be performed by measuring a reference sample [8] [9] or by directly measuring the tip's stray field distribution [10]. Together with the disentanglement of different contrast contributions, quantitative MFM is a powerful method to resolve nanoscale magnetic textures [Fig. 2b) and f)] [3]. Future developments should include calibration in an extended range of spatial frequencies to address multiscale problems, and field-dependent calibrations. However, note that for advanced tip calibration and deconvolution procedures [9] high signal-to-noise ratio data are beneficial. This is achievable with high-quality factor cantilevers, which are best operated under vacuum conditions [6], and requires the use of advanced MFM instrumentation that is not yet commercially available.

*Advanced Magnetic Probes:* High-resolution imaging is typically performed with high-aspect ratio tips with tip radii below 5 nm, coated with a few nanometers of ferromagnetic material on one side. However, the magnetic stray field of the tip can still be too large, leading to changes in the sample's micromagnetic state. In such a case, the magnetic moment of the tip must be reduced, which can however not be achieved by further lowering the tip's coating thickness. Instead, alternative approaches, such as the use of low-moment magnetic alloys or the design of more sophisticated tip coatings, consisting of two antiferromagnetically-coupled ferromagnetic layers can be explored to further advance the technique.

Generally, tips permitting an adjustment of their magnetic state are beneficial for MFM [6]. Although the magnetic state of high-aspect ratio magnetic tips can be reliably flipped, it would be advantageous to change the tip's magnetic state at higher frequencies, potentially even in the presence of the sample. This would aid in separating magnetic contrast contributions from those arising from the sample topography.

One promising approach is the use of tips coated with a superparamagnetic granular alloy (SP tips), which allows for the adjustment of the size and direction of the tip's magnetic moment with an external field. The tip moment is proportional to the external field and follows it without hysteresis [11]. This technique, known as alternating MFM (A-MFM), significantly improves spatial resolution even at ambient conditions, provided that the sample magnetization is stable in the applied oscillatory field. In addition, A-MFM permits mapping the gradient of the sample vector field along the direction defined by the applied ac-field driving the tip with high precision [11]. Moreover, concepts to measure the magnetic field value including uniform field component are presently explored.

Freitag et al. [12] have developed a new cantilever design that uses ultra-soft bending modes and higher flexural modes to simultaneously measure stray fields and their derivatives at the sample surface. This approach can improve surface analysis in materials science, nanotechnology, and biophysics.

*Imaging Dynamic Magnetic Processes:* As a scanning technique and because of the small signal detected, MFM imaging is limited to line rates of a few Hertz. Imaging dynamic magnetic phenomena thus appears challenging. However, side-band or A-MFM enables the imaging of temporally oscillating processes, such as the magnetic excitation of superparamagnetic particles [13] or high-frequency fields from a recording head [14]. Recently, time-averaged MFM has also been used to image spatially varying processes, including nanoscale domain wall movements oscillating at frequencies up to 1 kHz, with the potential for measuring at hundreds of kHz in the future [15].

**Current and Future Use of the Technique for Material Science**

Each of the following subsections describes typical current application fields and then points out future applications that become accessible through a wider availability of more advanced MFM instrumentation or a further development of the MFM technique as outlined in the previous section.

*Data Storage Applications*: The initial MFM development was boosted by the need to characterize magnetic recording media [Fig. 2g) (i)] [16], to investigate the impact of specific head designs on the bit structure and storage density, and study magnetization reversal processes in longitudinal and perpendicular hard disk media [Fig. 2g) (ii)]. MFM was essential for the development of advanced data storage methods such as heat or microwave assisted recording [17] and domain-wall-based racetrack devices [18]. More recent research topics address topologically protected non-collinear magnetic structures [Fig. 2h)] [19].

*Characterizing the Micromagnetic State of Thin Film Multilayers*: MFM has become the standard method for characterizing magnetic thin films and multilayers [20], unravelling the physics governing their micromagnetic state and its dependence on applied magnetic fields and temperature [Fig. 2b) and f)]. Patterning film materials introduces new anisotropies, functionalities, and applications. MFM is applied to studying closure domain structures and magnetic vortices [Fig. 2i)] [21], [Fig. 1c)], and could reveal monopole-like excitations in artificial spin ice (frustrated magnets) [Fig. 2j)] [22] and the spatial profile of a spin-nano-oscillator [Fig. 2k)] [23].

*Materials and System Design based on the Quantitative MFM Analysis of the Micromagnetic State*: Through quantitative MFM, all vector components of the sample's stray field can be measured [1][9] and the comparison to micromagnetic models enables quantitative materials research [3]. An example is the correlation of the areal density of pinned uncompensated spins at the interface between ferromagnetic and antiferromagnetic layers with the exchange-bias field [Fig. 2l)] [24]. More recent quantitative MFM work in combination with micromagnetic simulations was performed to reveal the various reversal mechanisms of 3d metallic multilayers exchanged-coupled to rare-earth-transition metal alloys for spintronic applications [Fig. 2f)] [25]. Regarding topological chiral structures, MFM could determine the different soliton types present in these systems [Fig. 2m)] [26]. Very recent MFM work also addressed 2D materials, where different magnetic skyrmion types were investigated at low temperature in van der Waals heterostructures [27].

Permanent magnetic materials [Fig. 2n)] have been studied by A-MFM which is suitable for distinguishing topographical and magnetic contrast contributions relevant of these sintered materials with magnetic degradation-free grain boundary fractured rough surface [Fig. 2o)] [11]. Moreover, to measure the absolute local field which can distinguish magnetic and non-magnetic grain boundary phases will become available in the future.

Combining MFM with further techniques like Kelvin probe microscopy to eliminate electrostatic contributions enables a quantitative characterization of current distributions via the detection of Oersted fields to characterize leakage currents in integrated circuits for device failure analysis [28] and the local sensitivity of sub-micron Hall-crosses [29]. MFM was also combined with in-situ modulating electric fields, to image both magnetic and ferroelectric domains in multiferroic systems [30]. More recently, combining MFM with piezo-response force microscopy and scanning transmission electron microscopy lead to the realization of magneto-electrical-optical coupling in $BiFeO_3$-based thin films [31]. MFM is therefore expected to play a key role in novel low power consumption magnetoelectric memories research.

*Manipulation of Sample States by the Tip Field*: From the beginning of MFM, the tip stray field has also been used to locally manipulate nanoscale magnetic objects or magnetization structures, for example to nucleate and manipulate vortices in superconductors [Fig. 2p)]. Recent work addressed the nucleation and manipulation of skyrmions in chiral materials [32]. With advanced scanning control, these activities are expected to have significant impact in future studies of magnetic nanoparticles [33], artificial spin ice [22], and reconfigurable magnonic crystals [34].

*MFM and Magnetic Exchange Force Microscopy (MExFM) in UHV* (see also chapter 5): MFM performed under UHV conditions permits the study of reactive surfaces or magnetic system which must be prepared on single crystalline surfaces, which may gain importance particularly if combined with atomic-resolution AFM or Kelvin probe microscopy techniques. Furthermore, in UHV, MExFM can characterize magnetism on the atomic scale to image, for example, the antiferromagnetic unit cell in NiO [35] or, more recently, the cycloidal spin spiral in one monolayer of Mn on W(110) [36] [Fig. 2q)].

*Dynamic Phenomena Studied by MFM*: Existing time-averaged and time-resolved MFM techniques allow studying fundamental aspects of nanoscale domain wall oscillations [Fig. 2r)] [15]. These are not only relevant in novel magnetic nanodevices such as domain wall resonators [37] and memory devices, but also in sensor materials based on a domain wall-based magneto impedance effect [38] and in classical soft magnetic materials for transformer applications. Further advances in imaging dynamic magnetic processes would arise if one could rapidly switch the magnetic state of the tip on a time scale down to a few nanoseconds using optical triggering, independent from the state of the sample. This will allow for the implementation of highly resolving stroboscopic imaging techniques, thus advancing the development of modern spintronic devices.

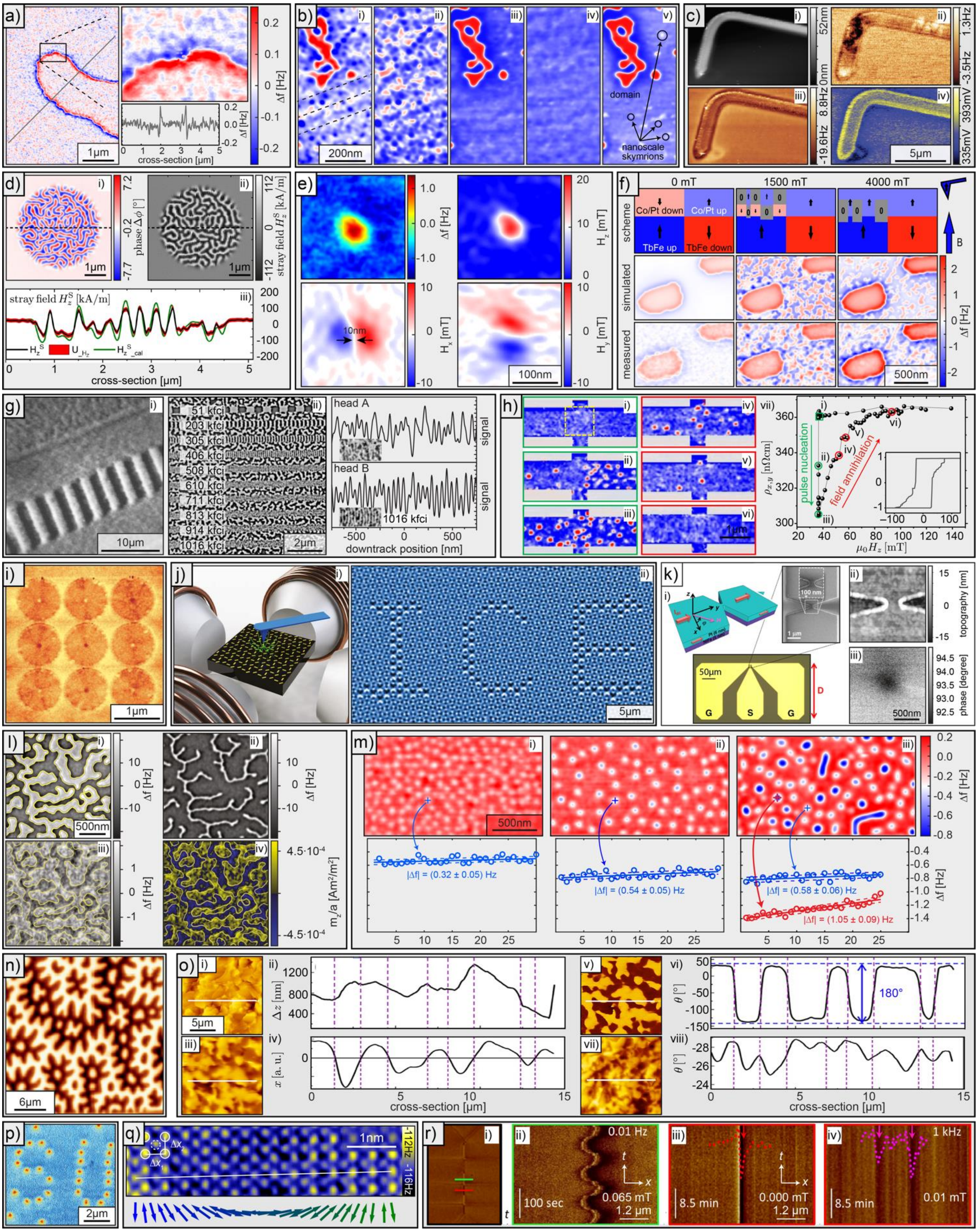

**Figure 2**:

a) **Advanced tip and cantilever design:** Overview image acquired with a vacuum MFM using an ultra-high quality factor cantilever shown together a zoomed image (shown right) acquired at the location of the rectangular box, and a cross-section. The red/blue frequency shift pattern arises from a local variation of the areal density of the magnetic moment corresponding to a variation in magnetic layer thickness by ± 1 atomic monolayer [ref. 6].

b) **Disentangling topography and magnetism:** a) Nanoscale skyrmions (panel v) in $SrIrO_3/SrRuO_3$ bilayers observed by MFM after a careful subtraction of the topographic (panel ii) and magnetic (panel iv) backgrounds from the raw MFM data (panel i) [ref. 3].

c) **Disentangling topography, magnetism, and Kelvin Contrast:** Co nanowire topography, Kelvin potential, and frequency shift image. The magnetic domain pattern could be revealed by compensating electrostatic force [ref. 2].

d) **MFM Tip Calibration**: MFM image of a patterned Co/Pt multilayered thin film (i) and thereof calculated stray field distribution at z = 60 nm (ii). Simulated and calibrated stray field data with expanded uncertainties are compared in the cross-section image (iii) [refs. 8 and 9].

e) **Quantitative MFM - from Δf to the vector components of the stray field:** MFM data of a skyrmion in a $[Ir(1)/Co(0.6)/Pt(1)]_{x5}$-multilayer measured 12.0 ± 0.5 nm above the surface (i). Vector components, $H_z$, $H_x$, and $H_y$, of the magnetic field obtained from the deconvolution of a using the tip calibration function [DOI: 10.1021/acs.nanolett.7b04802].

f) **Quantitative MFM – Simulation of MFM contrast from candidate micromagnetic structures:** Simulated (middle row) and measured (bottom row) MFM data of exchange-coupled double layers; the simulated MFM data was obtained by using the candidate magnetization structures shown in the top row [DOI: 10.1021/acsanm.9b01243].

g) **Data storage applications:** (i) 8µm wide track on a longitudinal magnetic recording medium (CoPtCr) [ref. 16]. (ii) Overview of tracks written by high-density magnetic recording head B. In contrast to head A, head B permits the writing of a 1016 kfci track (see track cross-sections right to the image) [DOI: 10.1016/j.jmmm.2004.10.048].

h) **Spintronic applications:** varying number of individual skyrmions in a Hall cross and corresponding transverse resistance $\rho_{xy}$ [DOI: 10.1038/s41565-017-0044-4].

i) **Magnetic vortices in nanostructures:** MFM data obtained on vortices in circular permalloy islands with 1 µm diameter and 50 nm thickness with different core polarities [ref. 21].

j) **Spin-ice**: experimental setup (i) and result of the local magnetic state manipulation (ii) of a magnetic charge ice [ref. 22].

k) **MFM on an operational spin hall nano-oscillator**: (i) Schematic (left top), SEM image of a nanoconstriction (right top), and optical microscopy image of the final device in a conventional design with top contact coplanar waveguides (with length of D) providing electrical access to the SHNO (bottom). topography (ii) and MFM image (iii) of an SHNO with a constriction width of 300 nm, operated at $H$ = 800 Oe and $I_{dc}$ = 6 mA [ref. 23].

l) **Mapping pinned uncompensated spins in antiferromagnet/ferromagnet exchange-biased samples:** MFM data obtained at 8.3 K on a Si/Pt(20)/ $Pt(2)[Co(0.4)/Pt(0.7)]_{20}/Co(0.6)/CoO(1)$ exchange biased sample (layer thicknesses in nm). panel i) after zero-field cooling showing the ferromagnetic domain pattern highlighted by the yellow lines. ii) domains in an applied field of 200 mT, and iii) saturated state. The yellow lines (outlines of the initial domain pattern) reveal that below a white domain there is a granular pattern with a dark contrast. iv) sub-monolayer areal density of the pinned uncompensated spins at the antiferromagnet/ferromagnet interface obtained from a quantitative analysis of the MFM frequency shift data depicted in iii) [ref. 24].

m) **Reproducible tip-sample distance control for quantitative MFM:** Determining the location of different skyrmion types (strong and weak |Δf|contrast - panel iii) within ferromagnetic /ferrimagnetic/ferromagnetic trilayers by performing MFM at the same distance and with the same (unaltered) tip on a subset of layers (panels i and ii) from the original trilayer [ref. 26].

n) **MFM on permanent magnet materials:** Domain contrast on bulk $Sm_2Co_{17}$ grain imaged with a high coercive ($\mu_0$Hc > 0.8T) $SmCo_5$ tip [DOI: 10.1039/c8nr03997f].

o) **A-MFM imaging of the topography and magnetic fields on rough surfaces using superparamagnetic tips:** Magnetic domain imaging of a very rough, fractured surface of Sr ferrite magnet without topographic crosstalk by A-MFM imaging of the topography and magnetic fields on rough surface using superparamagnetic tips i), ii) topographic image and its line profile of grain boundary fractured surface of Sr ferrite magnet. iii), vi) magnetic field gradient image and its line profile with fixed measuring direction. v), vi) magnetic phase image and its line profile. 180º phase difference corresponds to the up and down magnetic field direction. vii), viii) conventional MFM image and its line profile, with uncertain measuring direction [ref. 11].

p) **Manipulation of the micromagnetic state:** Manipulation of vortices in Nb thin film and observation at 5.5K [DOI: 10.1063/1.3000963].

q) **Magnetic Exchange Force Microscopy:** Atomic resolution exchange force frequency shift image of a Mn monolayer on W(110) measured at constant height relative to a reference tip-sample distance obtained by STM operated at $U_{bias}$ = -10mV and $I$ = -2nA reduced by 0.29nm. As a cantilever a tuning-fork sensor is used oscillated with an amplitude of 50pm. The bottom part shows a sketch of one half of the cycloidal spin spiral along the [1-10] direction [ref. 36].

r) **Imaging dynamic phenomena:** Domain wall oscillation in a permalloy Landau structure. (i) demagnetized state, (ii) Real-time imaging of the quasi-static domain wall oscillating with 0.01 Hz (iii) Static DW in the x–t plane, (iv) Time-averaged imaging of the DW oscillating with a frequency of 1 kHz ($\mu_0 H_{ac}$ = 0.10 mT) in the x–t plane [ref. 15].

## Acknowledgements

The A-MFM measurement system was developed by the support of JST (Japan Science and Technology Agency)/SENTAN Program in 2008-2015.

# 5. Spin sensing on the atomic scale with scanning tunneling microscopy and non-contact atomic force microscopy

*A.A. Khajetoorians[1], 0000-0002-6685-9307*
*N. Hauptmann[1], 0000-0003-4264-2493*
*S. Baumann[2] 0000-0002-3821-494X*

[1] Scanning Probe Microscopy Department, Institute for Molecules and Materials, Radboud University, Nijmegen, The Netherlands
[2] University of Stuttgart, Institute for Functional Matter and Quantum Technologies, Stuttgart, Germany

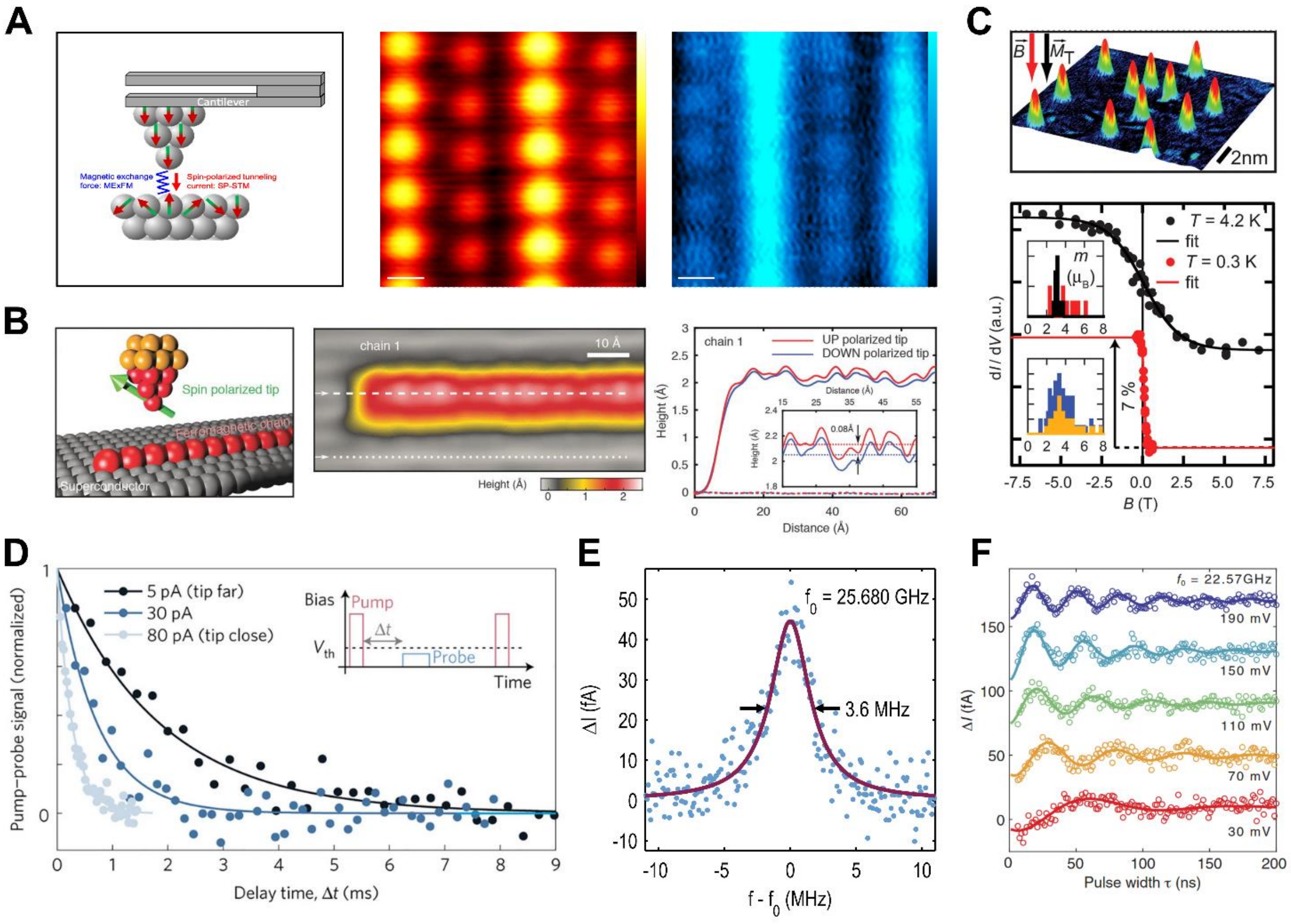

Figure 1:
A: Schematic and imaging of a spin spiral based on the bilayer of Fe/Ir(111), using a combination of SP-STM and MExFM with a tuning fork sensor (bulk Fe tip, T = 6.3K) [11].
B: Schematic and SP-STM measurements of the magnetic state of a self-assembled Fe chain on Pb(110) and its interplay with superconductivity (Cr tip with Fe cluster at apex, T = 1.4 K, B = 0 T) [24]
C: Single atom magnetization curve measurements of individual Co atoms on Pt(111), using SP-STM (Cr tip, T = 0.3 K image, curves: 0.3 K, and 4.3 K respectively). [4]
D: Pump-probe measurement of an individual Fe atom on a thin layer of MgO/Ag(100), illustrating the effect of the tip proximity on the relaxation time (T1) of the atomic spin. (Ir tip with Fe cluster, T = 0.6 K , B = 2 T) [16]
E: Measuring the electron spin resonance of an individual Fe atom on a thin layer of MgO/Ag(100), using ESR-STM. (Ir tip with Fe cluster, T = 0.6 K, B = 5.7 T (with $B_z$ = 0.2 T)) [7]
F: Coherent excitation of an individual TiH molecule on a thin layer of MgO/Ag(100), using pulsed ESR-STM (Ir tip with Fe cluster, T = 1.2 K, B = 0.9 T) [8]

## Current State of the Technique

Spin sensing on surfaces with atomic-scale resolution has been achieved using two different, but compatible methods: scanning tunneling microscopy (STM), and non-contact atomic force microscopy (nc-AFM) [1]. STM-based spin sensing methods rely on local detection of a spin-component of the tunneling current, whereas nc-AFM-based spin sensing methods rely on the detection of local forces related to magnetic exchange. Spin-polarized scanning tunneling microscopy (SP-STM) is the magnetic sensing technique based on scanning tunneling microscopy/spectroscopy that uses a probe with a non-zero magnetization. This leads to an additional contribution to the measured tunneling current that is sensitive to the projection of the tip and sample magnetization, analogous to a magnetic tunnel junction. The method usually requires an applied external magnetic field to measure changes of this projection to infer the change in magnetization. SP-STM has been used to image and manipulate magnetic order on a variety of metallic surfaces [2, 3]. Most recently, it has been used at sub-Kelvin to mK temperature to measure the magnetization of individual and coupled atoms [4, 5] as well as to measure spin dynamics [6-8].

Magnetic exchange force microscopy (MExFM) is a variant of nc-AFM which is based on measuring the short-range magnetic exchange force between a magnetic probe and sample [9]. This is in contrast to magnetic force microscopy (MFM), which is sensitive to long-range magnetic dipole forces. MExFM detects the local exchange force between the probe and sample, and therefore is a short-range method that can atomically resolve the magnetic ground states of materials, independent of current, but also be combined with SP-STM detection [10, 11]. MExFM can be also used to quantify the distant-dependent exchange forces between a tip and sample [10, 12, 13].

While both SP-STM and MExFM provide spatial resolution down to individual atoms, historically they suffered from poor time resolution (> microseconds). However, over the last decade, new capabilities have been developed to extend the spatial and temporal resolution, currently reaching all the way to the picosecond-range [14]. These include all-electrical pump-probe spectroscopy, and stochastic resonance spectroscopy [6, 15, 16]. Additionally, electron spin resonance techniques have been developed in order to coherently excite atomic spins on surfaces [7, 8]. In terms of magnetic sensing, the all-electrical based methods have been used to extract weak magnetic interactions, as well as extract the intraatomic hyperfine interaction of an individual atom [17, 18].

## Future Advances of the Technique

The future of magnetic sensing at the ultimate smallest length-scales involves pushing beyond simple imaging for example by adding new physical understanding by changing the detection schemes for MExFM or by extending the temporal resolution for SP-STM. Furthermore, integration with other methods will be a key advancement for either technique. Here we focus specifically on: (1) instrumental development of MExFM, (2) integration of spin sensing with optical time-resolved techniques, (3) linking local spin sensing with other methods in single measurement setups.

### *MExFM*

An advancement of the MExFM method would be its implementation to $^3$He temperature and below to access fragile quantum states of matter that emerge at low temperature. However, there are multiple hurdles toward this end based on signal detection. MExFM is based on a tuning fork design in order to facilitate simultaneous current detection, and requires signal amplification near the cold junction. But conventional electronics require higher temperature to properly operate, and typically cannot withstand high magnetic fields.

In addition, further developments need to be made on the magnetic probe used in the method. Up to date, primarily ferromagnetic probes have been used [9, 10, 19], for example 3*d* transition metals [9]. The use of ferromagnetic probes strongly limits the ability to detect robust antiferromagnets on surfaces, as well as to investigate the magnetic structure without additional perturbations driven by large stray fields. Thus, using antiferromagnetic probes for exchange force detection, with minimal stray fields and a much more robust stable magnetic structure, will be key. Moreover, further developments to measure the distance-dependent exchange force are necessary and depend on controllable functionalization of the tip. Here, tip functionalization, as done with CO molecules to enhance spatial resolution [20], will allow to tune the exchange mechanisms responsible for the measurement and to quantify the magnetic exchange energy landscape between different chemical terminations of the probe and the surface.

***Combinations with time resolution***

STM/AFM methods have been able to sense fast magnetic switching times. However, currently spin sensing is still limited to electrical detection, and thus in the range of 100 ps. Yet, there are possibilities to improve these magnetic sensing capabilities to extend the time resolution by orders of magnitude. In particular, methods using ultra-fast THz [14] or other fs-optical pulses as excitation pulses, to excite optical transitions, vibrational modes or phonons, combined with the magnetic sensitivity and high spatial resolution in STM promise new insight into ultra-fast processes in materials and can even give access to unknown, short-lived transient states.  Similarly, experiments that involve light detection, rather than excitation, allow for studying new material properties once magnetic sensitivity is added to the existing capabilities. The combination of electrical and optical excitation/detection will create unique possibilities for example to study the dynamics of coupled modes, such as magnon-phonon coupling. Furthermore, extending the use of frequency-resolved methods, such as stochastic resonance spectroscopy or ESR-SPM [7, 8, 15,28], will allow for new ways to map out driven dynamics in complex materials with sub-Angstrom resolution. Thereby short-lived excited states of matter, that are not easily detectable with established methods, can be probed and utilized.

***Hybrid measurements with spin sensing***

While spin sensing with STM/nc-AFM based methods boasts fantastic spatial and time resolution, there are extreme demands on the measurement conditions. Due to the sensitivity of the probe, samples are typically studied in ultra-high vacuum (UHV) environments and require atomically clean surfaces to probe, often restricted to single crystalline surfaces. These combinations make it extremely challenging to apply more advanced sample fabrication, such as systems that are lithographically defined. Therefore, one of the developments in the field includes being able to explore a vaster amount of samples, such as heterostructures synthesized *ex situ* and *in situ*. Moreover, there is a continuing trend in combining STM and nc-AFM based methods, with others, such as transport-based measurements. Future spin sensing developments in this realm will also involve computability with such methods, as well as integration with other methods reviewed in this roadmap.

**Current and Future Use of the Technique for Material Science**

The current state-of-the-art as well as many of the cutting-edge developments outlined here related to spin sensing with STM/nc-AFM-based methods will be extremely valuable for future fundamental characterization in a variety of material systems. A key advantage of these scanning probe methods over most other magnetic sensing techniques is their unprecedented spatial resolution which adds local information to the processes present in complex materials that are otherwise impossible to resolve. This local information can bring understanding to the influence of defects and other impurities in materials, for example in their role as pinning centers or photon emitters, or it can be used to resolve local structural rearrangements, e.g. due to correlated electronic states. Thus, this information can guide future developments of materials with specific functionality driven by their local composition and/or their local response to time-dependent excitation.

***Lower dimensional materials***

Understanding the local electronic structure of correlated states of matter and its interplay with spin is currently a vast field of research, examples of which are charge/spin density waves or Mott insulator phases. This particularly includes looking at the role of dimensionality in layered, van der Waals materials [21]. Currently, the local interplay of spin and charge carriers in such materials remains elusive, in both the absence as well as in the presence of a magnetic field. In this realm, local spin sensing based on combinations of STM/nc-AFM will be vital in understanding the role of the spin degree of freedom. Similarly, there are many open questions linked to the role of defects in these materials, and their influence on static spin ordering as well as magnetic excitations, such as magnons. Likewise, probing spin textures in novel lower dimensional superconductors, and imaging new phases of matter, will be of key interest. Furthermore, the role of dimensionality is becoming more and more relevant in 2D and 1D magnets, in understanding the role of various exchange mechanisms, as well as in less frequent types of ordering, such as ferrimagnetism.

***Complex magnetic states***

It is imperative to understand how complex magnetism emerges based on the interplay between structure and interactions, in order to create new generations of magnetic materials. For example, $\pi$-orbital magnetism, as derived from carbon-based materials like graphene, is extremely difficult to detect without local probe measurements [22]. Understanding the potential of magnetic ordering in these materials, and manipulating the spin degree of freedom is possible with the development of STM/nc-AFM based methods. On top of that, STM/nc-AFM based methods can allow for on-surface synthesis of such systems which opens up new possibilities for future design directions. In addition to $\pi$-orbital based magnetism, other types of magnetic order are emerging which necessitate a local characterization of the spin structure. This includes looking at new types of chiral ordering, e.g. skyrmion-based materials, as well as at magnetic states observed in frustrated magnets, such as self-induced spin glasses [2, 3]. Spin based imaging will also be relevant in looking at material systems with complex phase diagrams, for example in the case of 1D ordered systems which exhibit spin-charge separation [23], as well as systems described by multiple order parameters (such as multi-ferroics). Finally, spin sensing based on STM/nc-AFM methods will be vital in understanding spin in topological materials [24], especially in edge states [25].

***Atomic-scale spin dynamics***

The ability to resolve spin behavior at the atomic scale and with high temporal resolution opens up the pathway to understand atomically resolved spin relaxation dynamics as well as to quantify coherence times and ultimately coherent control of spins [6-8, 15]. The state-of-the-art plus future development of spin sensitive time resolved methods based on STM/nc-AFM may address multiple different material systems. With the further development of THz based STM/AFM interrogation [14], it may be possible to resolve magnon dynamics, or coupled magnon-phonon modes, at the atomic scale in a variety of ordered magnetic structures, with spatial resolution. With the advancement of stochastic resonance spectroscopy and electrical pump-probe spectroscopy [6, 15], it is also possible to probe the relaxation dynamics of coupled spin structures, for example those that exhibit complex relaxation pathways due to various Markov processes [26]. Using similar techniques, it may be possible to look at non-equilibrium dynamics in 1D and 2D spin structures on surfaces, for example those that are chiral, which self-assemble or are artificially constructed. Using a combination of relaxation dynamics and coherent dynamic techniques, based on continuous wave excitation, it may also be possible to understand local ordering in fragile states of matter, like quantum spin liquids as well as pursue coherent manipulation of quantum nanostructures that rely on spin-based read-out [27].

**Acknowledgements**
This project has received funding from the European Research Council (ERC) under the European Union's Horizon 2020 Research and Innovation Programme (grant agreement no. 818399 and 947717), NWO-VICI programme (VI.C.212.007), and from the Baden Wurttemberg Foundation Program on Quantum Technologies (Project AModiQuS).

# 6. Nanoscale Magnetic Resonance Imaging

*Alexander Eichler[1] and Christian L. Degen[1]*

[1] Laboratory for Solid State Physics, ETH Zürich, 8093 Zürich, Switzerland
Orcid accounts: https://orcid.org/0000-0001-6757-3442 and https://orcid.org/0000-0003-2432-4301

**Current State of the Technique**
Nanoscale Magnetic Resonance Imaging (NanoMRI) was suggested by theoretical physicist John A. Sidles in the early 1990s as a means to resolve biomolecular structures with three-dimensional atomic resolution [1,2]. At that time, Sidles worked at the Department of Orthopedics at the University of Seattle, and understood how useful it would be to image individual complex molecules, such as single proteins, on the atomic scale. His proposal, known as magnetic resonance force microscopy (MRFM), came shortly after the invention of the atomic force and magnetic force microscope (AFM and MFM, respectively, see Section 4). The vision of MRFM relies on mechanical force sensors to detect the weak magnetic signal of (ideally) single nuclear spins. Technically, its function is similar to that of a magnetic force microscope (see section 4), while using the pulsed radio-frequency techniques known from MRI to manipulate the spins and encode information.

Over the course of the past 30 years, MRFM has enabled impressive advances in NanoMRI. Experimental milestones include the detection of a single electron spin in 2004 [3], the 3D imaging of tobacco mosaic viruses at a resolution of 5-10 nm [4], and the demonstration of magnetic resonance diffraction with subangstrom precision [5]. By comparison, conventional MRI used in clinical medicine has a detection limit of order $10^{18}$ nuclei, limiting the technique to the millimeter lengthscale.

MRFM has not remained the only approach to NanoMRI. Over the past decade, other types of magnetic sensors have emerged that allow for the detection and mapping of magnetic fields at the nanometer scale [6]. Most prominently, these include NMR sensors based on single nitrogen vacancy (NV) centers in diamond, as well as the technique of electron spin resonance scanning tunnelling microscopy (ESR-STM), illustrated in Fig. 1 (see also Sections 2 and 5). The various NanoMRI techniques not only differ in their experimental approach and sensitivity, but also in their field of view and scope of use. The field of view is roughly determined by the physical size of the field gradient source. It ranges from >100 nm for a nanomagnet employed in MRFM to <10 nm for the atomic-sized sensors associated with an NV center or STM tip. It is therefore likely that the different techniques will find application for purposes that require either large imaging depth or optimal spatial resolution [6]. In addition, some techniques are compatible with room-temperature operation while others require cryogenic vacuum environment.

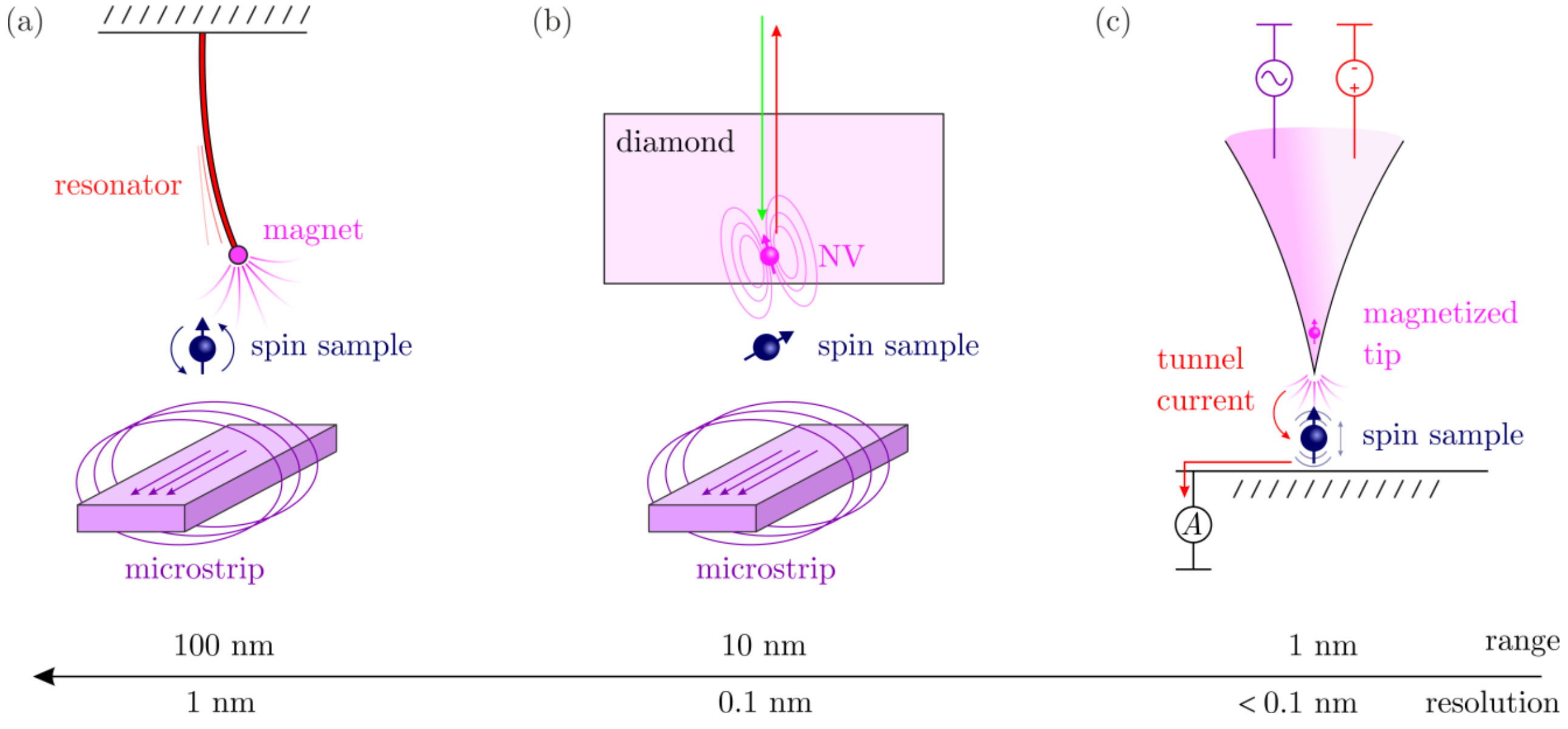

***Fig. 1: Popular NanoMRI techniques**: (a) MRFM uses a nanomechanical resonator together with a nanoscale magnet to generate a force on the spins in the sample, which is detected as a displacement. A microstrip serves as an antenna that generates radio-frequency pulses for NMR spin inversion [4]. (b) In NV-based NanoMRI, an electronic defect spin situated within nanometers from the diamond surface acts as the magnetic sensor [7-9]. The sensor spin is optically initialized and read out, as shown by green and red beams. NMR detection is achieved by a periodic modulation of the sensor spin at the nuclear Larmor frequency. The local gradient source for MRI is either provided by the electronic sensor spin itself or by an additional nanoscale magnet. (c) In ESR-STM, a tunnel current flowing between the sharp scanning tip and the conducting substrate serves as the signal. When the scanning tip is magnetized, the current depends on the alignment between the tip and surface moments, permitting detection of the surface spin orientation. Microwave pulses applied to the scanning tip can be used to generate an effective oscillating magnetic field permitting ESR of the sample spins [10,11].*

**Future Advances of the Technique**

In order to study the structure of complex molecules, NanoMRI will need to image magnetic signals with sub-nanometer resolution over distances from 10-100 nm. Approaching this goal requires advances in many directions including sensitivity, spatial selectivity, long-term stability of the apparatus, and sample preparation. The various NanoMRI subcommunities are currently pursuing different trajectories to reach this goal. To give an impression of the diverse aspects entering this process, we highlight below a few selected efforts:

*Surface functionalization* – As the technique operates on the nanometer scale, control of surface properties is paramount. For example, the coherence time and charge stability of NV centers is strongly impacted by imperfections on the diamond surface [12]. Likewise, the exact surface termination plays a crucial role in the intrinsic dissipation of nanoscale mechanical oscillators [13] and in non-contact friction effects as the oscillators are approached to the sample surface [14]. Much work is therefore invested into mitigating the harmful effects of surface imperfections on the sensor performance. Obviously, the exquisite atomic surface control in ESR-STM is a great advantage of that technique.

*Signal-to-noise ratio* – The central bottleneck of NanoMRI is detection sensitivity. This sensitivity can be improved both through optimized sensor geometry and by a reduction of sensor noise. Nanomechanical sensors are typically limited by thermomechanical force noise, which is why the most sensitive experiments are performed at cryogenic temperatures. Furthermore, mechanical sensors are designed with the aim of minimizing the damping coefficient $\gamma = \omega_0 m/Q$ where $\omega_0$, $m$, and $Q$ are the sensor's angular resonance frequency, mass, and quality factor, respectively. There are two main strategies to achieve low $\gamma$: on the one hand, long and thin cantilevers or nanowires have low $m$ and $\omega_0$ [6]. On the other hand, strings or membranes made from e.g. strained nitride have higher $m$ and $\omega_0$ but can possess enormously high $Q$ [6,15]. NV-based NanoMRI, by contrast, is limited by optical shot noise as well as the spin coherence time; here, the avenues towards improvement are a photonic structuring of the sensor, schemes to enhance the optical contrast and readout fidelity, as well as dynamical decoupling detection protocols [16].

*Compressed sensing and image reconstruction* – In most NanoMRI applications, the averaging time required per voxel is significant and leads to timescales of hours, days, or even weeks per image. Many strategies have been devised to reduce this time. A universal approach to speeding up the data taking is through "compressed sensing" [17]. In this method, the measurement apparatus samples only a random subset of the points on the sampling grid in real space. Since NanoMRI data is rarely acquired in real space, such undersampling can result in almost the same image quality as a fully sampled data set after the reconstruction step. Beyond data acquisition, efficient and robust routines for reconstructing the three-dimensional spin density from datasets will be further developed [18,19].

*Hyperpolarization* – Conventional MRI relies on the bulk magnetization that is generated in the sample by the large polarizing field of a superconducting magnet. Although the thermal polarization is very small (ca. 1 in $10^5$ nuclei), this is compensated by the large number of nuclear spins in a voxel (>$10^{18}$). By contrast, in NanoMRI, voxels only contain 1-$10^6$ nuclear spins, and the thermal polarization becomes minute. Consequently, many NanoMRI experiments rely on statistical polarization, which, however, has many drawbacks in terms of signal averaging and spectroscopy. A promising approach to increasing the sensitivity while maintaining the advantage of bulk polarization are hyperpolarization techniques, where nuclei are polarized via electronic spins or optically. Nuclear spin alignment can either be achieved by an external polarizing agent [20] or, for NV-NMR, by the optically-polarized NV center itself [19]. Future implementations of NanoMRI are expected to take advantage of nuclear hyperpolarization.

**Current and Future Use of the Technique for Material Science**

A grand goal of NanoMRI is the imaging of single molecules with true three-dimensional resolution, thus extending clinical MRI to the realms of structural biology and nanoscale materials research. Ideally, this imaging will have near-atomic resolution and the full chemical specificity known to NMR and MRI. One key advantage of NanoMRI over competing methods such as X-ray crystallography or cryo-electron microscopy is the fact that it is non-destructive, allowing to scan a single specimen various times without damaging it and obviating the need for crystallization or ensemble averaging. The results of such a scan will be complementary to computer-based methods aided by artificial intelligence, which recently have made encouraging progress. As in many other fields, simulations and direct measurements will contribute different information to provide a mature and powerful platform for structure determination.

Beyond structural biology, NanoMRI is expected to feature many interesting niche applications in various areas of materials science, spin physics, and quantum engineering. In materials science, the method can be used to study paramagnetic defects and spin diffusion in surface layers [21]. For example, magnetic defects contribute to 1/f noise in superconducting qubits, limiting their longer coherence time. Such undesirable surface defects are difficult to investigate with other methods, who lack the sensitivity and elemental resolution to clearly identify them at the nanoscale. Further, NanoMRI will enable the study of spin diffraction in crystalline and amorphous solids at the scale of lattice constants and inter-atomic distances [5]. This capability will be useful for studying spin interactions in periodic materials, investigate spin transport, and identifying mechanisms of decoherence and dissipation. Another interesting application is the imaging of multi-qubit spin registers in solid materials platforms [18]. The capability to directly image and characterize individual spin registers is valuable to build a solid understanding of the interaction in the host material, and to thereby improve the platform quality. Such imaging and control showcase the potential of NanoMRI for designing functional materials for quantum applications.

Finally, a most exciting and fruitful aspect of NanoMRI are the many technical improvements made to sensor platforms, such as pristine mechanical resonators, ultrapure diamond, or spin control and image reconstruction methods. These improvements are by themselves a driving force towards better materials systems and instruments. The exciting technologies that are being envisioned in the quantum sensing and nanotechnology communities depend on such advances to turn our improved understanding of nature at the atomic scale into concrete and useful applications.

**Acknowledgements**
The authors thank Fabian Natterer, Konstantin Herb, Nils Prumbaum, and John Abendroth for discussions.

# 7. Magnetooptical imaging

## 7a. Kerr microscopy

*Jeffrey McCord[1,3], Michael Vogel[2,3]*

[1] Department for Materials Science, Kiel University, Kiel, Germany [ORCID: 0000-0003-0237-6450]
[2] Department for Materials Science, Kiel University, Kiel, Germany [ORCID: 0000-0001-6253-5965]
[3] Kiel Nano, Surface and Interface Science (KiNSIS), Kiel University, Kiel, Germany

**Current State of the Technique**

Magneto-optical (MO) microscopy [1][2] is a polarised light microscopy technique for the characterization of magnetic materials and the imaging of magnetic domains [3]. It is suitable for general characterizing of materials with magnetic phases. Due to its straightforward nature, this traditional technique has become a commonly employed laboratory method for examining magnetic textures down to the nanoscale.

There are several MO effects used to image magnetic domain activity. In transmission and reflection, the MO Faraday and Kerr effect (MOKE) are two linear MO effects. The MO sensitivities are defined by the relative alignment of the magnetization to the plane of incidence and the orientation of the polarisation of the light, giving rise to three fundamental linear effects: the polar, longitudinal, and transverse MO effects (Figure 1a). The quadratic MO effect is the Voigt effect. Depending on the geometry, the various effects contribute simultaneously, the separation of which is done by advanced illumination schemes for wide-field microscopy [4][5][6]. Confocal scanning MOKE allows for selective sensitivity [7][8]. By this, magnetic vector information can also be extracted.

MO microscopy uses modified optical polarisation microscopes or similar setups [1]. It allows wide-field imaging or scanning confocal configurations (Figure 1b) with flexible scales from centimeters to hundreds of nanometres. The Abbe diffraction limit is the fundamental limit to spatial resolution. A basic MO setup uses a separate illumination and observation path, particularly suitable for large-scale sample MO imaging with in-plane magnetisation alignment. Other setups are based on modified polarised light microscopes using a single objective to illuminate and image the magnetisation state. This is suitable for high spatial resolution and high magnification imaging. Where wide-field schemes allow direct imaging of magnetic objects regardless of magnification or image size, scanning methods allow MO signal extraction with a higher MO signal-to-noise ratio. Due to the surface sensitivity of the MO Kerr effect, probing magnetic states of sub-nm thick films is standard. By relying on optics, flexible temporal resolution down to the sub-ps regime can be achieved with appropriate pulsed light sources, typically through stroboscopic imaging schemes. Example images are shown in Figure 1c.

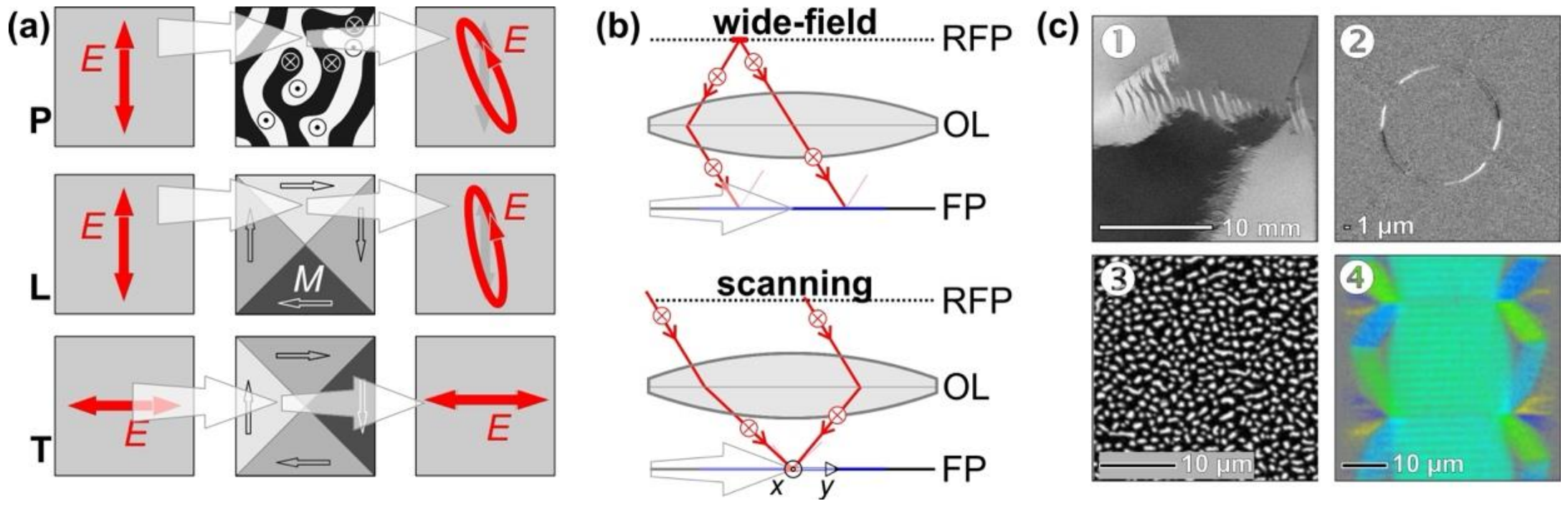

***Figure 1*** *a) Principle geometry of the polar (P), longitudinal (L), and transverse (T) MO Kerr effects. Exemplary magnetic domain structures are sketched. The MO interaction transforms the linearly polarised incoming light* $E_{inc}$ *into an elliptically polarised or amplitude-varying* $E_{refl.}$*. b) Illustration of two imaging modes with oblique plane of incidence:* ***TOP*** *wide-field mode where the light is focused in the rear focal plane (RFP) of the objective lens (OL) resulting in parallel illumination in the focal plane (FP),* ***BOTTOM*** *scanning mode characterized by a parallel beam focused in the focal plane (FP) scanned along the xy-plane c) (1) Large view MOKE image from a low anisotropy* $Ni_{81}Fe_{19}$ *(40 nm) film and (2) a circular magnetic nanowire (300 nm width)* $Co_{90}Fe_{10}$ *(0.8 nm)/*$Ni_{81}Fe_{19}$ *(28 nm). Reproduced from [1] with permission of IOP Publishing. (3) Skyrmion bubble formation in a thin film of* $Co_{40}Fe_{40}B_{20}$ *(1.4 nm)/MgO. (4) Component selective quantitative time-resolved domain imaging of a* $Co_{40}Fe_{40}B_{20}$ *(120 nm) film at a field frequency of 3 GHz (time resolution of 7 ps). Reprinted from [6], Copyright (2017), with permission from Elsevier.*

### Future Advances of the Technique

Future developments to improve signal-to-noise ratio (SNR) and spatial resolution include several needed adaptations of the MO imaging technique. It is expected that novel imaging modes will be required to advance. New functionalities are projected to become available.

- *MO contrast recovery*: While the SNR in scanning Kerr microscopy can be enhanced by lock-in amplification, the SNR for wide-field imaging is significantly lower. With the advantage of wide-field imaging in a single exposure, related image detection schemes with improved SNR are an obvious extension to wide-field Kerr microscopy. This will allow changing from the most commonly used differential imaging scheme with a magnetically homogeneous magnetic background image. Such imaging schemes require significant but worthwhile changes in microscope illumination and image detection. In-line polarisation variations based on real-time GPU image processing have already been demonstrated in this direction [9]. The fundamental goal is to obtain a low-noise wide-field magnetic micrograph without the need for additional images of a different magnetic state in a most efficient way. The addition of plasmon filter structures for MO contrast enhancement is another possible development avenue. AI and machine learning based methods for image noise reduction and local contrast enhancement can be implemented.
- *MO effects*: More intelligent illumination schemes will enable direct component selective MO imaging. Multiple simultaneously applied illumination sources with different polarisation states should allow switchable and single exposure imaging of e.g. pure and continuously configurable in-plane MO contrast independent of the actual planes of incidence. Furthermore, the possible use of full illumination aperture is expected to reduce unwanted effects from topographic structures. Such an implementation will also lead to a significantly better spatial resolution.
- *Improvement in spatial resolution*: Dielectric microelements, such as microspheres, can focus light in their vicinity. Due to a near-field interaction a so-called photonic nanojet (PNJ) forms. This produces a virtual image with enhanced resolution and magnification [10]. Incorporating them into MO imaging systems can improve spatial resolution significantly. A challenge for MO microscopy is to maintain the polarisation of the light. Alternatively, moving to shorter imaging wavelengths is another way to improve spatial resolution in MO microscopy. This requires a change to quartz optics and is more easily implemented in scanning Kerr microscopy setups in conjunction with CW or pulsed UV or soft-x-ray laser sources. Yet, imaging by PNJs appears to be the most promising technique for pushing the spatial resolution limit of wide-field MO imaging well below 100 nm. Direct magnetic scanning near-field optical microscopy may play a role in the future, with the first practical realisation for time-resolved MO imaging [11].

- *Time-resolved MO microscopy*: The ability to image fast magnetisation dynamics using stroboscopic imaging techniques is standard in scanning and adapted for wide-field MO microscopy. Yet, extending fast wide-field imaging modes to single-shot imaging of non-repetitive events has yet to be explored for timescales beyond the microsecond regime. Providing single-shot imaging for the full range of magnetisation dynamics down to at least the nanosecond regime will open a whole new field for imaging the magnetisation behaviour of magnetic materials and devices.
- *Conoscopic MO imaging*: Observation of the objective's back focal plane, or Fourier imaging, allows new domain imaging modes. In conoscopic mode, a range of magnetic texture information (e.g., lateral dimensions of magnetic domains and orientation) can be extracted from Fourier space images. Temporal resolution for imaging magnetization dynamics using stroboscopic imaging techniques is standard in MO microscopy. Yet, fast single-shot imaging is limited by the brilliance of the illumination source used within lab-based microscopy setups. For time-resolved imaging at timescales below the SNR limit, averaging in Fourier space would allow for extracting otherwise hidden domain structure parameters.
- *Integration with other magnetic microscopy techniques*: The possible high temporal resolution and the ability to extract magnetometric data with MO imaging offer substantial benefits in combination with different techniques, especially scanning microscopy techniques. Therefore, MO microscopy is expected to be further integrated with complementary techniques of higher spatial resolution for the simultaneous extraction of magnetic domain information by different techniques. Adapted optical setups will be developed for this purpose.

**Current and Future Use of the Technique for Material Science**

MO microscopy is becoming increasingly important in materials science. MO microscopy provides a straightforward method of identifying magnetic phases in ferrous alloys, complementing the research and development of advanced materials [12]. (See Figure 2 for examples from this section). The adaptation of MO microscopy to materials engineering is just beginning. With the growing availability of commercial MOKE microscopes, MO imaging is expected to be increasingly applied in the future. Research on magnetic materials for existing and future energy applications with the possibility of time-resolved operando device characterisation is only possible with MO microscopy, often assisted by MO indicator films [1].

Furthermore, the MO surface sensitivity enables studies of 2D crystallographic materials, with numerous new material systems being introduced. 2D magnets within the family of van der Waals bonded materials have been obtained experimentally, with properties markedly different from their bulk phase. The need of cryogenic temperatures and high magnetic fields to control the magnetism in such materials makes advanced high-resolution MO microscopy an ideal technique for investigation. Time-resolved scanning Kerr was already used to investigate spin/valley lifetimes in monolayer $WS_2$ [13]. MO microscopy will further contribute to the investigations of magnetic topological semimetals like magnetic Weyl semimetals (WSMs), which are of interest for novel spintronic devices. MO microscopy has so far proven to be effective for the investigation of spatial effects of WSM heterostructures. Non-trivial ways of purely spin-orbit torque-driven magnetisation switching, aided by interfacial chirality effects, can be easily studied and quantified by low-temperature MO microscopy [14]. The integration of high temporal resolution MO microscopy to study the dynamics and lifetimes of processes in semimetals prepares MO microscopy for future challenges in this area. It is anticipated that with improved MO sensitivity and spatial resolution, MO microscopy will continue to adapt to the various challenges of studying new classes of materials. With the move towards small device integration, advanced MO microscopy will become even more indispensable.

Magnetic materials provide a versatile platform for the formation of real-space non-trivial topological winding spin solitons like skyrmions. Improved MO microscopy will continue to be used to investigate skyrmionic bubbles [15]. The previously discussed advances in spatial resolution will be vital in studying the magnetization dynamics of single skyrmions and skyrmion strings. Together with the rise of sufficiently low damping materials, this will lead to insights into the resonance modes of skyrmions.

The observation of which will be within the reach of future MO microscopy. Also, antiferromagnetic (AFM) materials are candidates for future spintronic devices because of their scaling advantages due to minimised stray magnetic fields and ultrafast switching speeds. However, only non-collinear AFMs exhibit MOKE contrast [16]. The MO Voigt contrast and the linear dichroic material contrast allow spatial detection of the AF spin distribution [17]. As the symmetry of both effects is equivalent, a clear distinction between the effects is not straightforward. Current and future AFM materials studies will benefit from the proposed improvement in contrast sensitivity and advanced time-resolved imaging schemes. Second harmonic MO signal generation (Chapter 7b) offers a direct way of imaging fully compensated AFMs.

The spatiotemporal characterisation of functional magnetic material devices will remain at the heart of MO microscopy. In this respect, multiferroic or magnetoelectric (ME) materials will be studied, including piezoelectric/ferromagnetic composites with large ME coupling coefficients. MO microscopy can provide unprecedented insight into the local interactions in the multiferroic materials [18]. For ME sensing devices, the magnetic domain behaviour is of relevance due to domain wall mediated noise. As has been demonstrated for surface acoustic wave devices, high-resolution time-resolved MO microscopy can also provide understanding of electrical response characteristics [19]. Because MO microscopy allows quasi-static and time-resolved imaging of functional devices, even when covered in transparent layers. It is often the only choice for obtaining magnetic micrographs of incorporated materials.

In this context, time-resolved MO microscopy provides unique insight into the dynamic properties of a wide range of magnetic materials in bench-top setups, such as the study of collective magnetic excitations, e.g., various spin-wave-related phenomena for envisaged future devices. Of particular interest for future information processing are non-linear processes, in which degenerated phase-states were identified by advanced time-resolved microscopy schemes [20]. This demonstrates the adaptivity of MO microscopy to new scientific challenges. Another glimpse on future advances is investigating the ultra-fast behaviour of magnetic domains [21] by MO diffraction pattern analysis.

MO microscopy, especially with improved MO sensitivity and spatial resolution, will continue to adapt to the various challenges of studying new classes of materials and devices. Although it is a traditional technique, the extreme flexibility of MO microscopy means that it will continue to evolve.

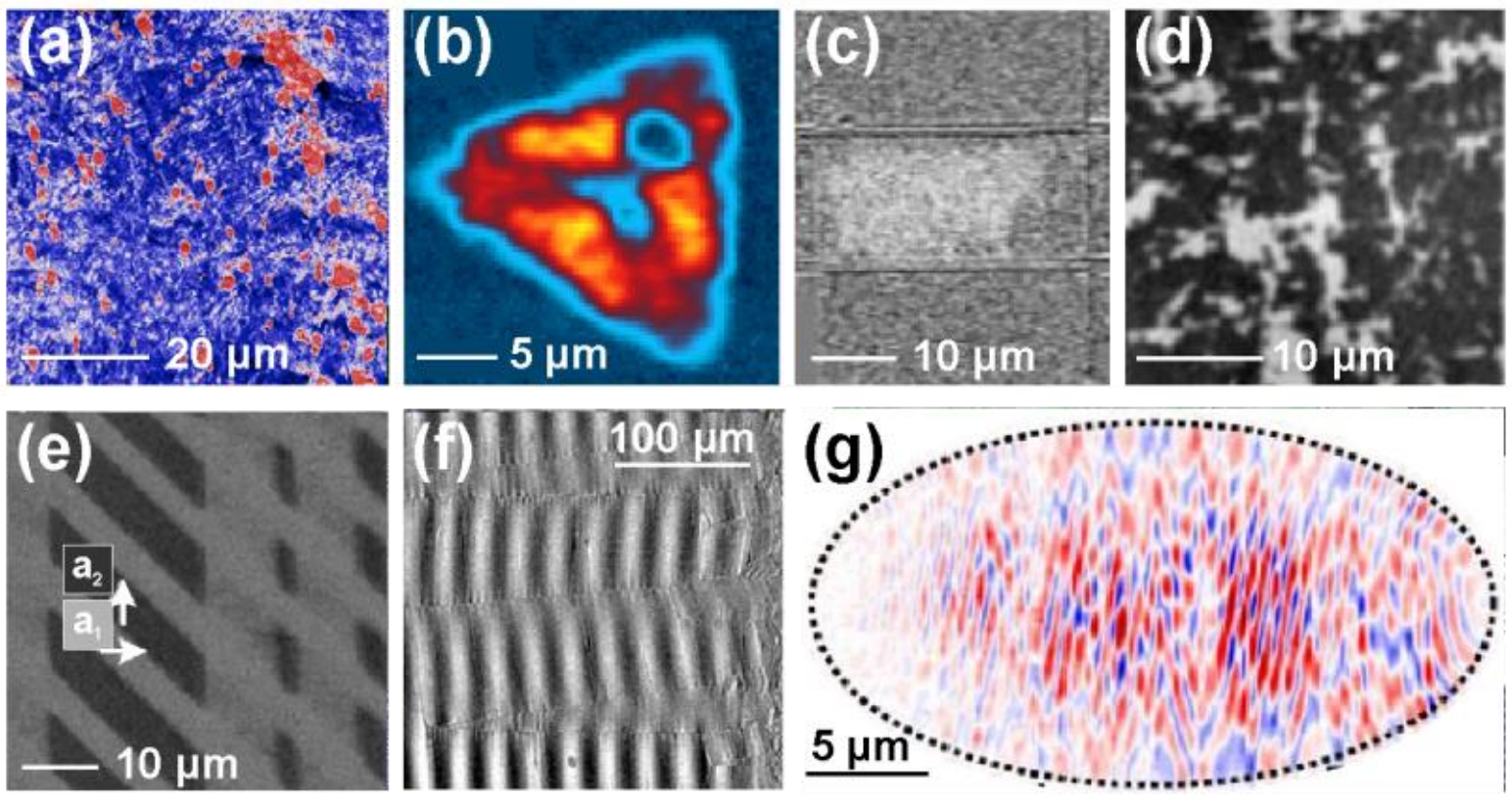

***Figure 2**: Example MO images from various materials and devices. (a) MO microscopy image of a deep cryogenic heat-treated steel surface in an applied magnetic field, where the (color-coded) blue regions correspond to martensite (adapted from [12]). (b) Scanning MOKE microscopy image of a $WS_2$ island, displaying a spatial map of spin density. A central core of low spin/valley density surrounded by regions of higher spin/valley density become visible (adapted from [13], with permission of IOP Publishing). (c) Spin-orbit-torque driven magnetization switching measurements in a $WTe_2$(80 nm)/NiFe heterostructure (apdapted from [14] with permission of Springer Nature). (d) Magnetic domains in an antiferromagnetic NiO(20 nm) film obtained from a MO microscope (adapted figure with permission from [17], Copyright (2019) by APS). (e) Remanent ferromagnetic domain distribution in a ferroelectric/ferromagnetic $BaTiO_3$/CoFe structure (adapted from [18] with permission*

*of John Wiley & Sons Copyright 2019). (f) Time-resolved wide-field image of a moving SAW, interacting with the magnetic domain walls in a quartz/FeCoSiB structure (adapted from [19]). (g) Spatially-resolved time-resolved measurements of non-linear spin waves in a $Ni_{80}Fe_{20}$ thin film element (adapted from [20]).*

**Acknowledgements**
*J.M. acknowledges funding through the German Research Foundation (DFG) through the DFG Heisenberg programme (Mc9/9-1, Mc9/9-2), and the Collaborative Research Centre CRC 1261 "Magnetoelectric Sensors: From Composite Materials to Biomagnetic Diagnostics". M.V. acknowledges funding by the US-DOE, Office of Science, MSED and by the US-DOE, Office of Science, BES, under contract no. DE-AC02-06CH11357*

# 7b – Second-Harmonic-Generation Microscopy

*Manfred Fiebig*[1]

[1] Department of Materials, ETH Zurich, Zurich, Switzerland, ORCID 0000-0003-4998-7179

**Current State of the Technique**

Optical second harmonic generation (SHG) denominates the frequency doubling of light in a material. In the electric-dipole description of the light field, it is expressed by $P_i(2\omega) = \varepsilon_0 \chi_{ijk}^{(2)} E_j(\omega) E_k(\omega)$, where $\mathbf{E}(\omega)$ and $\mathbf{P}(2\omega)$ describe the incident light at frequency $\omega$ and the frequency-doubled polarisation induced in the material, respectively [1]. As a multi-photon process, SHG requires intense light fields to be observed, so its first detection occurred in 1961 [2], right after invention of the laser. SHG is a highly symmetry-sensitive technique [3] and as such perfectly suited to study systems exhibiting ferroic order [4], which breaks point-group symmetries by definition [5]. For example, electric-dipole SHG is only permitted in non-centrosymmetric systems and is therefore often used for the investigation of inversion-symmetry-breaking ferroelectric states, which breaks inversion symmetry. The first SHG contributions from magnetic order were reported in 1991 [6], following predictions and precursor experiments from the 1960s [4]. Since SHG, in contrast to linear-optical techniques, does not couple to the magnetisation but to the symmetry of a magnetically ordered material, SHG is particularly efficient in the study of antiferromagnetic materials, where the magnetisation is zero [7].

The investigation of magnetically ordered materials with SHG is particularly advantageous when the specific degrees of freedom of optical techniques come into play [4]. (i) Polarisation-dependent SHG spectroscopy can distinguish between differently ordered sublattices of a material. In particular, the coexistence between magnetic and electric orders in a so-called multiferroic is accessible with the same SHG experiment and in a single experimental run, which gives direct access to the magnetoelectric coupling effects between the coexisting ordered states. (ii) Spatial resolution up to the limit of the wavelength of about 1 $\mu$m provides access to domain patterns. This permits to image the distribution of antiferromagnetic spin-reversal domains within seconds, which is unattainable with other techniques. Domain structures are mostly imaged in the near-surface bulk region, but interference techniques [8] and Cerenkov SHG [9] are also capable of three-dimensional imaging. Despite the wavelength-limited of the resolution, the ordered state of nanoparticles can be addressed. On the other hand, spatial resolution may be increased to, in principle, <100 nm by SHG near-field imaging [10]. (iii) Finally, with a temporal resolution up to the laser pulse length of a few femtoseconds, magnetisation-dynamical processes beyond the equilibrium of electron, spin, and lattice temperatures are resolved, which offers prospects in the direction of ultra-fast magnetic reading or writing.

On the theoretical side, mechanisms coupling the magnetic order to the nonlinear photon field were identified [4], and a general tool (#SHAARP) calculating the polarisation selection rules for SHG experiments on ordered materials was developed [11, 12].

**Future Advances of the Technique**

The most important future development in SHG microscopy is probably its evolution from a basic-research tool towards practical or even commercial applications. For example, in conceptual experiments SHG was used as an in-situ characterisation tool that tracks the emergence of functional properties like a spontaneous polarisation or magnetisation during the growth of epitaxial heterostructures [13]. It is thus possible to observe the threshold thickness at which the ordered state emerges, the interaction between the sequentially deposited constituents, and the formation of interim phases that, albeit disappearing again with the ongoing growth, determine the structure and properties of the finalised heterostructure. Thus far, such in-situ SHG (ISHG) probing was mostly applied to ferroelectric materials, but the application to magnetically ordered systems is imminent. Since it is possible to track the evolution of the magnetic order at any time during the growth process, the formation of undesired magnetic states can be avoided and targeted magnetic functionalities can be enhanced by feedback-controlled adaption of the growth conditions. In heterostructures combining magnetically and electrically active materials, their cross-correlation can be optimized in the same way and help to obtain novel multiferroics or enhanced magnetoelectric coupling effects.

It is foreseeable that the conceptual demonstration of ISHG will soon lead to the development of a prototype that can be flanged onto the growth chamber for sputtering deposition just like the electron gun for in-situ electron diffraction. In combination with a software like #SHAARP for signal analysis [12], this will open up ISHG probing to non-specialists in lasers and nonlinear optics. A commercial setup for ISHG maybe expected two to five years from now.

In parallel, the concept of probing a magnetic material during its composition (in situ) will be expanded towards probing a magnetic device during its operation (operando) by SHG. In magnetic materials, functionality is usually synonymous to modification, mostly switching, of a magnetic state, and because of the non-invasive, remote nature of SHG probing, this can be tracked with spatial, temporal, and sub-lattice resolution without destroying the working component.

In moving towards the technological-commercial implementation of SHG microscopy, the methods of nonlinear-optical probing of magnetic order will be developed further. The present focus on the leading-order electric-dipole contribution to SHG will be expanded towards processes including magnetic-dipole or electric-quadrupole contributions to the light field. This permits more selective addressing of the electronic transitions of a material and in particular the investigation of systems with centrosymmetric magnetic order [4, 14, 15]. The associated processes are significantly weaker than the leading electric-dipole contribution, so ways to counteract this inherent attenuation will be explored. This may involve resonant tuning of one or more of the light fields involved in the nonlinear excitation process to the electronic transitions of the probed material. Along with this, nonlinear processes other than SHG, like sum-frequency generation or third harmonic generation, may be involved [4]. Finally, the SHG process may be transferred from (near-) visible frequencies towards other photon-energy ranges. Terahertz SHG appears to be specifically promising as it allows to probe the elementary excitations in materials resonantly and thus very directly and effectively. Hence, significant extensions of SHG and other nonlinear optical methods are still foreseeable in the field of spectroscopy. In contrast, the limits in terms of temporal and spatial resolution may have been reached already. A laser pulse length of 100 fs seems to remain the optimal compromise between temporal and spectral resolution. While in terms of thickness, the complex ordered states of a single monolayer of a material can already be detected [16], the superior spatial resolution of optical-near-field experiments is diminished by the extreme complexity of near-field SHG techniques. Here, a compromise may be an improvement of the statistical tools for analysing the far-field SHG response from structures ordering on sub-resolution length scales.

SHG microscopy may be expanded by combining the previously mentioned degrees of freedom of optical techniques (i) to (iii) in a synergetic way. For example, ultrafast magnetization-dynamical experiments with SHG probing may be performed with optical resolution with the help of specially enhanced digital cameras ("EM-CCD") or improved statistical image analysis. This will permit to observe the time-dependent as well as the spatial dissipation of a non-equilibrium state of the spin system at the same time. In pump-probe experiments, intense excitation in the x-ray, (near-) visible, or terahertz range will be synchronized to SHG probing. The aforementioned combinations of techniques will especially give a boost to the study of spin dynamics in systems with compensated magnetisation, where SHG probing is the method of choice.

**Current and Future Use of SHG microscopy for Material Science**

One of the greatest strengths of SHG microscopy is spatially resolved imaging of spin-related structures with "hidden" order. These are foremost forms of magnetic order exhibiting zero net magnetization like antiferromagnetism. Because of a lack of magnetization to couple to, established linear-optical techniques, mostly the magneto-optical Kerr effect, are not applicable, and symmetry-sensitive SHG microscopy comes into play for imaging antiferromagnetic domains [7]. With the growing interest in antiferromagnetism for spintronics applications [17] or as neighbouring phase to other correlated states, convenient experimental access to antiferromagnetic domains will become ever more important because essentially the functionality of any ferroic material roots in the distribution and manipulability of its domains.

Currently, there is also a growing interest in novel forms of magnetically compensated ferroic order. Since their physical origin is still under investigation, domain imaging by SHG microscopy is all the more important since it addresses the ordered state via macroscopic symmetry properties with no need to know the microscopy. One of the new types of compensated magnetic order is ferrotoroidicity, the spontaneous alignment of magnetic whirls, which is interesting because of its inherent magnetoelectric effect. Other materials may exhibit spin-driven ferroaxial order, where magnetic screws break mirror symmetries, or magnetic-multipolar types of order. Broken spatial and/or temporal inversion symmetries in all these systems can act as a source for new magnetoelectric optical or transport properties [4].

The study of magnetoelectric correlations in multiferroics will remain a central theme of SHG microscopy. Magnetoelectric correlations are ultimately based on the coupling between magnetic and electric domains, and the only technique that can image both within the same experiment is SHG. There are also a number of spin states, however, whose correlations are not captured by the symmetry formalism used for ferroic systems and to which SHG microscopy may nevertheless couple. This includes dynamical order in the form of loop currents [18] or supercurrents [19] as they occur in superconducting materials. The conducting surface state of topological insulators may be accessed by SHG because of its capability to distinguish between surface and bulk contributions. Other systems with topological properties are skyrmions and Weyl semimetals, which are interesting for new electronics and computing technologies. First very preliminary characterisation experiments by SHG are presently reported for all these systems [14], so a rapid expansion of SHG from ferroic towards topological and spin-correlated systems at large is likely.

There is a good chance that oxide electronics will be revolutionised by the development and commercialisation of ISHG in the growth of epitaxial heterostructures with magnetoelectric functionalities. The current, mostly methodical experiments show that ISHG measurement can be used to control the growth process in many ways [4]. In feedback loops during the deposition, multi-domain formation can be promoted or suppressed. The deposition may be stopped at the point of emergence of magnetic or electric order to obtain particularly large susceptibilities. The influence of and coupling between the interfaces separating neighbouring constituents can be identified and used as an additional parameter to design functionality. Interim phases occurring during growth are detected, which helps to understand the relation between the deposition process and the final product. In addition, the impinging laser light itself may be used to modify the formation processes locally through the thermal and electronic excitations it introduces.

The connection of spatial and ultrafast temporal resolution in pump-probe imaging experiments will help to connect the relaxation processes between electrons, spins, and the lattice to spatial dissipation processes and gradient effects. In systems with multiple competing interactions, this can lead to better understanding and control of the subtle balance between states of similar energy. This may lead to an improved concept for ultrafast magnetization processes and, ultimately, towards all-optical magnetic phase control.

An inherent disadvantage of optical experiments around the visible range is that the associated photon energies of ~1 eV may be far away from the characteristic energy of the states that are actually probed. Magnetic excitations, for example, are quantified in magnons with energies of around 10-100 meV. Here the extension of SHG experiments into the terahertz range with either terahertz SHG or a combination of terahertz stimulation and visible-SHG probing may offer significantly more direct access to magnetic correlation dynamics. This may be applied in studies of the pseudogap of (especially unconventional) superconductors [18], in investigations of the competition between magnetic order and disorder near quantum-critical points, or in the generation of radiation-driven multiferroic-like states with dynamical magnetoelectric correlations [20].

# 8. X-ray Techniques

## 8a. – Magnetic X-ray microscopies in real space

*Peter Fischer[1,2], Aurelio Hierro-Rodriguez[3,4], Simone Finizio[5], Sarnjeet Dhesi[6]*

[1] Materials Sciences Division, Lawrence Berkeley National Laboratory, Berkeley CA 94720, USA (ORCID 0000-0002-9824-9343)
[2] Department of Physics, University of California Santa Cruz, Santa Cruz CA 95064, USA (ORCID: 0000-0002-9824-9343)
[3] Departamento de Física, Universidad de Oviedo, 33007 Oviedo, Spain (ORCID: 0000-0001-6600-7801)
[4] CINN (CSIC-Universidad de Oviedo), 33940, El Entrego, Spain (ORCID: 0000-0001-6600-7801)
[5] Swiss Light Source, Paul Scherrer Institute, 5232 Villigen PSI, Switzerland (ORCID: 0000-0002-1792-0626)
[6] Diamond Light Source, Chilton, Didcot, Oxfordshire, OX11 0DE, UK (ORCID: 0000-0003-4966-0002)

**Current State of the Technique**

Magnetic X-ray microscopy in real space can be done with **X**-ray **P**hoto**e**mission **E**lectron **M**icroscopy (X-PEEM) [1] and with X-ray **F**resnel **z**one **p**late (FZP) based systems, realized in magnetic **S**canning **T**ransmission **X**-ray **M**icroscopy (STXM) and full field **M**agnetic **T**ransmission **X**-ray **M**icroscopy (MTXM) [2-4].

There is complementarity as X-PEEM provides surface sensitivity, whereas STXM and MTXM allow for bulk studies up to the absorption lengths of X-rays (~150nm).

Prerequisites for both are polarized X-rays to use XMCD/XMLD as element-specific and quantifiable magnetic contrast for FM/FiM/AF/AM materials, high resolution and efficiency X-ray optics for STXM/MTXM, various X-ray detectors, i.e., 2D X-ray detectors for MTXM and XPEEM and X-ray point detectors, e.g., Avalanche Photo Diode APD) or fluorescence detectors for STXM, or sample current detection in STXM. Various detection schemes are adopted, i.e., X-ray reflection, X-ray fluorescence, secondary electron detection. Contrast enhancement and background reduction is achieved by modulating the X-ray polarization (left/right for XMCD and various angles for XMLD), comparing contributions from various spin-orbit coupled X-ray absorption edges, e.g., $L_3/L_2$, $M_5/M_4$, or normalization to magnetically saturated (flat) images.

A variety of substrates can be used, e.g., X-ray transparent membranes (mostly $Si_3N_4$) for absorption detection in MTXM/STXM, solids and conductors for X-PEEM or STXM in sample current detection mode, luminescence or phosphorescence for specific STXM applications.

Stroboscopic pump-probe schemes using the inherent time structure of SR sources allow for time-resolved magnetic microscopy of fully repeatable magnetization dynamics [5-7].

3D imaging of FM materials is performed through tomographic [8] (or laminographic [9] – for extended surface samples) imaging, by the acquisition of several projections at different orientation angles (ideally around more than one rotation axis) of the sample with respect to the probing X-ray beam considering that the XMCD contrast measures the projection of the local magnetization onto the photon propagation direction. To acquire the single projections transmission techniques such as STXM and MTXM are used but also X-PEEM in transmission (shadow) mode [9].

External parameters can be applied during the image recording: temperature, external magnetic fields (limited with X-PEEM), excitations (B-field, current, E-field).

In general, radiation damage is not an issue for magnetic systems of interest in real space x-ray microscopy, such as metallic films, multilayers or nanopatterned magnetic structures.

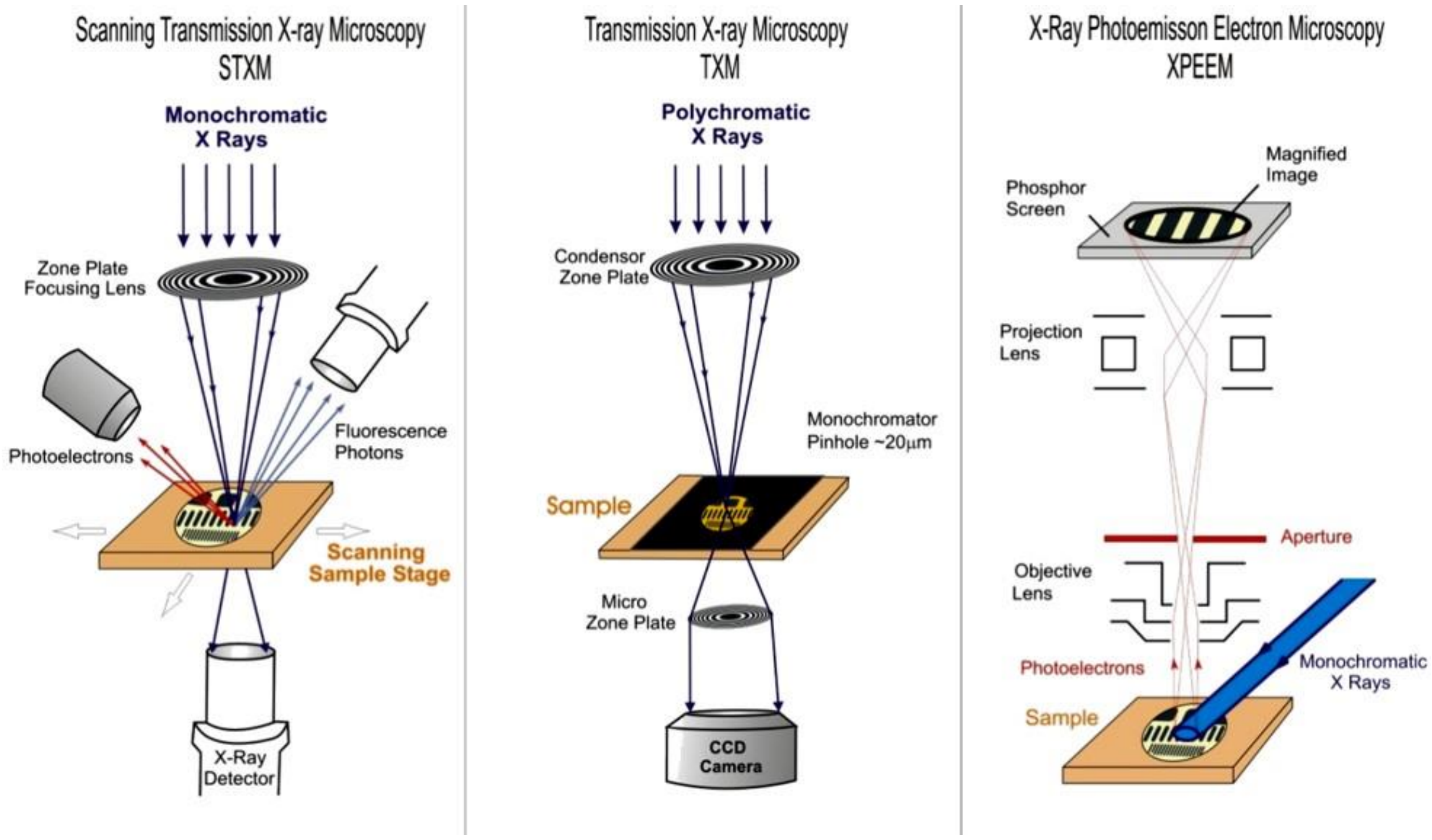

***Figure 1**: Schematics of the various direct imaging magnetic X-ray microscopes. From J. Stoehr and H. C. Siegmann, (2006) X-rays and Magnetism: Spectroscopy and Microscopy. In: Magnetism., vol 152. Springer, Berlin, Heidelberg. https://doi.org/10.1007/978-3-540-30283-4_10 reproduced with permission from SNCSC.*

**Future Advances of the Technique**

Several large-scale facilities worldwide are currently in the process of upgrading their synchrotron sources mostly to enhance the coherent flux, which opens new possibilities for magnetic X-ray microscopies. A few examples, which are not meant to be conclusive, showcase current trends.

- Future STXM at SLS

The Swiss Light Source (SLS) will undergo an upgrade to a diffraction limited source (DLS) during 2023 - 2026. SLS 2.0 will increase the photon flux at both the soft and hard X-ray energy ranges by several orders of magnitude, allowing for a reduction of measurement time, and to image samples with weaker XAS/XMCD signals. However, the increased photon flux yields an increased probability of multi-photon events at the point detector used in STXM imaging, leading to erroneous measurements [10]. In order to overcome this issue, the APD installed at the PolLux STXM will be replaced with a multi-sector APD, each sector with its separate amplifier and discriminator, thus reducing the probability that two photons interact with the same sector, producing an erroneous count, if compared with a single-sector APD.

- Future PEEM at DIAMOND

The Diamond-II upgrade will provide improved capabilities for magnetic imaging using coherent probes, enable an increase in the number of beamlines using a new lattice as well as increasing the photon flux at higher photon energies by increasing the energy from 3GeV to 3.5GeV. A new flagship Coherent Soft X-ray Imaging and Diffraction (CSXID) beamline will focus activity on coherent 3D tomography from ferroic materials and host a comprehensive portfolio of sample environments for imaging under magnetic and electric fields and at cryogenic temperatures. The beamline is designed to take full advantage of the increase in coherent flux and aims to provide state-of-the-art 3D magnetic vector tomographic ptychographic reconstructions in a fraction of the time that is currently possible. The upgrade will also replace the existing insertion devices on the XPEEM facility with fast polarization switching devices enabling considerably accelerated magnetic imaging of thin films and interfaces.

- Future developments with MTXM

There are no current plans worldwide for significant upgrades in MTXM. The MISTRAL beamline at ALBA is currently the only MTXM with a dedicated setup for performing magnetic tomography in novel 3D nanomagnetic structures, which is one of the frontiers in current nanomagnetism research [11, 12]. Developing novel optical designs based on reflective rather than diffractive optics seems to be an interesting alternative as it would increase the sample-optics distance and would allow to image buried interfaces. Operating at low T and at high magnetic fields would be essential for in-situ and in-operando studies of quantum effects in magnetic materials. Utilizing the increased coherence at future SR facilities, which so far has been considered counter-productive to MTXM might require significant dedicated efforts to advance conventional diffractive X-ray optical elements. [13]

Other noteworthy developments with complementary advanced magnetic spectromicroscopy techniques are

- SP-LEEM

Spin-Polarized Low-Energy Electron Microscopy (SPLEEM) is used for the study of surfaces and interfaces with the ability to detect the spin-polarization of the electron beam. Recent developments include sample stages to vary the sample's temperature down ~4 K. The strength of SP-LEEM are investigations during in-situ deposition of metallic and oxide films or multilayers. Important contributions by SP-LEEM include investigations of chiral spin textures [14]

- Momentum microscopes

Momentum microscopy (MM) [15] allows to investigate the electronic structure of solids by utilizing a combined PEEM/ARPES spectrometer to couple lateral information of about 50-100nm with the spectroscopic analysis. Sample information is obtained either in real space mode (PEEM) or in reciprocal space mode (ARPES and momentum microscopy) resulting in a constant energy map showing a real space image or a constant energy surface of the samples band structure (MM), or an energy dispersive image of a real or reciprocal space coordinate (ARPES). Coupling ARPES instruments with spin sensitive devices, e.g., Mott detectors allows to extract also the three components of the spin polarization vector with high energy and momentum resolutions.

**Current and Future Use of the Technique for Material Science**

The specific features of magnetic x-ray microscopies will continue to provide unique information to understand the underlying microscopic properties and behavior in magnetic materials. The elemental specificity represented in the spectroscopic fingerprint is probably the single most important feature and will be key to disentangle spatial and temporal distributions of spin and orbital magnetic moments and their dynamics in heterogeneous magnetic materials.

The sensitivity to the projection of magnetization onto the photon helicity in ferro(i)-magnetic systems and the angular dependence of linear dichroism in antiferromagnetic systems will allow to address the full three-dimensional character of spin textures including topologically modified systems, such as curved surfaces or core-shell nanoparticles. Revealing the depth dependence of spin textures, e.g., via reflection mode imaging will be critical to understand the role of interfaces, and modifications of magnetic interfaces in the presence of spin—orbit or asymmetric exchange interactions [16].

The increase of coherence at next generation light source, which will tremendously benefit x-ray scattering based techniques, is probably less relevant to direct real space imaging magnetic x-ray microscopies, although using phase sensitive x-ray optics might be an interesting avenue to pursue, harnessing a combination of amplitude and phase magnetic contrast methods.

The recently discovered altermagnetic systems [17] will most likely exhibit interesting topological spin textures, albeit initial magnetic imaging experiment will most likely address the anisotropic spin textures in k-space, however exploring a combination of XMLD and XMCD might we worth exploring.

The relatively large cross sections of dichroism effects and the quantifiability (magneto-optical sum-rules [18, 19]) will provide microscopic information and complement macroscopic high sensitive magnetization information, e.g. through SQUID magnetometry.

The inherent temporal structure of x-ray sources, which can range from the 10-100ps range at SR storage rings down to the fsec regime at XFELs can provide unprecedented insight into the dynamics of spin textures. In addition to fully repeatable dynamics, including standing and propagating spin waves [20], which can be also be readily simulated in micromagnetic simulations, it will be the uncorrelated spin fluctuations on the nanoscale [21] which can be anticipated to become a focus of future research. However, real space X-ray microscopies will be most likely not feasible to experimentally address those topics due to the high intensity of the FEL pulses, and their strong variability due to the SASE emission. Among the challenges would be enormous space charges in X-PEEM, radiation damage to the illumination optics in MTXM and STXM and enormous photon exposures to the sample with STXM.
Topics of increased interest will be in-operando studies of future real magnetic devices, including investigations of spin currents and spin accumulation in nanoscale magnetic materials [22].
Although the spatial resolution of X-ray microscope will most likely not surpass other techniques, such as SP-STM or EM-based microscopies, correlating magnetic x-ray microscopies with those techniques will enable a coherent understanding of magnetic phenomena across length and time scales and in multiple dimensions.
Combining tomography with time resolution seems to be a viable path to explore complex systems [23], notably artificial spin ice systems and other artificially created 3D magnetic nanostructures.
The implementation of variable temperature sample environment and much larger magnetic fields will be of high interest to investigate the microscopic origin of magnetic phase transitions.
Magnetic x-ray microscopies, notably when operating in x-ray transmission, which is the simplest detection mode with the least perturbation of the recorded signal requires x-ray transparent substrates similar to transmission electron microscopies. Until recently, those membranes used mostly non-epitaxial materials. However, recent developments in focused-ion-beam lithography enabled the fabrication of thin (<100 nm) membranes from single-crystalline substrates such as SrTiO3 and Gd3Ga5O12, opening the possibility to study epitaxial materials, e.g., multiferroic perovskites, magnetic garnets) with TXM techniques [24].

**Acknowledgements**
PF acknowledges support by the U.S. Department of Energy, Office of Science, Office of Basic Energy Sciences, Materials Sciences and Engineering Division under Contract No. DE-AC02-05-CH11231 (NEMM program MSMAG). AH-R acknowledges support by Spanish MICIN under grant PID2019-104604RB/ AEI/10.13039/501100011033 and by Asturias FICYT under grant AYUD/2021/51185 with the support of FEDER funds.

# 8b. Coherent X-ray imaging of magnetic systems

*Claire Donnelly[1,2a], Felix Büttner[3,4b], Ofer Kfir[5c], Wen Hu[6d], Sergey Zayko[7e]*

[1] Max Planck Institute for Chemical Physics of Solids, Dresden, Germany
[2] International Institute for Sustainability with Knotted Chiral Meta Matter (WPI-SKCM2), Hiroshima University, Hiroshima 739-8526, Japan
[3] Helmholtz-Zentrum für Materialien und Energie, Berlin, Germany
[4] Experimental Physics V, Center for Electronic Correlations and Magnetism, University of Augsburg, Augsburg, Germany
[5] School of Electrical Engineering, Iby and Aladar Fleischman Faculty of Engineering, Tel Aviv University, Tel Aviv 69978, Israel
[6] National Synchrotron Light Source II, Brookhaven National Laboratory, Upton, NY 11973, USA
[7] Max Planck Institute for Multidisciplinary Sciences, Göttingen, Germany

[a] [ORCID 0000-0002-9942-2419]
[b] [ORCID 0000-0002-6204-9948]
[c] [ORCID 0000-0003-1253-9372]
[d] [ORCID 0000-0003-4936-1483]
[e] [ORCID 0000-0001-9826-8627]

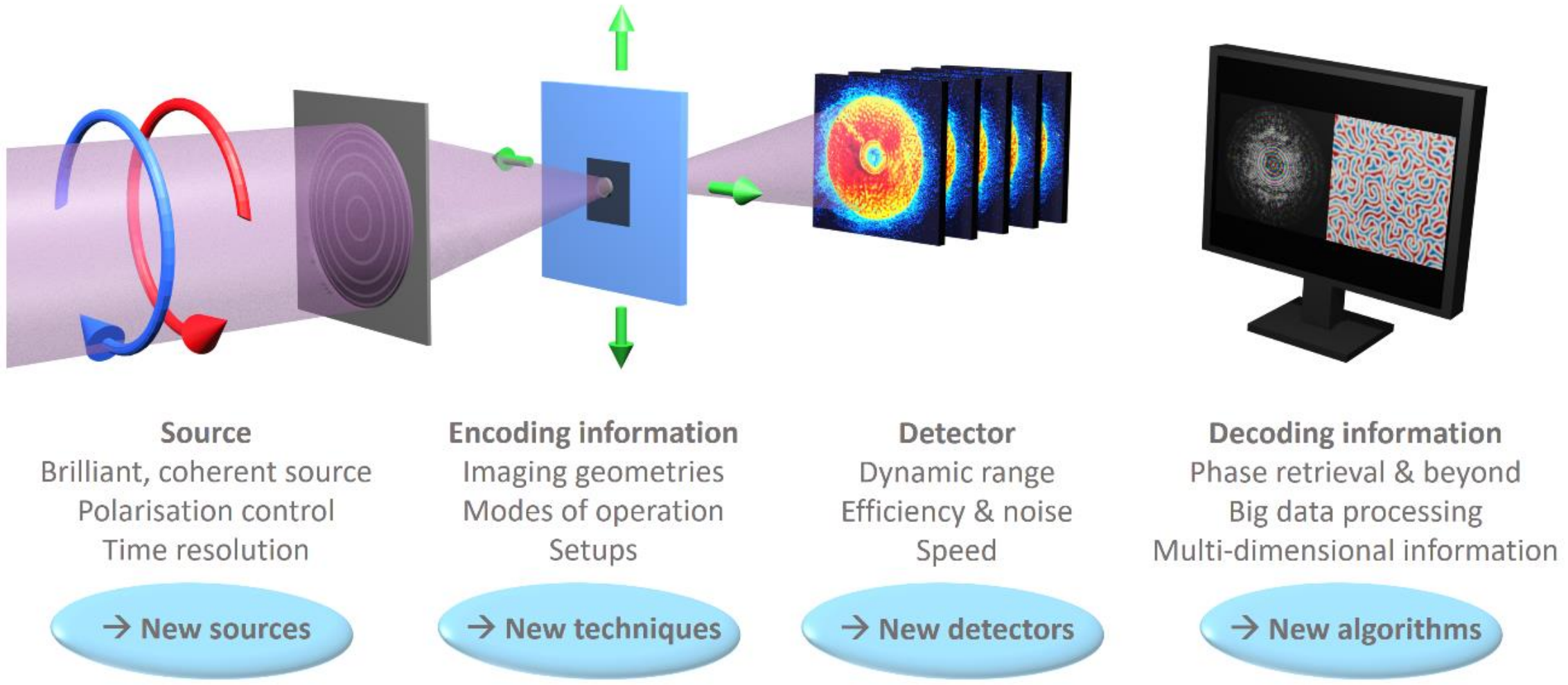

*Figure 1. Coherent diffractive magnetic imaging: experimental procedure and key areas of innovation.*

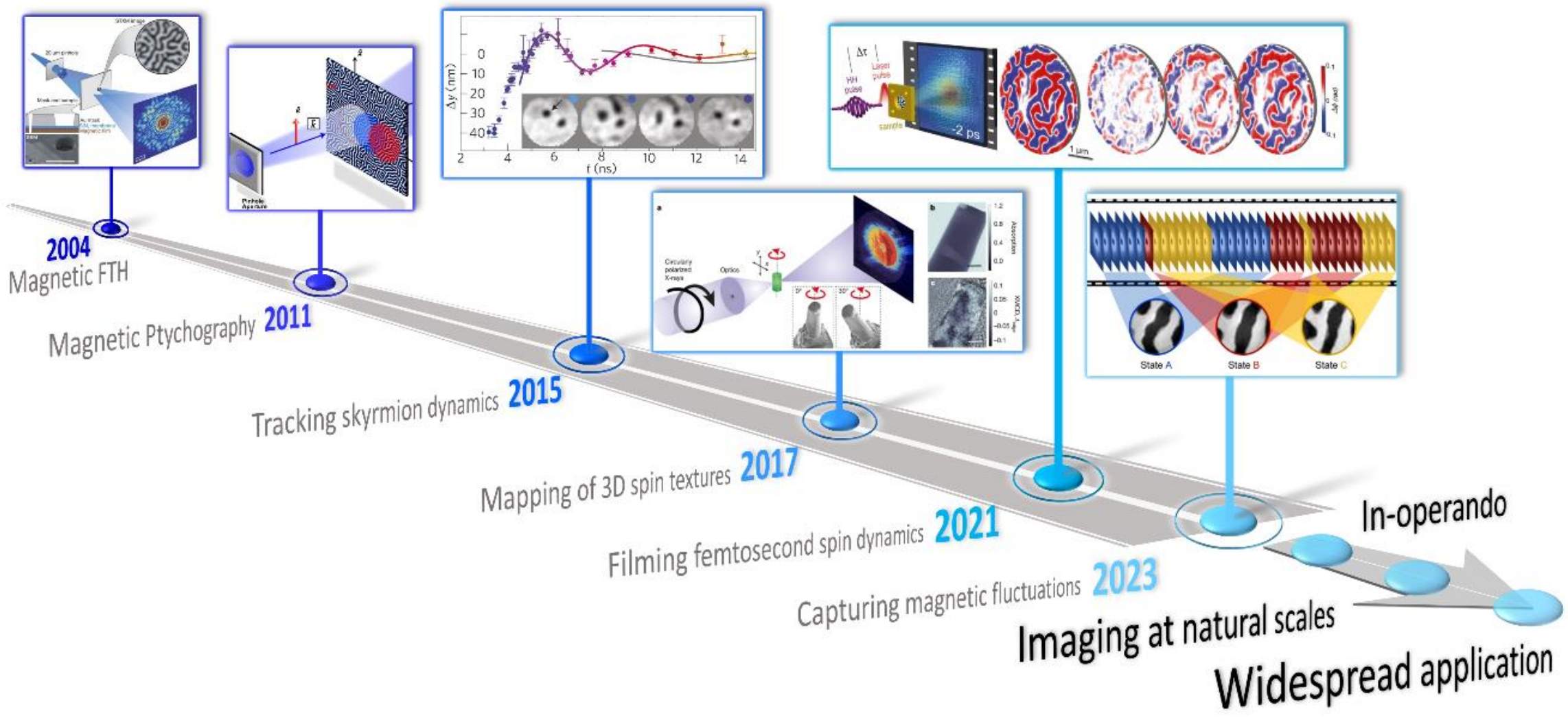

*Figure 2. Roadmap for coherent magnetic imaging, from the first demonstration of Fourier Transform Holography and ptychography, to the imaging of three-dimensional textures and spin dynamics. Reproduced from [3,5,8,9,14,18].*

## Current State of the Technique

Coherent X-ray magnetic imaging is a lensless approach to magnetic microscopy. Highlights of this approach are the simultaneous access to the element-specific phase and absorption contrasts and the promise of wavelength-limited spatial resolution in the X-ray regime. Its development sparked a new era of low-dose, high-contrast imaging with unique capabilities ranging from element specific phase contrast imaging[1–4] to spatial resolutions of 5 nm in two dimensions[5] (2D) and 10 nm in three dimensions[6] (3D).

The procedure for coherent magnetic imaging is shown in Figure 1. First, an (ideally brilliant) coherent source of polarized X-rays (synchrotron, FEL, or high-order harmonic generation (HHG)) is used to illuminate the magnetic sample. This illumination is micro-focused rather than nano-focused, which simplifies the need for advanced X-ray optics significantly. Spatial information is encoded in the coherent scattering distribution (high angles correspond to high resolution) collected on a 2D pixelated camera in the far field. Finally, the complex magnetic-refraction image is decoded from the scattering intensity algorithmically.

There are three main types of coherent imaging:

*Fourier Transform Holography (FTH):* The first experimentally implemented approach to coherent magnetic imaging was FTH[7], where the interference between the sample's exit wave and a reference wave (for example scattered from a point-like feature) physically encodes the phase information on the camera. Thus, a single Fourier transformation provides an image of the sample's exit wave, cross-correlated with the reference. Experimentally, apertures in an opaque (typically gold) layer form the sample's field of view and the reference hole.

*Coherent diffractive imaging (CDI):* Magnetic imaging by CDI captures the diffraction pattern in the far field and retrieves the sample's exit wave algorithmically. Here, the resolution is limited by the wavelength, the collected scattering angle, the signal above the noise, and the fidelity of the reconstructed phase. Oversampling (recording the diffraction pattern above the Nyquist frequency) and a well-defined transmission region ("support") are required to recover the phase of the complex-valued diffraction signal.

*Ptychography:* In contrast to CDI and FTH, ptychography[2,8] is a scanning technique in which an extended object can be imaged. An overlap between adjacent illumination positions relaxes the requirement for a well-defined field of view (FOV) and facilitates algorithmic retrieval of the amplitudes and phases of both the object and of the incident beam ("probe").

## Future Advances of the Technique

Technical developments of coherent sources and detectors as well as new reconstruction methods are driving advances in coherent magnetic imaging, leading to new science opportunities. Here we emphasize the diffraction-limited resolution, that provides a perspective for wavelength-limited resolution imaging. To date, both soft-X-ray ptychography[9,10] and CDI[5] have demonstrated magnetic imaging down to sub-10 nm, but the potential is far from being fully exploited . In the extreme-UV spectral range the combination of holography and CDI has pushed the spatial resolution to below the illumination wavelengths, reaching 16 nm spatial resolution at 21 nm wavelength[11]. Thus, coherent magnetic imaging with soft- and hard-X-rays[2] has the potential to access a few- or sub-nm resolutions in both 2D and 3D.

***Sources and detectors:***

Magnetic coherent imaging relies on synchrotron, FEL, and HHG radiation sources that deliver stable, bright beams with high flux, coherence and monochromaticity. Despite significant developments, the coherent flux remains a limiting factor. There is a massive global effort to address this issue, both at large-scale facilities (e.g., 4$^{th}$ generation synchrotrons and seeded FELs) and with the development of sources based on HHG and laser-plasma accelerators. We anticipate that the resulting gain in coherent flux, together with a stable, quantifiable incident wavefront, will soon enable widespread, high-throughput use of coherent X-ray magnetic imaging, even at sub-wavelength resolution and in higher dimension (tomographic) experiments[12].

Although X-ray large-scale facilities are leading in their potential spatial resolution, extreme-UV scattering from HHG beams[13] demonstrate the promise of coherent magnetic imaging by reaching comparable performances at a 10-fold longer wavelength. Moreover, these laser-based sources are becoming ubiquitous in the field of ultrafast laser science due their stable femtosecond-level accuracy and the method developments that a lab-scale setting with unrestricted beam-time allows for.

Detector improvements propel the prospects of coherent magnetic imaging through four main parameters: dynamic range, readout speed, quantum efficiency, and number of pixels. Quantitative high-resolution imaging requires capturing the dominant unscattered beam alongside the weaker ($10^{-6}$-$10^{-9}$) high-angle magnetic scattering that is comparable with the shot- and instrument noise. The quantum efficiency should be high and the instrumental noise low to discern the scattering. Higher numbers of pixels (e.g., 16 Megapixel) allow to retain the oversampling necessary for CDI while increasing the number of resolution elements within the field-of-view.

On top of the above, the frame rates of conventional cameras (CCD, CMOS) determine the speed of the data collection, benefiting dynamic measurements, and shortening the acquisition of ptychographic images. However, we believe that the rise of event-based detectors (e.g., from the MediPix collaboration) would change the fundamental function of the camera: first, time-tagged photon hits would add correlative, i.e., quantum information to the signal, in additional to the evaluation of classical-field scattering. Second, the figure of merit in event-based acquisition is the hit rate (Mhits/sec), making the concept of "frames" or "frame-rate" an arbitrary post-processing feature in such detectors.

***Algorithm developments***

A key aspect of coherent imaging is the decoding of information from the measured data with reconstruction algorithms. So far, the vast majority of magnetic coherent imaging has involved adaptations of non-magnetic approaches, because circular dichroism maintains the coherent imaging assumption[2,14]. Future advances could open up possibilities to disentangle multiple polarisation modes[15], extract noisy or weak signals from a non-magnetic background, and resolve multi-dimensional information from single-shot images[16]. Indeed, algorithms have allowed to discern correlative dynamics from holographic data[17]. Extending this to higher dimensions and alternative datasets will allow material case studies at a previously unimaginable level.

One of the most exciting directions involves the ongoing revolution of machine learning and artificial intelligence, which is impacting our entire society. First demonstrations of this for coherent imaging have demonstrated potential towards more rapid, robust reconstructions of multidimensional data via neural networks[18], and ptychography[19], allowing both significant increases in processing times, as well as the introduction of physics-guided reconstructions[20]. If these can be applied to magnetic imaging, and if the incident beam is sufficiently under control, there could be the possibility for quantitative extraction of material properties, transforming coherent imaging from an observation to a quantitative magnetometry technique.

**Current and Future Use of the Technique for Material Science**

Even though coherent X-ray magnetic imaging is relatively young, it has already contributed uniquely to several milestone discoveries in the field of magnetism. For example, in 2015, time-resolved X-ray holography was used to confirm the quasi-particle nature of magnetic skyrmions in $[Pt/CoB]_{x30}$ up to the GHz regime[21] via tracking the skyrmion with sub-3 nm precision and thereby discovering that the equation of motion of a skyrmion includes a mass term, whilst static holography has been used to image the 3D shape of skyrmion tubes[22] and identify skyrmionics cocoons[23]. Later, ptychography has allowed to resolve the actual profile of skyrmions in a similar $[Ta/CoFeB/MgO]_{x16}$ magnetic multilayer[24] , to map the 3D configuration of Bloch point singularities[12,25] as well as providing the first observation of magnetic vortex rings and torons[26], to name just a few. Moreover, in the ultrafast temporal range, coherent magnetic imaging for the first time unambiguously resolved the femtosecond dynamics of the domain walls during laser-induced demagnetization[11].

Indeed, the absence of lenses sets coherent imaging apart and removes associated challenges. For example, the intrinsic achromaticity of such purely far-field techniques makes them suitable for spectroscopic studies. With spatial resolutions below 5 nm, the imaging of magnetic domain wall profiles and spin textures at material defects are accessible[5], and even imaging of magnetic textures below the exchange length (e.g., Bloch points) is in reach. In combination with higher dimensional – spatial or temporal – imaging[12,27], this advance promises access to magnetism beyond current micromagnetic models.
Moreover, the phase contrast capability of coherent magnetic imaging offers reduced dose damage and higher penetration depth compared to traditional absorption-based imaging techniques[1], thereby extending soft X-ray imaging to dose-sensitive and micrometre-thick magnetic samples[4].

***New mechanisms targeted:***
Coherent imaging currently addresses ferromagnets, but it has a potential to impact other materials systems. For example, linear dichroism offers vectorial magnetic information of ferromagnets and imaging of antiferromagnets. Alternative geometries, such as coherent imaging in reflection, allow phase contrast imaging of antiferromagnetic domain walls[28] and their fluctuations[29] in natural as well as in artificial antiferromagnets[30]. Combining with coherent X-ray imaging such as reflection ptychography[31] or CDI would allow for resolving the details of the fluctuations of nanoscale textures. In ptychography, structured probes (e.g., a vortex beam[32]) may be extended to study spin-orbit effects in, for example, topological spin textures, polar vortices in ferroelectrics, and other materials. With the significant increase in coherent flux expected from the next generation of synchrotron sources, we envisage that the use of coherent approaches will expand significantly to include a wide variety of samples and physical phenomena.

### *Mapping dynamic behaviours*

Coherent X-ray magnetic imaging has also opened access to new territories in the temporal domain by providing a means to perform imaging at the fastest X-ray sources (both large-scale FEL, and table-top HHG). For an in-depth discussion of X-ray dynamic imaging, see Section 8c. Although at an early stage, the first discoveries highlight the potential of these developments for materials science research. Examples include the observation of spatial variations of ultrafast optical demagnetization[11,33] and of local materials inhomogeneities which mediate fluctuations between energetically degenerate stripe domain states[17]. Similar dynamic studies have been demonstrated for antiferromagnets[29]. We expect that coherent X-ray imaging of antiferromagnets, either by X-ray magnetic linear dichroism or phase contrast, will soon be extended to the ultrafast regime. This will allow imaging of switching processes in antiferromagnetic materials at their fundamental length and time scales and provide a previously unattainable understanding of the physics at play in this class of materials.

With the global rise of coherent X-ray facilities and the associated availability of instrumentation and data analysis support, we expect that coherent X-ray magnetic imaging will shift from singular highlight experiments to an important tool for the materials science community. Quantity and ease-of-use will enable quality by resolving standing challenges. Examples include the dynamical vector spin orientation profile of traveling magnetic domain walls and their correspondence with the 1D domain-wall model from the 1970s[34]; observing the generation dynamics of DMI- and frustration-stabilized magnetic skyrmions and their adherence with the Thiele equation that was developed for micrometre-scale, stray-field-stabilized bubble domains[35]; the imaging of irreversible processes, which are the dominant processes in real materials, either due to intrinsic (e.g., turbulences[36]) or extrinsic (e.g., inhomogeneities[37]) effects; and ultimately, also the imaging of magnetic materials at or near phase transitions.

### Acknowledgements

C.D. acknowledges funding from the Max Planck Society Lise Meitner Excellence Program. F.B. acknowledges funding from the Helmholtz Young Investigator Group Program. This research used resources of the National Synchrotron Light Source II, a U.S. Department of Energy (DOE) Office of Science User Facility operated for the DOE Office of Science by Brookhaven National Laboratory under Contract No. DE-SC0012704. O.K. gratefully acknowledges the Young Faculty Award from the National Quantum Science and Technology program of the Israeli Planning and Budgeting Committee and the Israel Science Foundation (grant 1021/22).

# 8c. Time-resolved x-ray imaging of magnetization dynamics on the nanometer scale

*Stefan Eisebitt[1,2], Simone Finizio[3], Bastian Pfau[1]*

[1] Max Born Institute, Max-Born-Str. 2a, 12489 Berlin, Germany [ORCID 0000-0001-7608-5061]
[2] Institute of Optics and Atomic Physics, Technische Universität Berlin, Straße des 17. Juni 135, 10623 Berlin, Germany [ORCID 0000-0001-7608-5061]
[3] Swiss Light Source, Paul Scherrer Institute, Forschungsstrasse 111, 5232 Villigen PSI, Switzerland [ORCID 0000-0002-1792-0626]

**Current State of the Technique**

Imaging of lateral magnetization distributions with sub-30 nm spatial resolution and the ability to add temporal resolution is provided by several x-ray imaging techniques. In general, sensitivity to the projection of the magnetization is provided via x-ray magnetic circular or linear dichroism (XMCD, XMLD), at suitable resonant transitions which typically are located in the soft x-ray energy regime. Detection with 2D detectors allows for full-field imaging, typically in transmission from thin film samples, of amplitude or phase contrast. The full-field imaging approaches can be distinguished by the mode of image formation into lens-based transmission x-ray microscopy (TXM) [1], interference-based Fourier-transform x-ray holography (FTH) [2] and coherent diffraction imaging (CDI) based on iterative phase retrieval. [3] See sections 8a and 8b for more details on these methods. Given the readout times of current 2D detectors and signal-strength limitations, image acquisition happens on a timescale of seconds in a quasistatic regime. [4] Temporal resolution in the nano-, pico- and femtosecond regime is derived from pump-probe schemes; see Fig. 1 for an overview. Most commonly, current pulses or AC currents are used for the study of magnetization dynamics at synchrotron-radiation (SR) sources, either to apply spin torques [5–7] or to be transduced into oscillating or pulsed magnetic fields. [2,8–10] The temporal resolution at SR sources is fundamentally limited by the electron-bunch lengths to at least several tens of picoseconds. At free-electron lasers (FELs), femtosecond temporal resolution can be achieved, typically in combination with laser-pulse excitation. [11] Here, single-shot pump-probe “snapshot” experiments are in principle possible given that a single x-ray pulse contains sufficient photons to create a high-resolution image after a pump event. [12] Typically, however, such an approach will severely perturb or completely destroy the sample in the soft x-ray regime. [12–14] Multi-shot imaging of a static ferromagnetic domain pattern by FTH at a FEL without sample damage has been reported at 2.25 MHz. [15]

Raster imaging can be implemented with fast point detectors in scanning transmission x-ray microscopy (STXM). Pump-probe imaging in STXM is mostly performed in an asynchronous fashion, with the actual pump-probe delay for each recorded photon count individually sorted according to their phase relation with respect to the pump. [16] By measuring the time-of-arrival of each x-ray photon, temporal resolutions beyond the width of the x-ray pulses can be achieved, but those are still ultimately limited to the tens of picosecond scale. [17] An additional advantage of measuring the x-ray photon time-of-arrival is that it becomes possible to image non-locked dynamics by post-processing the acquired time-of-arrivals. [18] Imaging of laser-induced magnetization dynamics after impulsive excitation by 2-ps (fwhm) laser pulses has recently been demonstrated in a STXM at 50-MHz repetition rate without observing sample damage at a laser fluence of 3.1 mJ/cm$^2$. [19] Ptychography (PTY) as an extension of CDI is an alternative scanning-based imaging modality. Relying on slower 2D detectors, only time-resolved experiments where the excitation is directly locked to the master clock frequency of the SR source have been performed so far. [20]

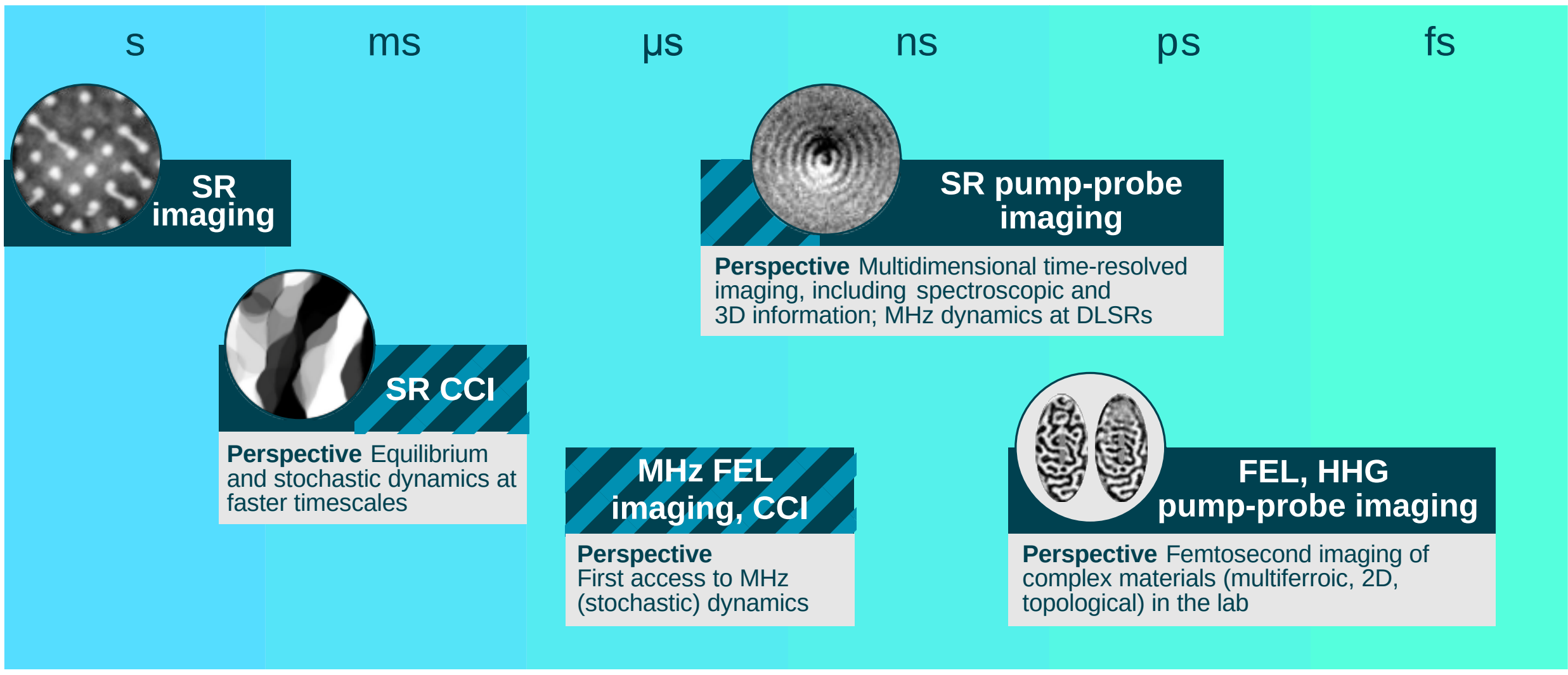

***Figure 1****:* Overview over the timescales reachable with different time-resolving techniques of x-ray magnetization imaging. Signal and detector constraints still limit the acquisition time of x-ray images to the second regime. Time-resolution on the relevant ns, ps and fs scale is only achieved with pump-probe methods, restricting the experiments to repeatable dynamics. New imaging methods, new sources, and new detectors open new research perspectives (shaded areas and text boxes), including investigating stochastic (non-repeatable) dynamics up to the MHz regime in the future. Inset images reproduced from Refs. 4, 25, 9, 11 (from left to right).

## Future Advances of the Technique

Many of the foreseeable advances of the techniques are driven by the increasing availability of x-ray sources with significantly increased brilliance and hence coherent photon flux, allowing to focus to smaller spot sizes, and – together with the improvements in computational power – allowing to harness the information content from interference patterns of x-rays transmitted through the sample. In this fashion, the increased availability of x-rays with polarization control at diffraction-limited storage rings (DLSRs) and FELs will push the spatial resolution in (static) imaging more routinely below the 10 nm limit, if sample damage is not an issue. At SR sources, the push to DLSRs is typically associated with bunch lengths typically exceeding 100 ps and hence a loss of temporal resolution over what is achievable today. On the other hand, the exploitation of sub-100-fs pulses at FELs for imaging of ultrafast magnetization dynamics has so far not really started, although the spatial resolution achievable should match well with the temporal resolution. Here, sample damage of solid samples, often in the form of thin films on membranes, is a severe issue for systematic studies with a focus on materials science. Studies with hard instead of soft x-rays and hence less absorption and radiation damage may be able to mitigate this, possibly paving the way for video-like imaging of dynamic magnetization phenomena with MHz repetition rates and non-perturbative pump-probe experiments accessing timescales down to fs.

Another exciting development is the ability to probe stochastic dynamics, e.g., magnetic fluctuations in thermal equilibrium, rather than deterministic dynamics accessible in pump-probe experiments. X-ray photon correlation spectroscopy (XPCS) analyses the magnetization fluctuations in Fourier space to measure characteristic timescales, typically, in the second to millisecond regime at SR sources. [21–23] Higher time-resolution is expected with the event of faster 2D detectors. Recently, XPCS found application also for studying nanosecond fluctuations of a skyrmion lattice at an FEL [24]. While XPCS can dissect the dynamics for different lengthscales, a real-space image is not retrieved. Coherent correlation imaging (CCI) overcomes this limitation, allowing video-like imaging of stochastic processes in a finite ensemble of possible states, where the photons required to generate an image can be collected at different temporal realizations of a state. [25] Exploiting this random repetition of states allows to stay away from destructive probe beam fluences to generate high resolution image frames in the video.

The x-ray source developments towards bright coherent beams do also take us into the third dimension: so far, almost all research with the imaging techniques described here has detected the projected magnetization in thin magnetic films, but in the recent years studies of magnetization vector within 3D structures have started to become possible. [26] It is a tremendous challenge to add temporal resolution on relevant temporal scales to 3D magnetization imaging given that tomography scans are required, but first steps have been taken, e.g., via laminography (Fig. 2). [27] Considering that the evolution of the vector field of the magnetization within a sample volume is almost a terra incognita today at pretty much any time scale, there is large potential especially for material science applications, with bright sources allowing to use hard x-rays with high penetration in spite of weak magnetic contrast. [26] While PTY has shown great advances in static 3D imaging, the lack of 2D detectors recording the time-of-arrival of the x-ray photons in the soft x-ray energy range with a temporal resolution below the repetition rate of the light source is, however, still a critical issue for the development of time-resolved PTY imaging.

Regarding imaging of ultrafast magnetization dynamics with temporal resolution in the fs up to tens of picosecond regime, the use of high-harmonic generation (HHG) in a laser laboratory is an exciting perspective. Given that HHG beams have a very high degree of coherence, they are suitable for time-resolved FTH, CDI, STXM and PTY. For the study of optically induced magnetization dynamics, they combine large flexibility in generating synchronized excitation pulses at various wavelengths with the access option outside of a large-scale facility. Time-resolved magnetization imaging has been performed via holographically assisted CDI exploiting Co M-edge (21 nm wavelength) XMCD, reaching 16-nm spatial resolution in static imaging of ferromagnetic domains. [28] From an imaging perspective, it is highly desirable to be able to work with XMCD-active resonances with much higher magnetic contrast in the future, such as the $N_{5,4}$ edges of the lanthanides, the $L_{3,2}$ edges of the 3d metals or even the lanthanide $M_{4,5}$ edges. However, it is a tremendous challenge to achieve the photon flux and polarization control at these photon energies required for imaging with HHG. Nevertheless, we note that first HHG-based time-resolved resonant magnetic scattering experiments at about 150 eV photon energy have been successfully carried out recently. [29] HHG-based time-resolved imaging can be expected to move to progressively higher photon energies in the next years, allowing to reach further core-level resonances of interest in ultrafast magnetism.

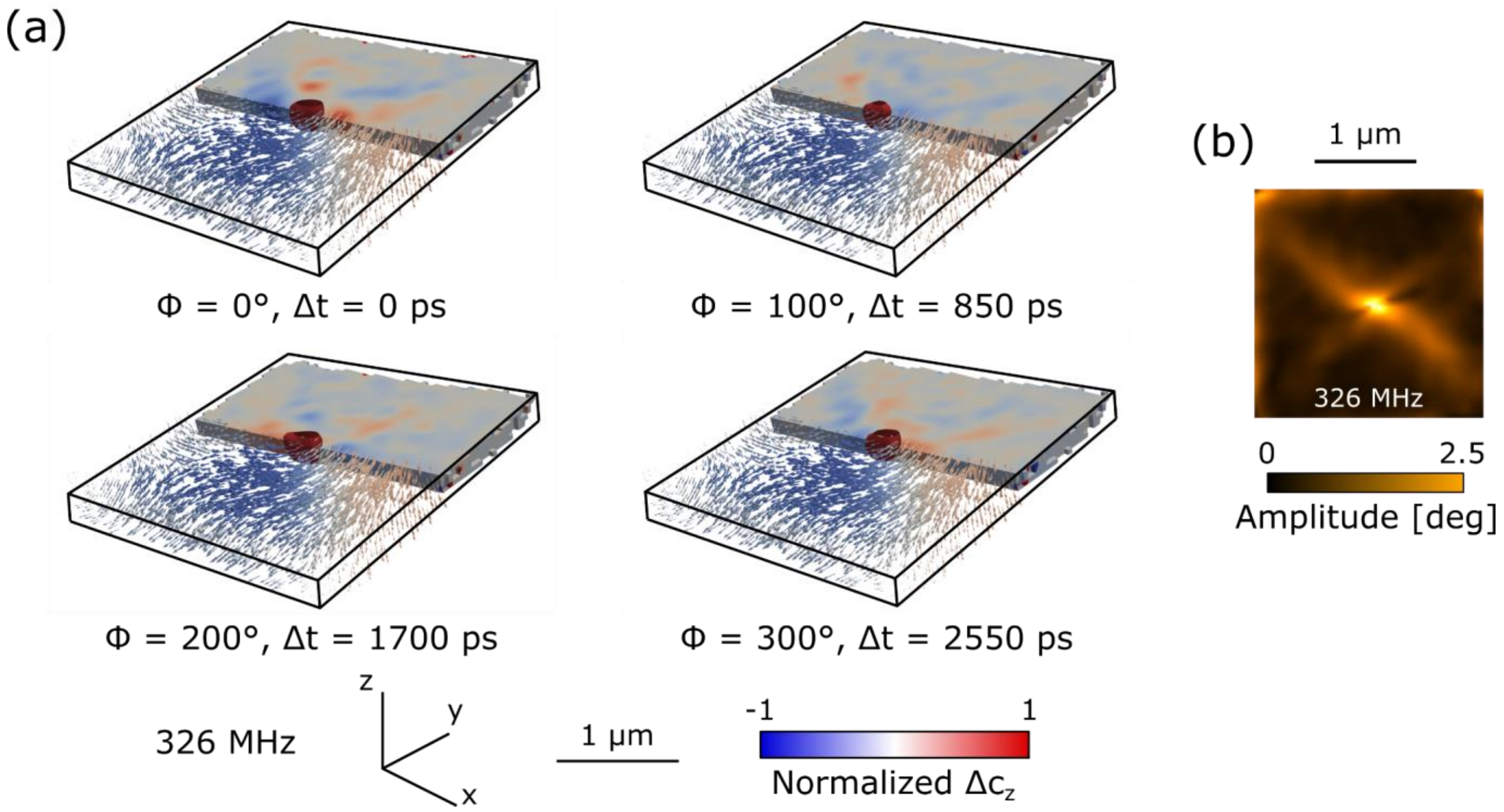

***Fig. 2:*** *(a) Snapshots of a time-resolved X-ray laminography image depicting the gyration of a magnetic vortex core in a 150-nm thick CoFeB microstructured square excited by an oscillating in-plane magnetic field at 326 MHz. The position of the vortex core is marked by the red isosurface of the curl of the magnetization. (b) Top-view of the microstructured CoFeB square showing the localization of the vortex core gyration. Adapted from Ref. 23 with permission from ACS Publications.*

**Current and Future Use of the Technique for Material Science**

The dominant contributions from time-resolved high-resolution imaging of magnetization dynamics to materials science are expected to lie in the field of (quasi) 2D materials, thin films and heterostructured systems based on such ingredients. Given that many existing and proposed functional devices for storage, computation and sensing are realized in thin-film technologies, and given that generating a particular device function typically goes along with a lateral patterning of components on the nanometer scale together with (transient) local changes of the magnetization state, this focus on "2D materials science" is hardly a restriction. We envision that the following (partially overlapping) areas will benefit, in particular, from progress in time-resolved magnetization imaging on the nanometer scale:

**All-optical switching (AOS)** of magnetization states via application of single or multiple light pulses is studied intensively as it a potentially very fast and energy-efficient means of controlling magnetization, e.g., for data-storage purposes. Considering spatio-temporal dynamics, the femtosecond dynamics can be expected to occur on the shortest lengthscale of neighbouring atoms, with spin textures on the nanometer or even mesoscale exhibiting slower dynamics in the picosecond or even nanosecond regime. The ultrafast processes involved include transport of hot electrons and spins, also in the lateral dimension. [11]

Intrinsically very fast dynamics on the interatomic scale is present in **antiferromagnetic and multiferroic systems**: Here, contrast can be generated by XMLD probing of the magnitude and direction of the Néel vector. Furthermore, the insensitivity of the imaging techniques discussed here to external electric and magnetic fields enables studies of device prototypes in operation. The atomic selectivity in resonant imaging is an additional asset to understand how dynamics relates to function in these compounds. Many of these systems are grown epitaxially on single-crystalline substrates, and we note that significant progress has been made in locally thinning such samples to soft x-ray transparency. [30,31]

Atomic selectivity is of great relevance also in the area of **2D magnets and heterostructures built up from 2D components**. The ability to stack van-der-Waals bonded 2D systems onto each other has fuelled the imagination of how to realize spin-based devices in such geometries. In the direction normal to the individual 2D constituents, spin currents may be generated, e.g., by the spin Hall effect or exploiting spin-momentum locking in suitable materials, possibly via optimized optical pulses. [32] The effect of spin injection into different layers, including, e.g., spin accumulation at interfaces or the generation of spin-orbit torques can nicely be “dissected” via imaging the local magnetization within the constituent layers separately.

Focusing more on lateral dynamics, **magnonic devices** in addition to spintronic devices can be expected to be understood in more detail on the basis of time-resolved studies with nanometer spatial resolution. [9,10] In passing, we note that x-ray-based reciprocal-space techniques may also be very valuable to address a frequency-momentum region not accessible by visible light.

In recent years, **topological spin structures in real space** and their dynamics have attracted considerable interest. So far, only few aspects of magnetic skyrmion dynamics have been accessible to study via time-resolved imaging, [2,5,6] while some insight, in particular, in the optically triggered generation and decay of such structures have been enabled by time-resolved reciprocal-space studies with x-rays. [33] In particular to understand dynamics such as nucleation, topological transformation, or interaction with defects, time-resolved real-space imaging of the transformations of magnetization textures will be key to an understanding of the fundamental processes, hopefully paving the way to also exert control in applications. This is also true for the vast field of domain-wall motion, where lateral heterogeneity is critical to the dynamics and where x-ray imaging already has made unique contributions to understand changes in the spin structure during motion. [34]

Understanding **rare failure events** (or even **rare success events**) at the odd time when something happens differently than it typically does, is a classical case for time resolved imaging – but, so far, has been elusive for processes at the nanometer level. Given the technological importance of understanding such events and optimizing devices, true “video-mode” imaging rather than pump-probe experiments may be very instrumental. As far as the dose required for magnetic imaging with a single exposure per frame goes, microsecond dynamics of magnetization processes should become accessible at high-repetition rate at x-ray FELs. However, studies may remain limited by dose-driven damage. The observation of **stochastic dynamics** including (thermal) fluctuations is a related field as far as the need for video-mode imaging goes. There is hope that combined correlation and imaging approaches like CCI can be pushed to the relevant spatio-temporal range for the respective problem. [25]

Finally, the observation of **magnetization dynamics in 3D** is an experimental challenge, with first steps recently taken. [26,27] In 3D, topologically nontrivial structures different from the ones that can exist in 2D are possible, with a hopfion being one example. It would be very interesting to observe such objects and their dynamics, which will be challenging even when only considering the size of the required data sets. While pump-probe schemes are most promising in this respect, the application of suitable pump events may represent another formidable challenge depending on the material or device to be studied.

# 9 - Scanning Electron Microscopy with Polarization Analysis (SEMPA)

*Robert Frömter[1] and Mathias Kläui[2]*

Institute of Physics, Johannes Gutenberg University Mainz, 55099 Mainz, Germany

[1] [ORCID 0000-0002-3399-5628]
[2] [ORCID 0000-0002-4848-2569]

**Current State of the Technique**

Scanning Electron Microscopy with Polarization Analysis (SEMPA or spin-SEM) [1] is a powerful magnetic imaging technique that was developed already in the 1980s in Japan [2], USA [3], Germany [4], and Switzerland [5]. The method relies on measuring the spin polarization of the secondary electrons (SE) that are locally emitted from the surface of a ferromagnet (FM) while scanning it with a focused primary electron beam. The vector of spin polarization of the SE is oriented opposite to the local magnetization of the FM within the very small information depth of less than 1 nm from the surface. The spin polarization of the SE is measured either by high-energy Mott scattering [2], [5] or by low-energy electron diffraction (SPLEED) at a W(001) single crystal [6].

By simultaneously measuring four symmetric scattering directions two components of the spin-polarization asymmetry are obtained simultaneously, usually corresponding to the two components of the magnetization within the surface plane of the sample, as illustrated in Figure 1. The missing out-of-plane component can be detected either permanently, by a 90° electrostatic deflection of the detector axis, or subsequently, by an electromagnetic spin rotator [7], or simply by tilting the sample [8], [9]. Due to the required vacuum environment, a wide range of sample temperatures can be easily accessed. Under optimum conditions a spatial resolution of 3 nm has been demonstrated imaging bit boundaries in a perpendicular recording medium [10]. Beam damage is not an issue on metallic samples for the used parameters, as most of the primary-beam energy is distributed into a large volume beneath the surface during the cascade process.

Owing to the very high surface sensitivity of SEMPA, dedicated UHV systems are needed to preserve the SE spin polarization at the sample and to operate the SPLEED detector. While the early systems were custom built, a full-featured commercial UHV SEMPA instrument has been available from Scienta Omicron GmbH since the early 2000s, making the technique available to a wider community.

Exploiting the single-electron detection scheme of the SPLEED detector and adding a fast time-to-digital converter, it is possible to subdivide the recorded single-electron counts into successive time frames that are synchronized with a periodic external excitation of the sample, like an alternating current in a microstrip underneath giving rise to a modulated Oersted field. In this way, a movie of periodically driven magnetization dynamics can be recorded at full duty cycle, i.e., without sacrificing counting efficiency [11], [12]. Even without further in-vacuum optimizations of the spin detector, a time resolution of 700 ps has been demonstrated [11].

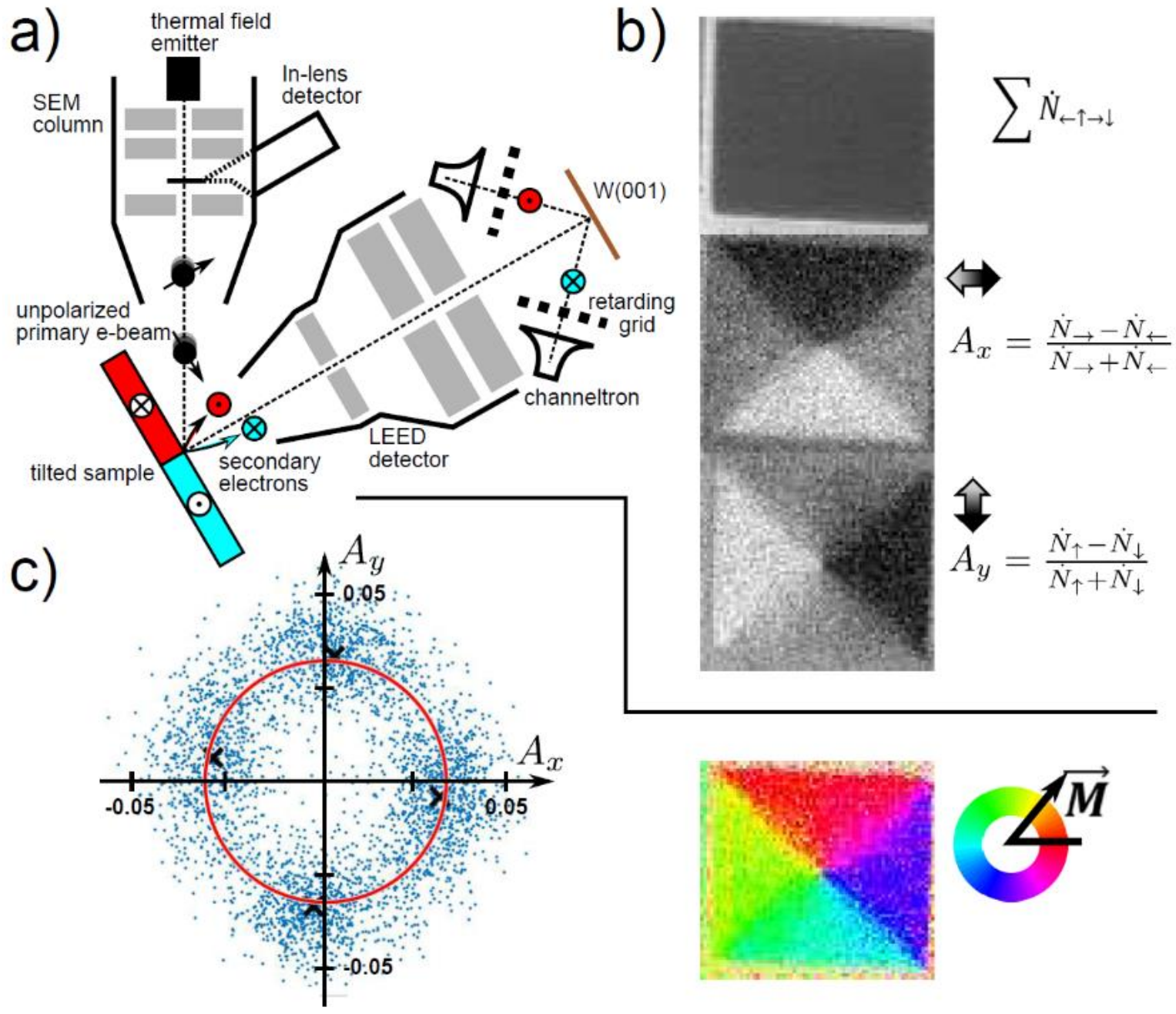

***Figure 1**: a) Schematic of SEMPA. b) Dataset from a 4 × 3 µm² $Ni_{80}Fe_{20}$ rectangle. The topography is obtained by summing the counts from all four detector channels. Asymmetry values $A_x$ and $A_y$ represent the two in-plane magnetization components. c) Color-coded in-plane magnetization map obtained using the scatter plot of the ($A_x$, $A_y$) pairs from all pixels of the image and calculating the respective angles from the centre (courtesy of D. Schönke).*

**Future Advances of the Technique**

Even though SEMPA may appear as a mature method, there are consistently new technological developments in fast electronics, electron optics, and detectors, leading to new advanced capabilities. Most notable is the already mentioned addition of time resolution. The presently obtained value of 700 ps is not limited by the detection electronics or the microchannel plate electron multiplier – both would allow for a time resolution well below 100 ps. It is determined by the time jitter of the electrons passing through the SPLEED detector, where especially the passages at very low kinetic energy at the sample surface and at the retarding grid contribute. By further miniaturization and using a faster filter, the total flight time and thus the jitter can be significantly reduced in an optimized detector design.

The above-stated maximum documented spatial resolution of 3 nm appears very competitive with other methods. However, it has been obtained using a unique UHV-SEM column that is equipped with a specifically optimized spherical and chromatic aberration corrector, allowing for 2 nm beam diameter at a working distance of 10 mm, energy of 20 keV and beam current of 1 nA [10]. Systems without aberration correction routinely obtain magnetic resolutions of 10 to 15 nm, which is a consequence of the required relatively large working distance and the high desired beam current, both increasing the lens aberrations. With today's progress in machine learning for image analysis, the focusing and astigmatism correction of the electron beam can be automatized, allowing for routine operation under optimized beam conditions, even on low-non-magnetic contrast samples.

There are also new developments concerning spin detectors. A significantly higher electron reflectivity at a similar spin sensitivity can be obtained when using the exchange interaction in scattering from an Fe layer at 6.3 eV instead of spin-orbit coupling in the W-SPLEED at 104.5 eV. Such a detector is presently marketed by FOCUS GmbH for SEMPA under the brand FERRUM [13]. We are presently characterizing the prototype of a SPLEED detector using Au/Ir(001) that also promises improved sensitivity and service time, while maintaining the two-component vector imaging capability. An improved detector efficiency enables higher spatial resolution, the study of lower-contrast samples, or reduced measurement times.

Given the reported progress in imaging on ambient-condition-stable samples, such as using capped surfaces or magnetic oxides, it could be interesting to revisit the potential for SEMPA as an add-on feature to standard SEM systems. While a Mott detector will operate under such reduced vacuum conditions (typically $5\times10^{-7}$ mbar), earlier attempts suffered from the build-up of cracked hydrocarbons on the sample surface, greatly reducing the SE spin polarization over extended exposure times [14]. A hydrocarbon-free vacuum and in-situ UV ozone cleaning may be applicable to mitigate such problems.

A shortcoming of SEMPA for certain applications, such as imaging of spin structures that are stabilized by external fields, is the apparent limitation to zero magnetic field. However, it has been demonstrated that in-plane fields of up to 100 mT can be applied during SEMPA imaging with appropriate corrections of the disturbing effects [15]. Locally confined out-of-plane fields of some mT can very likely be applied by using microcoils, as shown in in Ref. [16], and compensated for in a similar way as with the in-plane fields.

**Current and Future Use of the Technique for Material Science**

The main challenge in carrying out successful SEMPA investigations is the very low efficiency of all presently available detectors for the spin of free electrons. The contrast-to-noise ratio in the SEMPA images is dominated by the Poisson noise of the single-electron-counting statistics. Therefore, it increases linearly with the spin polarization of the detected SE and the spin sensitivity of the detector, but it increases only with the square root of the measurement time and the primary beam current [6]. The latter is usually at variance with spatial resolution due to lens aberrations, so it cannot be arbitrarily increased. To obtain the same image quality at half the spin polarization requires four times longer acquisition times. As a consequence, over the first decades many experiments were carried out on in-situ prepared uncapped ferromagnetic 3*d*-metal layers, where measurement times of several minutes were sufficient for good contrast, owing to the high spin polarization of the low-energy SE of up to 50 % for the case of iron. There are still today relevant scientific questions that can be addressed in this way, like the measurement of the Dzyaloshinskii–Moriya interaction strength in epitaxial Co/Ir(111) [8], or studying the metamagnetic transition of epitaxial FeRh(001) films [17].

To image the magnetization of ex-situ prepared samples two methods were used from early on: in-situ Argon-ion sputter cleaning and in-situ dusting [18], [19] with an ultrathin iron layer or a combination of both. Both methods can in principle restore the full spin-polarization of a 3*d* ferromagnet, however they will also modify the magnetic properties at the surface. While this is usually no problem with bulk samples, it may strongly influence ultrathin films and multilayers; which can even be exploited for depth profiling [20]. Sputter cleaning has been successfully applied for instance in a variety of hard-disk-drive related studies [10], [21], multiferroic materials [22], [23], exchange bias systems [24], and chiral domain walls in ferrimagnetic alloys [25].

Many magnetic systems that are in the focus of present-day physics and technology provide SE with less spin polarization than the 3d metals and their alloys, or are too sensitive to allow for sputter cleaning or dusting. Here, at present the only way is to increase the measurement time, which - due to issues with stability - usually requires drift correction based on automatic pattern matching. The correction can be carried out either on the fly, or post-acquisition on an image stack, and is usually done based on the non-magnetic sum image. In this way, measurement times up to several days for a single SEMPA image have been realized. Drift correction is also a necessary procedure for time-resolved SEMPA, as the measurement time scales linearly with the number of frames in a movie.

The ability to measure systems with reduced SE spin-polarization enables access to studying systems without any in-vacuum pretreatment, directly after inserting from atmosphere, provided the FM surface has been covered during deposition with a protective layer thin enough to retain some SE spin-polarization and tight enough to prevent oxidation. E.g., a monolayer of graphene will protect a clean Ni(111) surface, while retaining 1/3 of the original spin polarization from Ni [26]. In a systematic study using a sputter-deposited Pt wedge on 1.1 nm Co we could show that a cap layer of 1.1 nm Pt is sufficient to block oxidation, while retaining ¼ of the SE spin polarization from Co [27]. This allows, e.g., for studying chiral domain walls in sputtered multilayers [9], [28], as well as more recently topological spin structures such as merons and antimerons in layered synthetic antiferromagnets [29]. Due to its surface sensitivity, SEMPA is especially well suited to study synthetic antiferromagnets, whereas depth-integrating methods obtain only very little contrast due to the mutual compensation of the layers. By virtue of its bilayer-stepped interface, even the antiferromagnet $Mn_2Au$ could be imaged via its coupling to a permalloy layer that was protected by 2 nm of highly spin-transparent $SiN_x$ [30].
There are even magnetic materials where the surface is intrinsically stable against oxidation and that can be directly imaged in SEMPA, like the half-metallic perovskite manganite $La_{0.7}Sr_{0.3}MnO_3$ [31].

Being set up in a standard UHV lab environment, SEMPA offers unlimited access times in contrast to synchrotron-based X-ray methods. The capability for imaging untreated, ex-situ prepared, capped samples is a strong new feature, since it allows for direct interchange and comparison with other spatially resolving as well as integrating methods. A novel route to understand magnetic systems better, is thus the study of systems using complementary techniques. As previously shown, a comparison of magnetic imaging with for instance NV microscopy that detects the stray field can give additional insights into a sample's physics [32], as both techniques determine complementary quantities and have different depth and lateral sensitivities. In particular, given the very strong surface sensitivity, SEMPA will be a valuable tool in such studies where one combines SEMPA with for instance a more bulk sensitive or a 3D-imaging method.

**Acknowledgements**
This work has received funding from the European Research Council (ERC) under the European Union's Horizon 2020 research and innovation programme (Synergy Grant No. 856538, project "3D-MAGiC"). It has also been supported by the Deutsche Forschungsgemeinschaft (DFG, German Research Foundation), grant TRR 173 - 268565370 (project A01, B02) and grant SPP 2137 Skyrmionics - 403502522.

# 10. Transmission electron microscopy (TEM) Techniques

## 10a – Lorentz transmission electron microscopy

Fehmi Sami Yasin[1], Benjamin J. McMorran[2], Shinichiro Seki[3], Xiuzhen Yu[4]

[1] Center for Emergent Matter Science (CEMS), RIKEN, Wako, Japan [ORCID: 0000-0001-9382-7565]
[2] Department of Physics, University of Oregon, Eugene, Oregon, USA [ORCID: 0000-0001-7207-1076]
[3] Department of Applied Physics and Institute of Engineering Innovation, University of Tokyo, Tokyo, Japan [ORCID: 0000-0001-6094-8643]
[4] Center for Emergent Matter Science (CEMS), RIKEN, Wako, Japan [ORCID: 0000-0003-3136-7289]

**Current State of the Technique**

For decades, LTEM has been used to study magnetic configurations in thin magnets. In LTEM (right-hand-side (RHS), Fig. 1), a plane wave of electrons is transmitted through the magnetic field or vector potential $\left(\boldsymbol{A}(\boldsymbol{r}_\perp, z)\right)$ distribution, imparting a phase (Lorentz deflection in the classical picture) onto the passing electron wavefronts of the form $\phi_m(\boldsymbol{r}_\perp) = \frac{e}{\hbar}\int_L \boldsymbol{A}(\boldsymbol{r}_\perp, z) \cdot d\boldsymbol{r}$ [1]. The distorted wavefronts form measurable contrast after propagating some distance, revealed by defocussing $(\Delta f)$ the image. This allows for phase retrieval using the transport-of-intensity equation and an over-, under-, and in-focus image [2], although recent improvements require only a single defocussed image on some samples [3].

Consider the magnetic field distribution in a proper screw spin propagation with period $\lambda_{hel}$ along the y-axis within a flat specimen with thickness $t$, which can be expressed as $\boldsymbol{B}_\perp = B_0 \sin\left(\frac{2\pi}{\lambda_{hel}} y\right)\hat{x}$. The phase difference between the origin and $y$, $\Delta\phi_m(\boldsymbol{r}_\perp) = \frac{e}{\hbar}\int\int \boldsymbol{\nabla} \times \boldsymbol{A} \cdot \mathrm{d}\boldsymbol{S}$ (here $\boldsymbol{S} = \hat{x}$ ) can be rewritten using Stokes theorem to obtain $\phi_m(y) = \frac{e}{\hbar}\int_{-\frac{t}{2}}^{+\frac{t}{2}}\int_0^y \boldsymbol{B}_\perp \,\mathrm{d}y\,\mathrm{d}z = -\frac{e\,\lambda_{hel}}{h} B_0\, t\cos\left(\frac{2\pi}{\lambda_{hel}} y\right)\hat{x}$. The simulated image near the bottom RHS of Fig. 1 illustrates a defocussed image from the above field distribution. This powerful technique and phase retrieval algorithm were used to observe magnetic skyrmions in real space for the first time[4] (bottom RHS of Fig. 1).

Scanning TEM (STEM) mode offers an alternative optical setup, and two types of Lorentz STEM imaging are outlined on the left-hand-side (LHS) of Fig. 1 including differential phase contrast STEM (DPC-STEM) [5] and interferometric 4D-STEM [6], [7]. By measuring beam shifts across a segmented annular detector or imaging interferograms on a pixelated grid detector (e.g., charged couple device), DPC-STEM and interferometric STEM directly measure the in-plane magnetic induction and phase, respectively. DPC-STEM-measured in-plane magnetic induction maps of an antiskyrmion square lattice and elliptically shaped skyrmion present in the recently discovered $(Fe_{0.63}Ni_{0.3}Pd_{0.07})_3P$ [8] are shown below the DPC-STEM schematic while an interferometric 4D-STEM-measured magnetic induction map of Néel-type skyrmions in a room temperature two-dimensional (2D) polar magnet [9] is shown below the 4D-STEM schematic. Recent advances in detector acquisition speed and dynamic range have resulted in an explosion of LTEM work studying spin texture dynamics [10]–[12], and the rise of Lorentz 4D-STEM magnetic imaging. 4D-STEM was recently used by Nguyen *et al.* to resolve antiferromagnetic magnetic ordering in $Fe_2As$ [13].

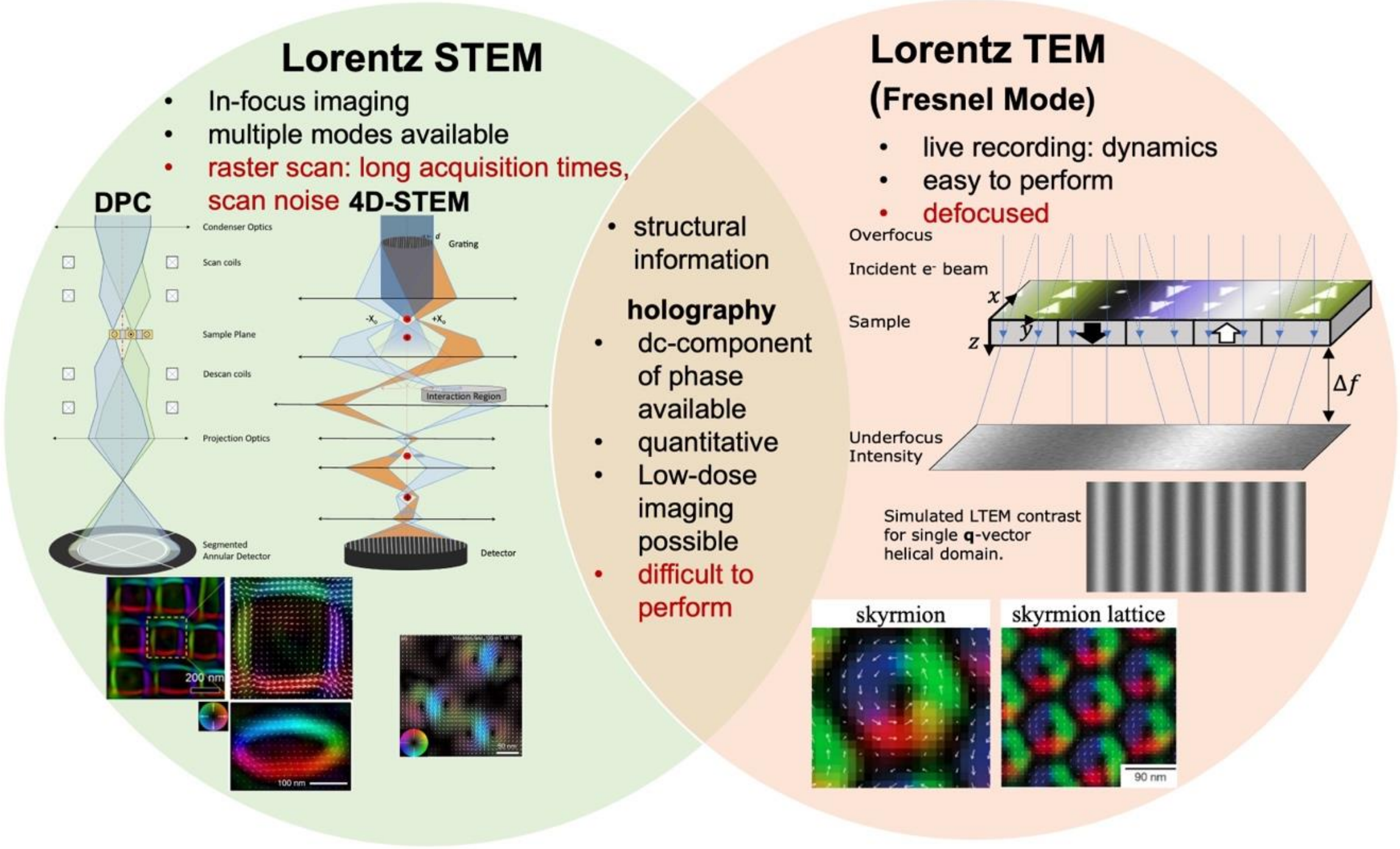

***Figure 1**: Overview of Lorentz (S)TEM techniques used for magnetic imaging along with advantages (black text) and disadvantages (red). Schematics include (left to right) differential phase contrast (DPC) STEM and a variant of 4D-STEM called STEM holography. DPC-STEM-measured magnetic induction maps of square shaped antiskyrmions and elliptical skyrmions shown below the DPC-STEM schematic. Reprinted figure S4e-h with permission from [Karube, K., Peng, L., Masell, J., et al., Nature Materials **20**, 335-340 (2021), Springer Nature.] 4D-STEM-measured magnetic induction maps of tilted Néel-type skyrmions shown below the 4D-STEM schematic Reprinted figure 4d with permission from [Zhang, H., Shao, Y. T., Chen, R., et al., Physical Review Materials, **6**, 044403 (2022).] Copyright 2022 by the American Physical Society. Schematic of LTEM shown on the right-hand side with a defocused LTEM micrograph simulated from a magnetic helical domain shown below. Magnetic induction maps of a skyrmion and skyrmion lattice calculated using TIE are shown at the bottom. Reprinted figure 1e-f with permission from [Yu, X. Z., Onose, Y., Kanazawa, N., et al., Nature **465**, 901-904 (2010), Springer Nature.]*

**Future Advances of the Technique**

LTEM has been a crucial tool for imaging magnetic materials, with several promising directions that could further establish LTEM as an indispensable technique for magnetism research. One major advancement is the real-time imaging of dynamic magnetic phenomena such as magnetic domain wall motion and spin texture transformations. This is achieved by holding the LTEM image at a constant defocus, rendering magnetic contrast on the detector that may be observed via live recording while applying an external stimulus such as magnetic field, electric current or heating. Recent developments in direct electron detector technology have significantly improved the frame rates and electron-dose efficiency. For example, the EMPAD-G2 can achieve tens of thousands of frames per second (fps) and offers dynamic ranges capable of single electron detection up to 180 pA per pixel [14]. Optically pulsed electron sources and fast beam blankers (choppers) [15] enable ultrafast TEM [16] and even ultrafast STEM, also known as 5D-STEM [17]. These types of advancements enable real-time imaging of dynamic magnetic events while applying external stimuli. Additionally, advancements in TEM holder technology, including holders combining several features such as heating/cooling, electric/thermal current, and the addition of piezo probes for applying controlled mechanical pressure on samples, have expanded the possibilities for varied magnetic responses and experimental designs that better represent real-world device conditions. For example, a recent experimental design allows for in-situ magnetic imaging and electronic transport measurements, enabling the direct correlation between the anomalous Hall effect and real-space spin textures for the first time [18].

These detector advancements have also enabled 4D-STEM holography and ptychography techniques that have, in some cases, remained dormant for over 40 years [7]. When configured in Lorentz mode, these techniques allow for multi-modal imaging, simultaneously mapping the magnetic field, electrostatic field, strain, and polarization. Ptychography was recently used to beat the usual diffraction limit and phase resolution in Lorentz mode [19]. This unprecedented ability is expected to advance further with the adoption of phase-structured electron beams using phase plates, diffraction holograms, and programmable apertures as well as the development of TEMs optimized for magnetic imaging with aberration correction and improved sample positioning for higher resolving and magnifying power in LTEM.

Another promising advancement in LTEM is the progress in three-dimensional (3D) imaging using tomography holders and reconstruction algorithms. Traditional 3D magnetic imaging requires the acquisition of 2D projections of the 3D magnetic field at many tilt angles which can be used to back project the full 3D magnetic field. It is typically labour-intensive, requiring fine alignments of the field of view for every tilt angle along two perpendicular tilt axes. However, machine learning-driven software has begun automating much of this process, reducing the human effort involved in data acquisition. Additionally, the development of multibeam techniques such as multibeam electron diffraction [20] may enable the imaging of 3D dynamic magnetic phenomena when configured in Lorentz mode. With simultaneous acquisition of multiple tilt angles, these techniques allow for real-time viewing during the application of external stimuli, further enhancing the understanding of 3D magnetic features.

Finally, the integration of artificial intelligence (AI) into LTEM experiments has the potential to revolutionize the user experience and enable the design of previously unthinkable experiments. AI can be utilized for real-time electronic aberration correction, sample and beam drift correction, and real-time magnetic induction mapping, to name a few operative procedures. In post-processing, AI can assist in sifting through large data sets which are often generated in 3D LTEM imaging and 4D-STEM experiments and help correlate the magnetic real space imaging with other analytical techniques such as transport measurements, strain mapping, X-ray microscopy, or pump-probe ultrafast microscopy, providing a comprehensive understanding of the magnetic structure within materials and its response to external stimuli.

**Current and Future Use of the Technique for Material Science**

As mentioned earlier, one of the most powerful capabilities available in LTEM is the real space measurement of spin textures and their response to external stimuli in live time. One example is the recent discovery and real space observation of magnetic (anti)merons [21], spin textures pictured in Fig. 2a that twist around half the unit sphere from their cores to their border, granting them half integer topological charge, $N_{(anti)meron} = \frac{1}{4\pi}\int\int \boldsymbol{n}\cdot\left(\frac{\partial\boldsymbol{n}}{\partial x}\times\frac{\partial\boldsymbol{n}}{\partial y}\right)dx\,dy = \pm 1/2$ . They contrast with magnetic skyrmions, whirling spin textures also pictured in Fig. 2a that traverse the entire unit sphere once from core to edge, with a corresponding integer topological charge, $N_{Skyrmion} = -1$. Using LTEM, Yu, et al. could track the merons as they warped in real space while under an increasing magnetic field towards the final skyrmion lattice state illustrated in Fig. 2b.

As more magnetic imaging-specific TEMs are developed with an eye towards improving the magnifying power and correcting more lens aberrations, real space observations of nanometric spin textures such as the square skyrmion lattice shown in Fig. 2c [22] and their dynamics are expected to become more commonplace. Advancements in detector speed and sensitivity will also enable previously unobtainable measurements of magnetic spin texture dynamics such as the electric current driven motion of magnetic skyrmions and their bunches shown in Fig. 2d [23]. With current technology, LTEM is limited to acquiring still images between current pulse applications since the spin texture motion is many orders of magnitude faster than the camera readout speeds. While such frame rates may not be obtainable in the near future, and indeed the most promising path may lie in ultrafast pump-probe LTEM methods currently under development, magnetic thin plates have already shown under certain conditions to host spin texture responses slow enough and/or in a steady state deformation to view via live recording. Such an electric current-induced skyrmion response was recently measured and quantitatively characterized [11]. LTEM clearly shows the spin transfer torque-driven elliptical deformation and stretching of skyrmion domain walls, illustrated in Fig. 2e.

One magnetic imaging-specific STEM has already been developed at the University of Tokyo. MARS, or Magnetic field free Atomic Resolution STEM, enables the direct observation of atoms in a magnetic field free environment. This achievement was made possible by the design of a new objective lens that maintains the resolving power of the microscope while isolating the sample from any magnetic fields and including aberration correction. Using this TEM configured for DPC-STEM, Shibata, et al. resolved the atomic magnetic field distribution inside antiferromagnetic haematite ($\alpha$-$Fe_2O_3$), shown in Fig. 2f [24].

2D materials have also seen a surge of interest due to their emergent properties such as superconductivity and magnetism that are measurable using LTEM [25]. While magnetic thin plates generally avoid electron beam knock-on damage, 2D materials composed of lighter elements may not. To avoid such radiation damage, LTEM imaging may be performed at lower beam energies.

In conclusion, magnetic imaging using LTEM is an active and exciting field with several promising directions for future advancements. Real-time imaging of dynamic magnetic phenomena, 3D imaging, integration of AI during experiments and post-processing, and improvements in resolution and aberration correction are some of the areas that hold great potential for unlocking new insights into the behavior of magnetic materials. These advancements promise to enable researchers to design and conduct experiments that were not previously feasible, opening new avenues for understanding and manipulating magnetic structures at the nanoscale.

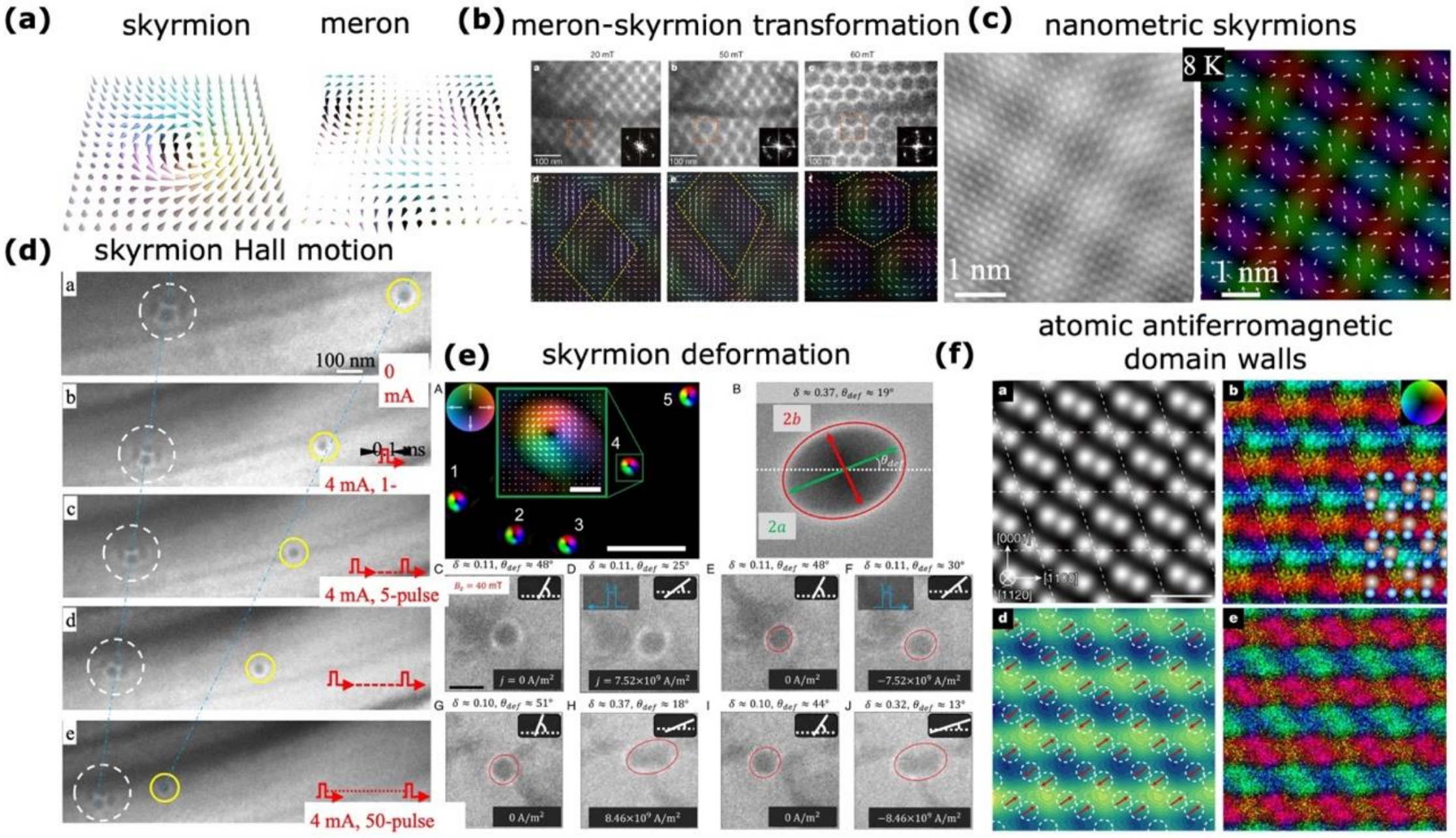

***Figure 2**: **a**, Magnetization schematic of a skyrmion (left) and meron (right). **b-e**, (b) Magnetic field-induced transformation from meron to skyrmion lattice, (c) nanometric square skyrmion lattice, (d) electric current-induced skyrmion bunch and isolated skyrmion Hall motion, and (e) spin transfer torque-induced elliptical skyrmion deformation measured using LTEM. **f**, Atomic resolution field free DPC-STEM-measured magnetic field distribution inside antiferromagnetic $\alpha$-$Fe_2O_3$. (b) reprinted figure 2a-f with permission from [Yu, X. Z., Koshibae, W., Tokunaga, Y., et al., Nature **564**, 95-98 (2018).]. (c) reprinted figure 4e, h with permission from [Khanh, N. D., Nakajima, T., Yu, X. Z., et al., Nat. Nanotech. **15**, 444-449 (2020), Springer Nature]. (d) reprinted figure 4a-e with permission from [Yu, X. Z., Morikawa, D., Nakajima, K., et al., Science Advances **6**, 25 (2020)]. (e) reprinted figure 2 with permission from [Yasin, F. S., Masell, J., Karube, K., et al., Proceedings of the National Academy of Sciences **119**, 41 (2022).] (f) reprinted figure 3a-b, d-e with permission from [Kohno, Y., Seki, T., Findlay, S. D., et al., Nature **602**, 234-239 (2022), Springer Nature].*

**Acknowledgements**

This work was supported in part by Grants-In-Aid for Scientific Research (A) (Grant No. 19H00660) from the Japan Society for the Promotion of Science (JSPS) and CREST program (Grant No. JPMJCR1874, JPMJCR20T1) from the Japan Science and Technology Agency (JST), Japan.

# 10b. Electron Holography

*Axel Lubk[1,2], Daniel Wolf[1]*

[1] Leibniz Institute for Solid State and Materials Research Dresden, Dresden, Germany
orcid.org/0000-0001-5000-8578
[2] Institute of Solid State and Materials Physics, TU Dresden, 01062 Dresden, Germany
orcid.org/0000-0003-2698-8806;

**Current State of the Technique**

In the general context of magnetic imaging, electron holography (EH) refers to a set of transmission electron microscopy (TEM) techniques that reconstruct the Aharonov-Bohm phase shift $\varphi_{\mathrm{mag}}$, imprinted on the beam electrons' (beam direction $z$) quantum wave function during transmission through a thin magnetic sample. Today's electron microscopes allow the recording of two-dimensional (2D) phase maps $\varphi_{\mathrm{mag}}(x, y)$ in various magnetic-field-free imaging modi (referred to as Lorentz (S)TEM, see Section 10a), namely off-axis EH (Fig. 1a), differential defocus and focal series inline EH, as well as differential phase contrast that provide a spatial resolution approaching the sub-nm regime [1] and a phase resolution beyond $2\pi/1000$ [2]. Under kinematical scattering conditions these maps can be quantitatively evaluated in terms of projected in-plane magnetic flux densities via their directional derivatives, i.e., $\partial_x \varphi_{\mathrm{mag}} = e/\hbar \int_{-\infty}^{\infty} B_y dz$ and $\partial_y \varphi_{\mathrm{mag}} = -e/\hbar \int_{-\infty}^{\infty} B_x dz$. Differential defocus EH (also referred to as TIE method) retrieves that information from two mutually defocused Lorentz LTEM images [3]. Focal series EH analyses a whole set of defocused images, while off-axis EH utilizes the interference between the electron wave passing both the object as well as a (field)-free reference region [4]. Differential phase contrast deviates from the above interferometric techniques in that local deflections of a focused electron beam are measured while scanning the beam over the sample (STEM), which may then be translated into the Aharonov-Bohm phase exploiting the semiclassical relationship between the wave front (phase) and classical trajectories (deflection). EH has been mainly employed for studying nanoscale magnetic field configurations governed by micromagnetics, including, e.g., nanoparticles [5], nanowires [6], domain walls [6], [7], skyrmions [8], [9], but also assembled 3D MRAM devices [10]. That includes the in-situ application of additional stimuli notably magnetic fields (mostly in $z$-direction, typically up to several 100 mT) and temperature (ranging from 4K to 1000K). Combining EH and electron tomography enables the reconstruction of magnetic vector fields in 3D (Fig. 1b) [6], [11]. To reconstruct the flux density vector field $\boldsymbol{B}$ being subject to $\operatorname{div} \boldsymbol{B} = 0$ at least two tilt series around mutually perpendicular axes are necessary. This approach has been previously used to study 3D magnetization textures such as transverse and longitudinal vortices in magnetic nanowires [6], [11] as well as Bloch skyrmion tubes in FeGe helimagnets [12].

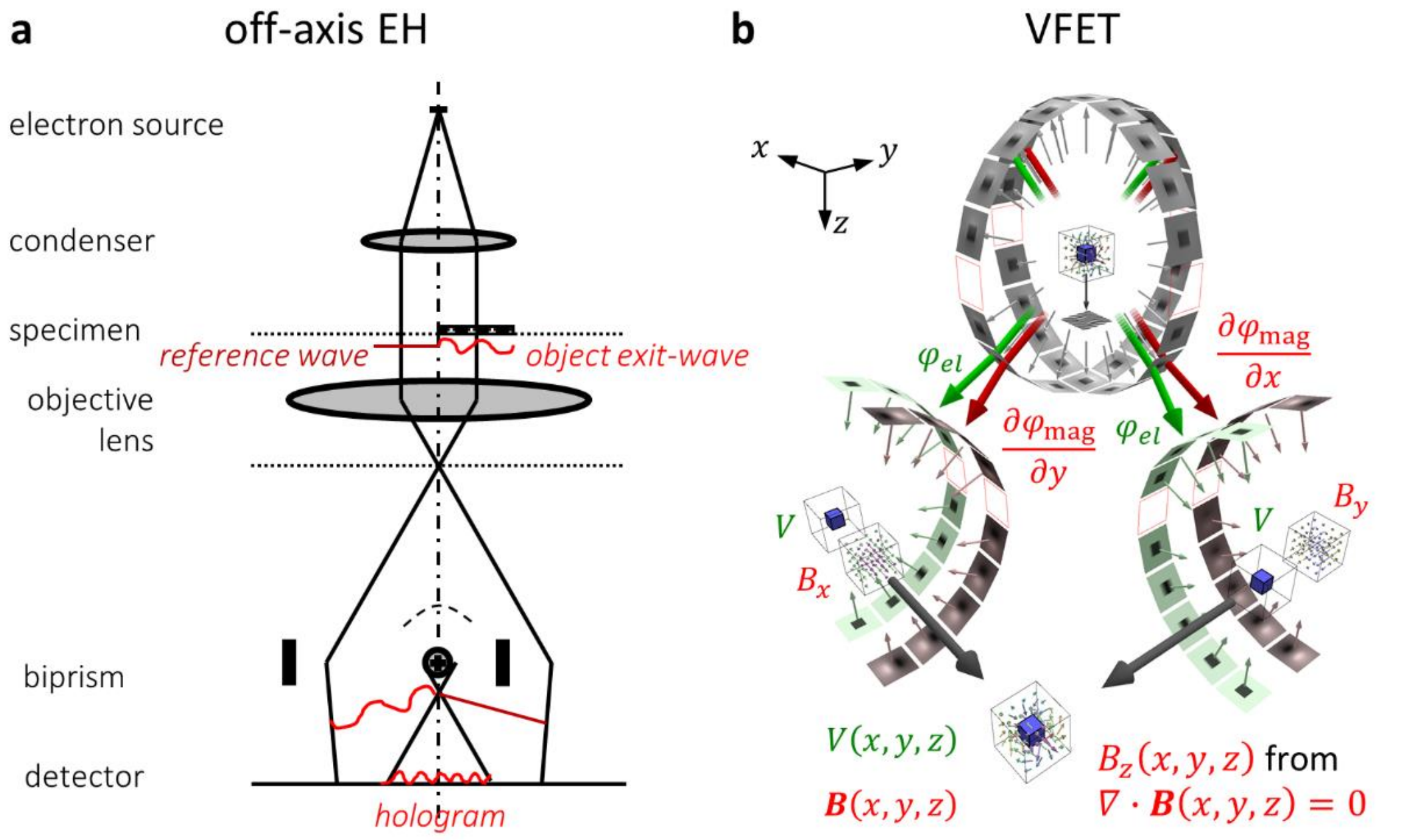

***Figure 1**: Scheme of (a) electron holography (here: off axis) and (b) vector field electron tomography employing two perpendicular tilt-axis. 360° tilt series are required to separate electric and magnetic phase shifts, which behave different under space inversion.*

**Future Advances of the Technique**

Over the last decade EH techniques have seen drastically improved precision and accuracy by employing high-brightness cold-field emission guns, improved aberration-corrected field free electron optics, automated long-term hologram series acquisition involving drift correction, improved detector technology, such as direct electron detection, and increased overall instrumental stability [2]. This quest will continue, e.g., with the arrival of next generation detectors and the increasing implementation of fully-fledged auto-alignment of the whole TEM column. Considering that spatial resolution in electron holographic magnetic imaging is noise-limited to a large extend (the magnetic phase shift is small compared to the dominating electrostatic contribution) this will further push the resolution to atomic length scales provided by the optics of the microscope.

Notwithstanding, the trend toward improved in-situ capabilities, better integration with other experiments and methods, and increasing the time-resolution [13] will probably the dominating theme in the next couple of years. Another emerging trend concerns the increasing implementation of automated evaluation of characteristic magnetic quantities, such as exchange stiffness, magnetocrystalline anisotropies, or topological charge, profiting from modern machine learning algorithms. The remainder of this section attempts a short outlook into these impeding advances of EH.

In-situ TEM techniques combine the optical power and versatility of state-of-the-art transmission electron microscopes with miniaturized instrumentation, which allow to manipulate and stimulate the magnetic sample within the small space available in TEM. With the advent of MEMS, 3D printing and other nanofabrication and micromachining technologies the last decade has seen a fast progress and proliferation of both commercial and custom-made sample holders (see e.g. [14] for the latter) for in-situ TEM that are compatible with the sample stages of the main TEM manufacturers and hence usable by a large community. These holders allow the application of, e.g., temperature, currents, electric and magnetic fields, light, strain, liquids, and gases. While all these capabilities may be fruitfully combined with EH to study transient and response phenomena (e.g., magnetostriction, magnetoionics) particularly strong impact may be expected from advances concerning the application of magnetic vector-fields, cryogenic temperatures, and sample rotation around three independent axes, in particular if these are combined. The reason is the rich phase diagram and phenomenology of mesoscale frustrated 3D magnetic configurations, such as magnetic skyrmions, which become accessible by such in-situ TEM studies. Magnetic vector-field TEM holders, which incorporate a beam deflection correction, are currently under development by various labs as are continuous flow liquid Helium sample holders, which facilitate long-term cooling and counter-heating in wide temperature range starting from approximately 5 K. Here the cooling would also mitigate some beam damage effects (i.e., radiolysis) hampering the study of beam sensitive magnetic materials. Three-axis 360° rotation holders as required for optimal magnetic vector-field tomography that require redesign of the objective lens in order to gain more space around the sample are also in their early development stages.

A second related development focuses on a better integration of EH with other experiments in the sense that all measurements are carried out at the same TEM sample, preferably in-situ but also ex-situ. Notable example for in-situ experiments is the combination of EH with current and entropy transport measurements employing modified electric biasing holders [15]. Such experiments allow a direct correlation of observed magnetic field configurations with longitudinal and transverse magnetotransport and hence the zoo of Hall and Nernst effects. Possible ex-situ registered measurements pertain, for instance, to XMCD based imaging methods facilitating the mapping of larger sample regions.

Significant improvements of the time-resolution of EH are closely linked to the ongoing development and proliferation of a variety of ultrafast TEM (UTEM) techniques, e.g., employing laser-pulsed electron guns, ultrafast beam deflection or very fast detectors in combination with ultrafast sample manipulation, e.g., via short current pules For example, the rotation of a magnetic vortex state in a thin permalloy film has been recently imaged using laser-pulsed ultrafast LTEM and a radio-frequency biasing sample holder, which has been synchronized with the gun [16]. Emerging ultrafast cavity and MEMS blankers fitted in existing TEM columns will further broaden the accessibility of such experiments operating in the sub-ns time resolution domain.

Although EH has been proven a useful technique for resolving magnetic field configuration on the nanoscale it typically requires comparison to simulations (e.g., micromagnetic modelling) to reveal the magnetization texture or structure underlying the magnetic flux density. The foremost reason is that EH is blind with respect to magnetic field strength $H$, corresponding to the irrotational part of the magnetization in magnetostatics. Consequently, to reveal the whole magnetization distribution, $H$ has to be provided by additional means, such as simulations. Therefore, ongoing research efforts elaborate on combining EH, micromagnetic (or other) modelling, and machine learning to retrieve the full magnetization, particular micromagnetic energy terms or characteristic magnetic parameters such as the exchange stiffness.

**Current and Future Use of the Technique for Material Science**

EH will remain a mainstay in the study of nanoscale magnetic field configurations determined by the competition of symmetric and antisymmetric exchange, magnetocrystalline anisotropy, Zeeman energy and dipolar interactions. Here the previously described advances will allow a significant extension of the accessible phase diagram spanned by temperature, external magnetic field, etc., particularly benefitting the study of the large class of anisotropic magnets, low temperature magnetic phases, multiferroics and topological magnetic materials. Improved vector field reconstruction incorporating simulations will address a larger class of topological non-trivial magnetic solitons, such as skyrmion tubes, braids [17], chiral bobbers [18] and Bloch points [19], including their topological charges, notably those of Néel-type exhibiting significant magnetic charges and hence field strengths $H$. Improved VFET will also help in revealing the impact of effective magnetic interactions such as anisotropies and antisymmetric exchange emerging in curved magnetic thin films [20]. Combination of EH and transport measurements, i.e., the measurement of longitudinal and transversal transport on well-determined magnetization structures, has the potential to shed new light on the origins and amplitude of anomalous and topological magnetic and thermal Hall effects. The combination of cryogenic EH and electron tomography as well as other TEM techniques such as spectroscopy may also spark a revival of high-resolution studies of magnetic field vortices in type-II superconductors such as high $T_c$ oxides.

Further advances in signal and spatial resolution bear the prospect of studying 2D magnetic materials as well as magnetic configurations emerging at surfaces and interfaces, e.g., in complex oxides, where symmetry breaking and a stronger localization of electrons can lead to an increased tendency for magnetic order, strong uniaxial anisotropies as well as antisymmetric exchange, resulting in a wealth of complex magnetic states. Further resolution improvement is also instrumental for the application of EH techniques on antiferromagnets facilitating microscopic insights into antiferromagnetic order and field distribution. That notably includes characterization of uncompensated spins at lattice defects or interfaces. A bit further down the road waits the detection of magnetic monopoles in spin ice materials, e.g., of the pyrochlore class.

Finally, development of ultrafast magnetic imaging (both LTEM and EH) will open the door to the quantitative study of transient and non-equilibrium magnetic states and hence the wide field of spintronics. That includes the characterization of the motion and pinning of magnetic domain walls in nanostructures, such as Bloch points or transverse vortex domain walls, of various types of skyrmions, by also spin thermalization at magnetic phase transitions, etc.

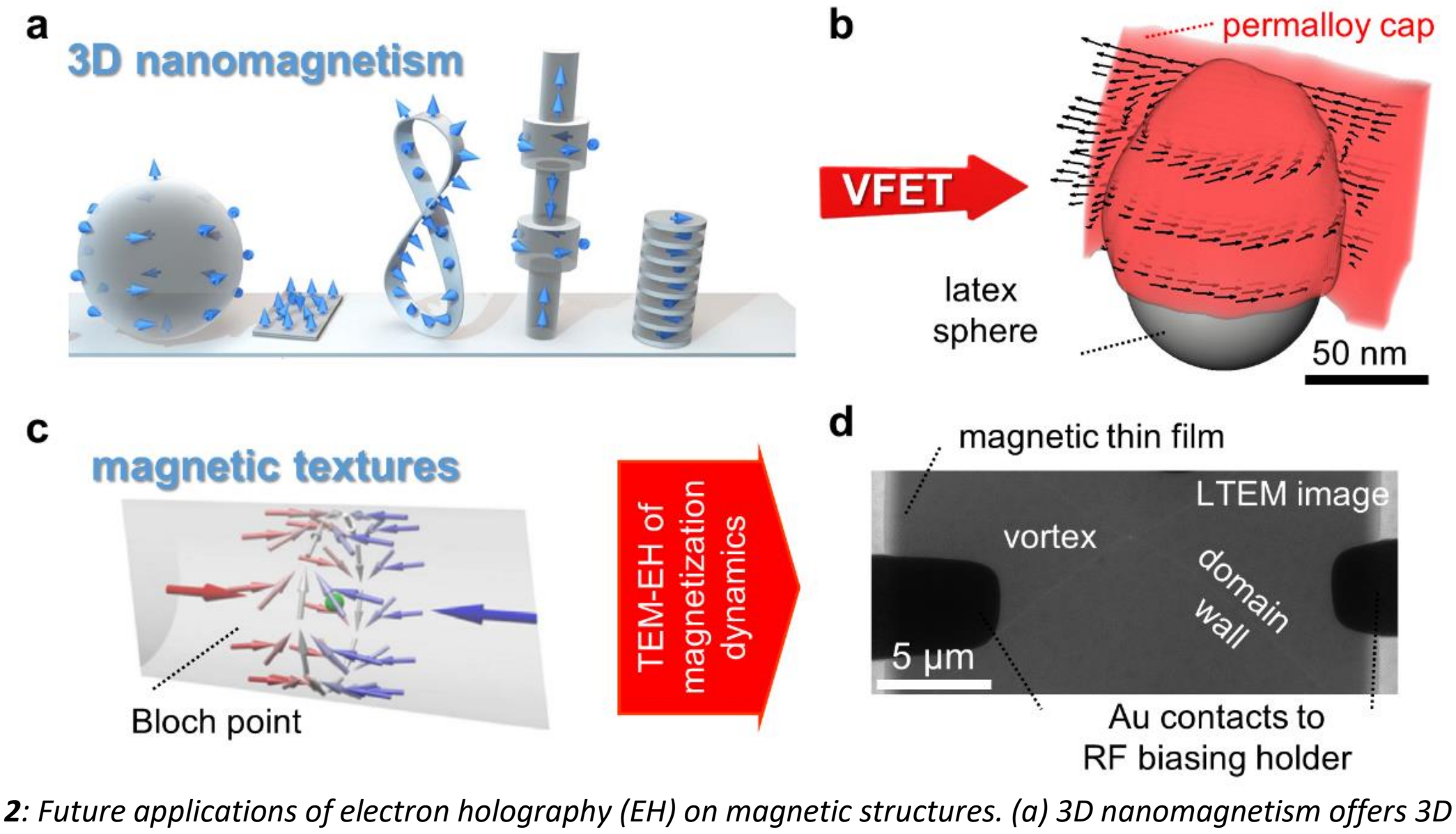

***Figure 2****: Future applications of electron holography (EH) on magnetic structures. (a) 3D nanomagnetism offers 3D magnetic spin configurations that can be reconstructed by 3D EH, i.e., vector-field electron tomography (VFET). (b) shows possible realisations of 3D nanomagnets including a reconstruction of curvilinear magnetism by VFET. The permalloy cap (red) is segmented from an experimental electron tomogram, from which the magnetization (black arrow plot) is simulated (see Ref.* [20] *for the details). (c) Simulated magnetic spin configuration of a Bloch point domain wall which can reach velocities >600 m/s driven by spin-transfer torques in cylindrical nanowires. (d) Ultrafast sample stimulation setup for measuring velocities of domain walls. (a) and (c) are reproduced with permission under CC4.0 license from Ref.* [19].

**Acknowledgements**

*The authors acknowledge financial support of the Deutsche Forschungsgemeinschaft (DFG) by the Collaborative Research Center SFB 1143 (project-id 247310070).*